\documentclass[prb,superscriptaddress,longbibliography,twocolumn]{revtex4-2}
\usepackage{graphics,graphicx, array, afterpage}
\usepackage{subfigure}
\usepackage{float}
\usepackage{qcircuit,amsmath}
\usepackage{grffile}
\usepackage[export]{adjustbox}
\usepackage{lineno}
\usepackage{xcolor}
\usepackage{longtable}
\usepackage{braket}
\usepackage{array}
\usepackage{multirow}
\usepackage{enumerate}
\usepackage{amsmath}
\usepackage{amssymb}
\usepackage{esint}
\usepackage{amsthm}
\usepackage{times}
\usepackage{multirow}

\usepackage[colorlinks, citecolor=blue]{hyperref}
\hypersetup{linkcolor=magenta,urlcolor=blue,citecolor=blue,pdfstartview={FitH},urlcolor=blue}

\begin{document}

\title{Steady-state topological order}% Force line breaks with \\

\author{Xu-Dong Dai}
\thanks{These authors contributed equally to this work.}
 \affiliation{ Institute for
Advanced Study, Tsinghua University, Beijing,  100084, China }

\author{Zijian Wang}
\thanks{These authors contributed equally to this work.}
 \affiliation{ Institute for
Advanced Study, Tsinghua University, Beijing,  100084, China }

\author{He-Ran Wang}
 \affiliation{ Institute for
Advanced Study, Tsinghua University, Beijing,  100084, China }

\author{Zhong Wang} 
\altaffiliation{ wangzhongemail@tsinghua.edu.cn }
\affiliation{ Institute for
Advanced Study, Tsinghua University, Beijing,  100084, China }

\begin{abstract}

We investigate a generalization of topological order from closed systems to open systems, for which the steady states take the place of ground states. We construct typical lattice models with steady-state topological order, and characterize them by complementary approaches based on topological degeneracy of steady states, topological entropy, and dissipative gauge theory. Whereas the (Liouvillian) level splitting between topologically degenerate steady states is exponentially small with respect to the system size, the Liouvillian gap between the steady states and the rest of the spectrum decays algebraically as the system size grows, and closes in the thermodynamic limit. It is shown that steady-state topological order remains definable in the presence of (Liouvillian) gapless modes. The topological phase transition to the trivial phase, where the topological degeneracy is lifted, is accompanied by gapping out the gapless modes. Our work offers a toolbox for investigating open-system topology of steady states.

\end{abstract}
\maketitle

\tableofcontents

\section{Introduction}\label{sec:intro}

The exploration of new phases beyond the Landau paradigm has been a significant theme in condensed matter physics. Among other progresses, quantum phases with topological order have been extensively explored \cite{wen1990topological,wen1990ground,wen2017colloquium}. Topological order in closed systems can be characterized by a number of complementary features, including robust topological degeneracy of ground states \cite{wen1990ground}, the emergence of deconfined gauge fields \cite{trebst2007breakdown,hamma2008adiabatic,tupitsyn2010topological}, and topological entanglement entropy (TEE) \cite{kitaev2006topological,levin2006detecting}. Apart from fundamental significance, topologically ordered phases also serve as promising candidates for quantum memory and quantum computing \cite{kitaev2003fault}. Nevertheless, realistic quantum systems are always coupled to the environment, and this inevitable coupling may undermine the topological order, thus hindering the self-correcting mechanism \cite{dennis2002topological,fan2023diagnostics,bao2023mixed,lee2023quantum,wang2023intrinsic}.

\begin{figure}[htb]
 \centering
\includegraphics[width=1.0\linewidth]{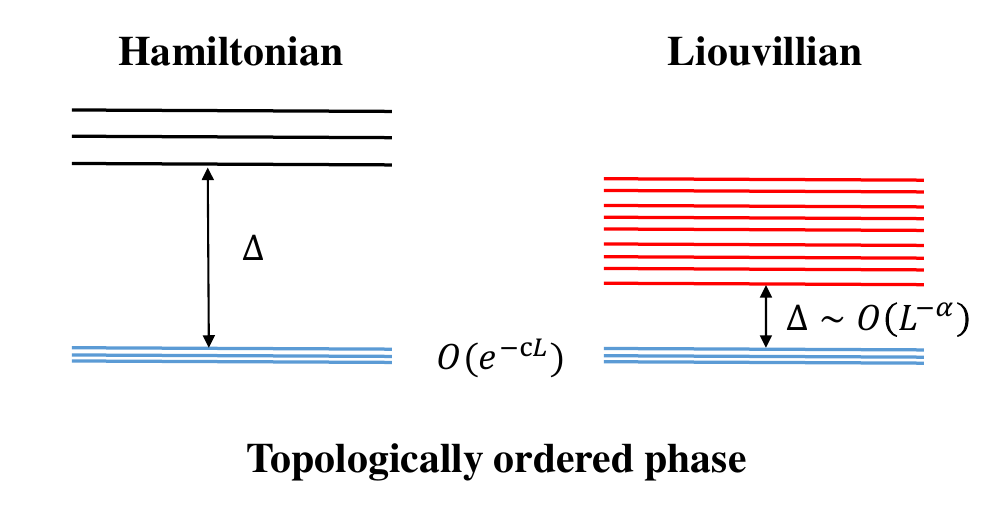}
\caption{Gap in the topologically ordered phase. Left panel: In the closed systems, the energy splitting of the degenerate ground states (blue lines) is of $O(e^{-cL})(c>0)$ order and there is a finite gap between the degenerate ground states (blue lines) and excited states (dark lines). Right panel: In open quantum systems, the (Liouvillian spectrum) splitting of degenerate steady states (blue lines) is of $O(e^{-cL})(c>0)$ order. Above the topologically degenerate steady steady states, the Liouvillian gap scales as $\Delta\sim L^{-\alpha}$. Here $L$ is the system size and $\alpha$ is a certain exponent.}
\label{fig:degenerate_gapless}
\end{figure}

Recently, the interplay between strong correlations and quantum dissipative effects has drawn considerable interest \cite{kraus2008preparation,diehl2010dissipation, diehl2011topology, kastoryano2011dissipative, reiter2016scalable}. Contrary to the conventional viewpoint that dissipation destroys quantum coherence, exotic phases of matter may emerge from the interplay between dissipative dynamics and quantum entanglement. One of the most intriguing questions is whether new phases with topological order can be realized in dissipative systems. For such systems, the steady states play a vital role as all initial states evolve into them under long-time evolution, and therefore it is natural to investigate the topological properties of these steady states. From this viewpoint, in a recent Letter (Ref. \cite{wang2023topologically}), we have constructed models with topologically ordered steady states, which exhibit dissipative topological order (DTO). In the present paper, we systematically characterize DTO by complementary features of steady states, including topological degeneracy, quantized topological entropy, and a dissipative deconfined gauge field. Our results highlight a significant difference between the ground-state topological order in closed systems and steady-state topological order in open systems. Whereas a finite energy gap above the degenerate ground states is a prerequisite for the definition of ground-state topological order, we show that (Liouvillian) gapless modes above the degenerate steady states should be allowed for steady-state topological order (See Fig. \ref{fig:degenerate_gapless} for an illustration). Despite the vanishing of Liouvillian gap in the thermodynamic limit, the topological degeneracy of steady states remain definable via the size dependence. In the topologically ordered phase, the splitting between topologically degenerate steady states is exponentially small with respect to the system size, while the Liouvillian gap between the steady-state subspace and the rest of states decays algebraically as the system size grows \footnote{Here we make the implicit assumption that that the system has translation symmetry, which excludes special cases with non-Hermitian skin effect, where there can be a finite Liouvillian gap even if the relaxation time is divergent. In other words, $\alpha$ in Fig.\ref{fig:degenerate_gapless} can be zero in such special cases.}. The topologically trivial phases without steady-state topological degeneracy, on the other hand, have a finite Liouvillian gap in the thermodynamic limit.  We also show that the steady-state topological order manifests itself in the algebraically slow relaxation of the system. Moreover, the steady-state topological order has a subtle dependence on spatial dimensions. Specifically, we find that the topological degeneracy of steady states is fragile in 2d, while it is robust in 3d and higher dimensions.
 
The paper is structured as follows. In Section \ref{sec:review}, we present a brief review of the toric code model and introduce the formalism for open quantum systems. Subsequently, in Section \ref{sec:2dmodel},  we construct two Liouvillian models, designed to realize the topological degeneracy of steady states in two dimensions. We obtain the exact form of topologically degenerate steady states on 2-torus and analyze their fragility under local perturbations. Utilizing the exact form of steady states, we calculate the topological entropy via two well-established schemes, the Levin-Wen's scheme \cite{levin2006detecting} and the Kitaev-Preskill's scheme \cite{kitaev2006topological}. In our Liouvillian model, the former scheme correctly produces a quantized topological entropy in the topologically ordered phase, and zero value in the trivial phase. In comparison, the latter scheme suffers from an ambiguity and exhibits non-universal behavior. Thus, we conclude that the Levin-Wen's scheme is more suitable for extracting topological entropy in open quantum systems. In Section \ref{sec:3dmodel}, we generalize our two models to three dimensions and obtain the topologically ordered steady states on 3-torus. In contrast to 2d case, we find that robust topological degeneracy in open quantum systems can only be realized in three or higher dimensions. Furthermore, our model can be viewed as $Z_{2}$ gauge theory, which unveils a deconfinement-confinement phase transition. In Section \ref{sec:relaxation}, we show that the topological degeneracy of steady states is always accompanied by the long relaxation time of Liouvillian dynamics, which typically implies the vanishing of the Liouvillian gap. We then elucidate the definition of topological degeneracy for such gapless Liouvillians. 

\section{Review on toric code and Lindblad master equation}\label{sec:review}
The toric code model, initially proposed by A. Kitaev as a candidate for fault-tolerant quantum computation \cite{kitaev2003fault}, has been widely recognized as a prototype model for investigating topological order. Despite its simplicity, the toric code model has fascinating properties including topological degeneracy of ground states, anyon excitations, and robustness against local perturbations. We use the toric code model as an example to introduce important universal features of topological order. We begin by presenting an overview of the toric code model on 2-torus and 3-torus. Subsequently, we introduce the framework of open quantum systems. 

\subsection{Toric code on 2-torus}\label{sec:2torus}
\begin{figure}[htb]
\centering
\includegraphics[width=0.9\linewidth]{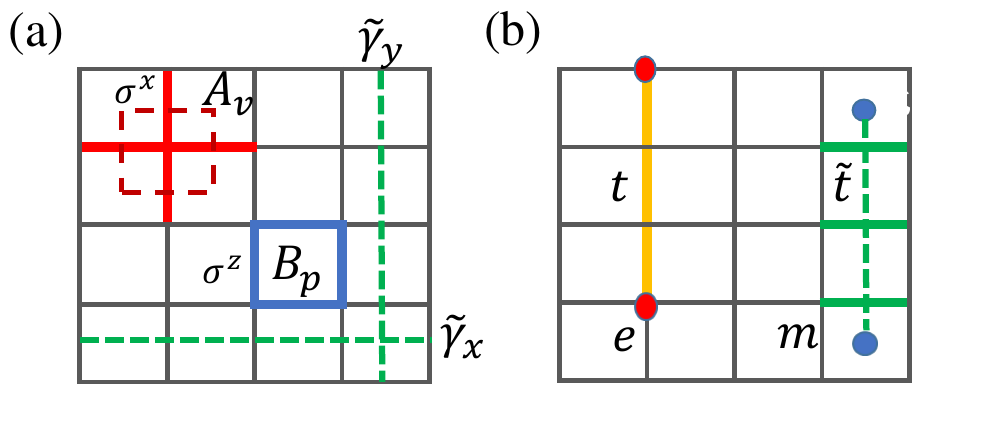}
\caption{(a) $A_v$ flips all the spins on the links (red solid bond) attached to the vertex $v$, which form a closed loop (red dashed line) on the dual lattice. $B_{p}$ is the product of $\sigma^{z}$ (blue solid bond) around a plaquette. $\tilde{\gamma}_{x}$ and $\tilde{\gamma}_{y}$ are non-contractible loops (green dashed line) on the dual lattice. (b) Particle-like $e$ excitations on vertices (red solid circle) with $A_v=-1$ and they are connected by a string $t$ (yellow solid line). $m$ excitations on plaquette (blue solid circle) with $B_{p}=-1$ and they are connected by a string $\tilde{t}$ (green dashed line) on the dual lattice.}
\label{fig:2dexcitation1}
\end{figure}

The Hamiltonian of the toric code model is
\begin{equation}
\begin{aligned}
H=&-\sum_{v}A_v-\sum_{p}B_p;\\ 
A_{v}=&\prod_{l|v\in \partial l }\sigma^x_l,\ B_p=\prod_{l|l\in \partial p}\sigma^z_l.
\label{toric code}
\end{aligned}
\end{equation}
Here, Pauli matrices act on the spin-$1/2$ degrees of freedom, which are located on the links of the 2d periodic square lattice.  $A_{v}$ and $B_p$ are called the vertex and plaquette operators respectively (See Fig. \ref{fig:2dexcitation1}(a) for an illustration). Since the above operators commute with each other, i.e., $[A_{v},B_{p}]=0$, $[A_{v},A_{v'}]=0$, $[B_{p},B_{p'}]=0$, the model can be exactly solved by finding common eigenstates of all those operators.
%this model is exactly solvable. 
When we work in $\sigma^{z}$ eigen-basis,  $B_{p}$ is diagonal. While $A_{v}$ can flip spins around the vertex $v$, it leaves all the $B_{p}$ unchanged. For later convenience, we define the stabilizer group generated by all the vertex operators $A_{v}$ \cite{hamma2005ground}:
\begin{equation}
\begin{aligned}
G=\{g\ |\ g=\prod_{\{v\}}A_v,\ \{v\}=\{v_{i},v_{j},\cdots\} \},\label{eq:groupG}
\end{aligned}
\end{equation}
where $\{v\}$ is a set of  vertices and $G$ includes all possible products of vertex operators. The ground states of toric code on 2-torus are four-fold degenerate. For any one of the ground states $|\psi\rangle$, we have $A_{v}|\psi\rangle=|\psi\rangle$ and $B_{p}|\psi\rangle=|\psi\rangle$ for all vertices and plaquettes. The condition $B_{p}=1$ means that links with $\sigma^z=-1$ flipped by $A_{v}$ form closed loops on the dual lattice. The ground state is just an equal-weight superposition of all possible dual loop configurations in each topological sector which is classified by non-contractible Wilson loops  $W_{x}$ and $W_{y}$. The four degenerate ground states can be written as
\begin{equation}
 \begin{aligned}
|\psi^{\mu \nu}\rangle &= W_{x}^{\mu}W_{y}^{\nu}\frac{1}{\sqrt{|G|}}\sum_{g\in G}g|\Uparrow\rangle;\quad \mu,\nu = 0,1\\
 W_{x}&=\prod_{i\in\tilde{\gamma}_x}\sigma_i^{x}, \quad  W_{y}=\prod_{j\in\tilde{\gamma}_y}\sigma_j^{x},\label{eq:2dgs}
 \end{aligned}
\end{equation}
where $|G|$ is the size of group $G$, and $|\Uparrow\rangle=|\uparrow\cdots\uparrow\rangle$ is defined as the state with $\sigma^z=1$ for all spins. $\tilde{\gamma}_{x}$ $(\tilde{\gamma}_{y})$ is a non-contractible loop on the dual lattice in the $x$ $(y)$ direction [Fig. \ref{fig:2dexcitation1}(a)]. Such non-contractible loops can only exist on a manifold with a nonzero genus. The ground state degeneracy depends on the topology of the manifold and is known as the topological degeneracy. Remarkably, such topological degeneracy is robust against arbitrary local perturbations.

In the 2d case, there are two kinds of particle excitations: $e$ and $m$ particles, corresponding to vertices with $A_v=-1$ and plaquettes with $B_p=-1$ respectively. They are topological excitations in the sense that a single $m$ ($e$) particle cannot be created by any local operators; they can only be created or annihilated in pairs via the string operator $S^{x}_{\tilde{t}}=\prod_{l\in \tilde{t}}\sigma^{x}_{l}$ ($S^{z}_{t}=\prod_{l\in t}\sigma^{z}_{l}$),  as shown in Fig. \ref{fig:2dexcitation1}(b). The self statistics of $m$ and $e$ particles is bosonic, but they have nontrivial mutual statistics: When moving one $e$ particle around an $m$ particle for one circle, an extra phase factor $e^{i\pi}$ is attached to the wavefunction. 

\subsection{Toric code on 3-torus }\label{sec:3torus}
\begin{figure}[htb]
\centering
\includegraphics[width=0.95\linewidth]{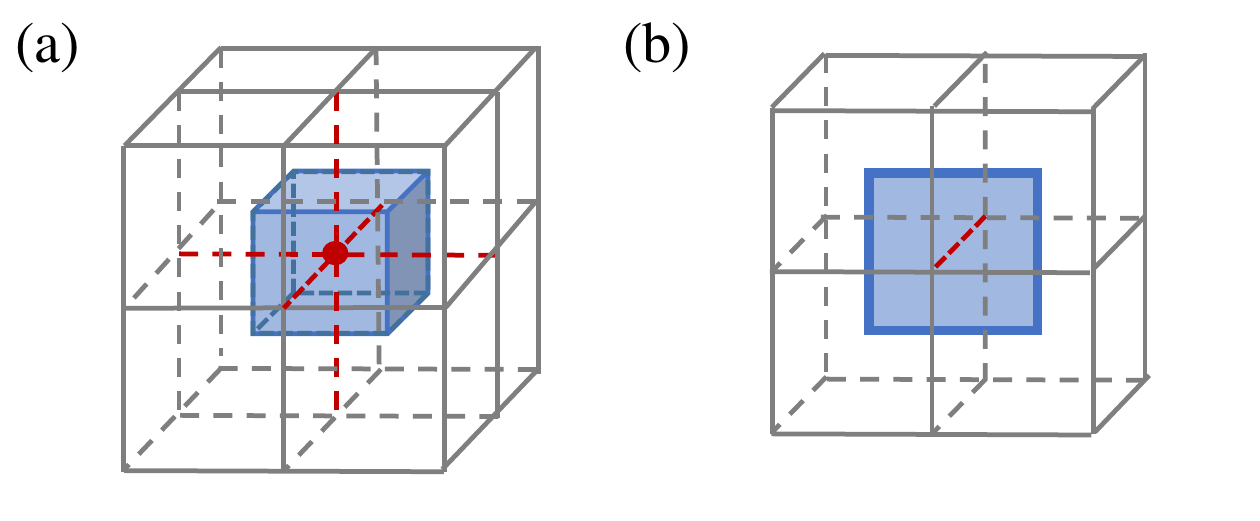}
\caption{(a) $A_v$ flips all the spins on the links (red dashed line) attached to the vertex $v$ (red solid circle). On the dual lattice, it acts on the six faces of a cubic (blue cubic) which form a closed membrane. Products of $A_v$ can create larger closed membranes. (b) Loop-like excitations $m$ at the center of plaquettes with $B_p=-1$. Flipping one spin (red dashed line) would change the sign of the four plaquette operators $B_p$ around it, which also form a plaquette (blue solid line) on the dual lattice. Generally, the $m$ excitations form loops on the dual lattice.}
\label{fig:3dexcitation1}
\end{figure}

In this section, we briefly review the generalization of Kitaev's toric code model on the 3d periodic cubic lattice \cite{hamma2005string,nussinov2008autocorrelations,reiss2019quantum}. The Hamiltonian takes the same form as Eq. \eqref{toric code}. In 3d, the plaquette operator is the same as that in the 2d case, consisting of Pauli $z$ operators on four links. The vertex operator now is the product of Pauli $x$ operators on six links sharing the same vertex. All the terms in the Hamiltonian mutually commute, and the ground states satisfy $A_v=1, B_p=1$. To construct the ground states, we start with the configuration $|\Uparrow\rangle$ where $B_p=1$ for all plaquettes, and then project it to the gauge sector with $A_v=1$ by applying $\prod_{v}\frac{1+A_v}{\sqrt{2}}$. This state can be easily visualized on the dual lattice. Since the operator $A_v$ flips all the links around the vertex $v$, on the dual lattice, it can be viewed as flipping all the faces of the dual cube [see Fig. \ref{fig:3dexcitation1}(a)], which can be considered as an elementary closed membrane. Therefore, the ground state is an equal-weight superposition of closed-membrane states on the dual lattice: $|\psi\rangle\sim \sum_{C}|C\rangle$, where $C$ represents the configuration of the contractible closed membrane. Generalizing the definition of the stabilizer group $G$ in Eq. \eqref{eq:groupG}  to 3d, the closed membrane state has the following form:
\begin{equation}
    |C\rangle=\prod_{\{v\}}A_{v}|\uparrow\uparrow\cdots\uparrow\rangle=g|\Uparrow\rangle.\label{eq:closedmembrane}
\end{equation}
Here $\{v\}$ denotes a set of vertices and $|\Uparrow\rangle$ is the state with all spins up. The explicit expression of the ground state is
\begin{equation}
\begin{aligned}
|\psi^{\{\mu_{i}\}}\rangle 
&= V_{xy}^{\mu_{1}} V_{yz}^{\mu_{2}}V_{zx}^{\mu_{3}}\prod_{v}\frac{1+A_v}{\sqrt{2}}|\uparrow\uparrow\cdots\uparrow\rangle\\
&=V_{xy}^{\mu_{1}} V_{yz}^{\mu_{2}}V_{zx}^{\mu_{3}}\sum_{g\in G}\frac{1}{\sqrt{|G|}}g|\Uparrow\rangle\\
&=V_{xy}^{\mu_{1}} V_{yz}^{\mu_{2}}V_{zx}^{\mu_{3}}\sum_{C}\frac{1}{\sqrt{2^{n}}}|C\rangle,\label{eq:groundstate}
\end{aligned}
\end{equation}
where $n$ is the number of vertices and $|G|=2^{n}$. $V_{xy}=\prod_{l\perp xy}\sigma^x_l$ creates a non-contractible membrane in the $xy$ plane ($V_{yz}$ and $V_{zx}$ are similar.). $\{\mu_{1},\mu_{2},\mu_{3}=0,1\}$ is the parity of the non-contractible membrane, which gives the 8-fold degeneracy. 
In 3d, the $e$ particle excitations are similar to the 2d case, that is, they are hardcore bosonic particles located at the end of the open strings [see Fig. \ref{fig:2dexcitation1}(b)]. However, the $m$ excitations (still defined as plaquettes with $B_p=-1$) are no longer point-like particles, but rather form loops along the boundary of the open membrane on the dual lattice, referred as the loop exciations. The excitation energy for a $m$-loop is proportional to the loop length. Also, one can never create a single $e$ or $m$ excitation. By applying a single flip of $\sigma^{z}$, one can create $m$ excitations in quadruplets which give the smallest loop on the dual lattice [see Fig. \ref{fig:3dexcitation1}(b)].

\subsection{Lindblad master equation}\label{sec:Lindblad}
Generally, quantum systems are inevitably coupled to the environment, which decoheres the pure quantum state to a mixed-state density matrix. %and we should tackle the evolution of the density matrix instead of the pure state. 
With Markovian approximation, the dynamics of the density matrix is described by the Lindblad master equation \cite{lindblad1976generators}:
\begin{equation}
    \frac{d\rho}{dt} =\mathcal{L}[\rho]= -i[H,\rho]+\sum_{a}\left(L_{a}\rho L^{\dagger}_{a}-\frac{1}{2}\{ L^{\dagger}_{a}L_{a},\rho\}\right),\label{eq:Lindblad}
\end{equation}
where $L_{a}$ is the quantum jump operator which gives the dissipative effect, $a$ denotes a specific quantum channel, and $\mathcal{L}$ is called the Liouvillian superoperator. Owing to the dissipative nature, all eigenstates of $\mathcal{L}$ have eigenvalues with a non-positive real part. Specifically, there must be at least one steady state $\rho_{ss}$ satisfying
\begin{equation}
    \frac{d\rho_{ss}}{dt}=\mathcal{L}[\rho_{ss}]=0.
\end{equation}
As the name suggests, any initial state will evolve to a steady state after long-time Liouvillian dynamics. In the following, we are crucially curious about the Liouvllians with more than one steady state, while the degeneracy is protected by topology. The exact meaning of topological degeneracy in open quantum systems will be specified later.

%for any initial state, after the long-time evolution of $\mathcal{L}$, it would evolve into a steady state $\rho_{ss}$, which satisfies
%\begin{equation}
 %  \frac{d\rho_{ss}}{dt}=\mathcal{L}[\rho_{ss}]=0.
%\end{equation}

\section{Open systems with topologically degenerate steady states on 2-torus}\label{sec:2dmodel}

In closed quantum systems, the ground state of Hamiltonian is important because it dominates the low-temperature physics. As a counterpart, in open quantum systems, the steady state plays an equally significant role because it dominates the physics after long-time evolution. In the following sections, we focus on constructing Liouvillians to realize topological degeneracy in the steady-state subspace, and explore the emergence of topological order in dissipative systems on the two-dimensional manifold.

\subsection{Model-1}\label{sec:model1}

\subsubsection{Topologically degenerate steady states}\label{sec:2dss}
As the first example, we design a purely dissipative model with $H=0$ and the following three kinds of quantum jump operators:
\begin{equation}
\begin{aligned}
%L_{l}&=\sigma^{x}_l\left(1-\frac{1}{2}\sum_{p:l\in \partial p}{B_p}\right)\left(1-\frac{1}{4}\sum_{p:l\in \partial p}{B_p}\right), \ {B_p}=\prod_{l\in\partial p }\sigma^{z}_{l}\\
L_{m,l}&=\sigma^{x}_lP(\sum_{p|l\in \partial p}B_p),\\
L_{z,l}&=\sqrt{\kappa_{z}}\sigma^{z}_{l},\\
 L_{v}&=\sqrt{\kappa_{v}}A_v,\label{eq:2dclassical}
\end{aligned}
\end{equation}
where $\kappa_{z}$ ($\kappa_{v}$) is the dissipation strength being uniform on all links (vertices).   $\sum_{p|l\in \partial p}B_p$ is the sum of two plaquette operators sharing the same link $l$. The operator $P(\sum_{p|l\in \partial p}B_p)$ is diagonal on the $\sigma^z$ eigen-basis. It depends on $\sum_{p|l\in \partial p}B_p$ as: 
\begin{equation}
P(x)\equiv\left\{
\begin{aligned} 
0,\quad &x>0;\\
q_{1},\quad &x<0;\\ 
q_{2},\quad &x=0. 
\end{aligned}
\right.   \label{eq:projection} 
\end{equation}
Here, $q_{1}$ and $q_{2}$ are some arbitrary positive numbers that give the relative amplitudes of the corresponding spin-flipping process. As we will show in the next section, the qualitative behavior of the Liouvillian does not depend on the specific values of $q_{1,2}$. For simplicity, we take $q_{1}=1$ and $q_{2}=1/\sqrt{2}$ \cite{glauber1963time}.
\begin{figure}[htb]
\centering
\includegraphics[width=1.0\linewidth]{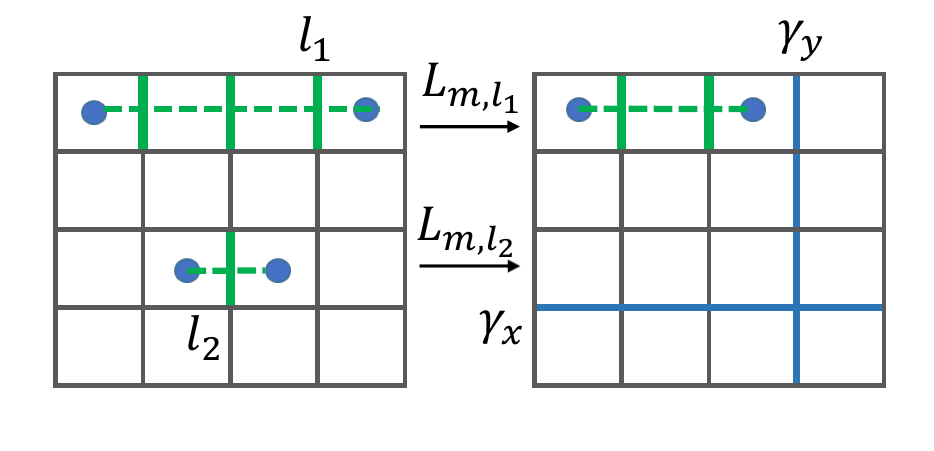}
\caption{ Action of $L_{m,l}$. Acting on the link $l_{1}$ $(l_{2})$, $L_{m,l}$ can move (annilihate) $m$ particles. $\gamma_{x}$ ($\gamma_{y}$) is the non-contractible 't Hooft loop along the $x$ (y) direction (blue line on the right panel).}
\label{fig:2dexcitation2}
\end{figure}

In Fig. \ref{fig:2dexcitation2}, we can see that
for any link $l$ there are two adjacent plaquettes, and the sum of $B_p$ on these two plaquettes can take three possible values: $2,0,-2$. We design $L_{m,l}$ to ensure that only the spin on link $l$ with $\sum_{p|l\in\partial p}B_{p}\leq 0$ can be flipped, while the local states with positive $\sum_{p|l\in\partial p}B_{p}$ are annihilated. Recall that negative $B_p$ correspond to $m$-particle excitations, therefore  $L_{m,l}$ can only move or annihilate $m$ particles, but never create them.
%which makes the number of  $m$ particles non-increasing because $L_{m,l}$ can only move or annihilate $m$ particles.
Due to this property, all kinds of closed loop (on the dual lattice) states $g|\Uparrow\rangle\langle \Uparrow|g'$ are commonly annihilated by all $L_{m,l}$, i.e., $L_{m,l}(g|\Uparrow\rangle)=0$, since there are no $m$ particles in such states.
%This constraint makes all kinds of dual loop states $g|\Uparrow\rangle\langle \Uparrow|g'$ the dark states of $L_{m,l}$, i.e., $L_{m,l}g|\Uparrow\rangle=0$. 
Then, taking the dephasing operators $L_{z,l}$ into consideration, all off-diagonal elements are damped, and only diagonal terms $g|\Uparrow\rangle\langle \Uparrow|g$  persist. Finally, $L_{v}$ mixes all closed loop states, leading to the four-fold degenerate steady states on the 2-torus:
\begin{equation}
\rho^{\mu\nu}=\frac{1}{|G|}\sum_{g\in G}W_{x}^{\mu}W_{y}^{\nu}g|\Uparrow\rangle\langle \Uparrow|gW_{y}^{\nu}W_{x}^{\mu};\quad \mu,\nu=0,1,\label{eq:2dss0}
\end{equation}
where $W_{x}$ ($W_{y}$) is the non-contractible loop operator defined in Eq. \eqref{eq:2dgs}. 
%Like the degenerate ground states in Eq. \eqref{eq:2dgs}, since there are four different topological sectors of loop states that are distinguished by non-contractible loops on the torus, we have four different steady states $\rho^{00}, \rho^{01}, \rho^{10}$, and $\rho^{11}$. 
We denote the full Liovillian superoperator with the aforementioned three kinds of quantum jump operators ($L_{m,l}$, $L_{z,l}$, and $L_{v}$) as $\mathcal{L}^{0}$. It is straightforward to check that $\mathcal{L}^{0}[\rho^{\mu\nu}]=0$.      

Due to the dephasing operators $L_{z,l}$, all off-diagonal elements of the density matrix vanish in the long-time evolution. Therefore, in the long-time limit, it is sufficient to restrict the density matrix to the diagonal subspace of $\sigma^z_v$, i.e., $\rho=\sum_{\alpha}p(\alpha)|\alpha\rangle\langle \alpha|$, where $\alpha$ stands for spin configurations in the $\sigma^z$ eigen-basis.
In the diagonal subspace, we can further map the density matrix to a vector:  %all the steady states are diagonal in the $\sigma^z$ basis. This means that all the off-diagonal elements of the density matrix will vanish in the long-time evolution and only the diagonal part will evolve into the steady state subspace. Since only diagonal elements contribute to the steady state, we can restrict the density matrix to the diagonal subspace, i.e., $\rho=\sum_{\alpha}p(\alpha)|\alpha\rangle\langle \alpha|$. Then it is straightforward to map the diagonal matrix to a state in vector space.
\begin{equation}
\rho=\sum_{\alpha}p(\alpha)|\alpha\rangle\langle \alpha|\rightarrow|\rho\rangle=\sum_\alpha p(\alpha)|\alpha\rangle,
\label{eq:map1}
\end{equation}
Correspondingly, the Liouvillian superoperator $\mathcal{L}^{0}$ can be mapped to the following operator in the vector space:
\begin{equation}
\Gamma^{0}=\sum_l(\sigma^x_l-1)P^{2}(\sum_{p|l\in \partial p}B_{p})+\kappa_{v}\sum_v(A_v-1).\label{eq:map2}
\end{equation}
%Since in the diagonal space $\sigma^{z}|\alpha\rangle\langle\alpha|\sigma^{z}=|\alpha\rangle\langle\alpha|$, then $L_{z,l}$ cancels.
Consequently, the long-time evolution of $\rho$ obeys the classical Markov dynamics generated by $\Gamma^0$, which satisfies $\sum_\beta\Gamma^0_{\alpha\beta}=0$. That is, the dynamics can be described by the following classical master equation:
\begin{equation}
\frac{d}{dt}p(\alpha)=\sum_{\beta\neq \alpha} (\Gamma^0_{\alpha\beta}p(\beta)-\Gamma^0_{\beta\alpha}p(\alpha)).\label{eq:markov}
\end{equation}
The long-time dynamics is essentially classical, and it is intriguing to see that topological degeneracy also exists in classical dynamics. The steady states $\rho_{ss}$ are mapped to the zero-energy ground states of the non-Hermitian ``Hamiltonian" $-\Gamma^0$: $-\Gamma^{0}|\rho_{ss}\rangle=0$, and the low-lying Liouvillian spectrum is determined by the low-energy spectrum of $-\Gamma^{0}$. We will adopt such quantum-to-classical mapping when studying the properties of steady states under perturbations.

\subsubsection{Fragility of topological degeneracy: a perturbative analysis}\label{sec:2dperturbation}
The model has four-fold degenerate steady states, analogous to the ground states of the toric code model. The steady states are mixed states composed of all kinds of closed-loop configurations created by $A_{v}$ on the dual lattice. However, as we will show in this section through a perturbative analysis, the degeneracy of these steady states is sensitive to local perturbations. 
In Hermitian systems, the topological degeneracy of the ground state is robust since the different degenerate ground states cannot be locally distinguished, and can only mix via highly non-local operations. Using degenerate perturbation theory, one can show that the degeneracy can only be lifted perturbatively up to the order proportional to the linear size of the system, resulting in an exponentially small energy splitting between these states. However, in open quantum systems, this argument may not hold since $\Gamma^{0}$ is non-Hermitian, of which the left and right eigenstates are not Hermitian-conjugate anymore. As demonstrated later, local operations can mix one right steady state with a left steady state in a different topological sector, and thus lift the degeneracy by a finite gap.

Consider the simplest kind of perturbation by adding another quantum jump operator $ L_{x,l}$,
\begin{equation}
 L_{x,l}=\sqrt{h}\sigma^{x}_{l},
\end{equation}
where $h$ is the perturbation parameter. For the effective Markov generator in the diagonal subspace (Eq. \eqref{eq:map2}), this corresponds to the perturbation $\Gamma=\Gamma^0+\delta\Gamma$ with $\delta\Gamma=h\sum_l(\sigma^x_l-1)$. When acting on loop states, $\sigma^{x}_{l}$ flips the spin on link $l$, creating two $m$ particles with $B_{p}=-1$. Then, all closed loops would be opened by $\sigma^x_{l}$ and mixed by $L_{v}=\sqrt{\kappa_{v}}A_{v}$. When the perturbation term creates a pair of $m$ particles, $L_{m,l}$ detects the excitation and moves it to another plaquette.  Therefore, a pair of $m$ particles excited from $\rho^{00}$ might move along the non-contractible loop $\tilde{\gamma}_{x}$ and then annihilate, which makes $\rho^{00}$ evolve into another degenerate steady state $\rho^{10}$. We can roughly see that the four degenerate steady states would mix under the perturbation.  Next, we confirm this statement in the first-order perturbation.% $_{L}\langle \rho^{0,\mu'\nu'}|\delta\Gamma|\rho^{0,\mu\nu}\rangle_{R}$. 

Different from the perturbation theory for the Hermitian Hamiltonian, for the Liouvillian, we have to know the left steady states $\rho^{0}_L$ of the Liouvillian operator $\mathcal{L}^{0}$, which is defined as the steady states of the adjoint Liouvillian operator $\tilde{\mathcal{L}}^{0}$:
\begin{equation}\label{eq:111}
\tilde{\mathcal{L}}^{0}[\rho^{0}_L] \equiv\sum_j L^{\dagger}_j\rho^{0}_L L_j-\frac{1}{2}(L_j^{\dagger}L_j\rho^{0}_L+\rho^{0}_LL_j^{\dagger}L_j)=0.
\end{equation}
Accordingly, we use $\rho^0_R$ to represent the steady states of the Liouvillian operator without perturbations.

Since $L_{z,l}=L_{z,l}^\dagger$, the dephasing effect works as well in $\tilde{\mathcal{L}}^{0}$. In the long-time limit, the off-diagonal elements of $\rho^{0}_{L}$ are all damped out, and we can restrict the evolution into the diagonal subspace. Adopting the quantum-to-classical mapping as Eq. \eqref{eq:map1} and Eq. \eqref{eq:map2}, Eq. \eqref{eq:111} is transformed to
\begin{equation}
(\Gamma^{0})^{\dagger} |\rho^{0}_L\rangle=0.\label{leftss}
\end{equation}
The left steady state can be generated by the adjoint Liouvillian evolution starting from the corresponding right steady state in the same topological sector:
\begin{equation}
|\rho^{0,\mu\nu}_{L}\rangle=\lim_{t\rightarrow\infty}e^{(\Gamma^{0\dagger})t}|\rho^{0,\mu\nu}_R\rangle.
\end{equation}
Therefore, the left steady states also have four-fold degeneracy,
\begin{equation}
\begin{aligned}
|\rho^{0,\mu\nu}_R\rangle &= \frac{1}{|G|}W_{x}^{\mu}W_{y}^{\nu}\sum_{g\in G}g|\Uparrow\rangle,\\
|\rho^{0,\mu\nu}_L\rangle &= W_{x}^{\mu}W_{y}^{\nu}\sum_{g\in G}g(|\Uparrow\rangle + \alpha(m_{2})|m_{2}\rangle +\alpha(m_{4}) |m_{4}\rangle+\ldots),\label{eq:leftss}
\end{aligned}
\end{equation}
where $|m_{2n}\rangle$ represents the states with $2n$ $m$ particles, and $\alpha(m_{2n})$ is the associate coefficient. The left and right steady states are bi-orthogonal:
\begin{equation}
\langle \rho^{0,\mu'\nu'}_L|\rho^{0,\mu\nu}_R\rangle =\delta_{\mu \mu'}\delta_{\nu\nu'}.  
\end{equation}
Here, it is difficult to write down the exact form of the left steady states, but we can extract the coefficients of states $|m_{2}\rangle$ with different open loop configurations, contributing to the first-order perturbation of the steady states.  In contrast to $\Gamma^{0}$ in Eq. \eqref{eq:map2},  $\Gamma^{0\dagger}$ can move and create $m$ particles, hence, the number of $m$ particles is non-decreasing.  Starting from a closed loop state $g|\Uparrow\rangle$,  $\Gamma^{0\dagger}$ can create open loops. Since the left and right steady states are four-fold degenerate, the first-order effective Hamiltonian $\Gamma^{(1)}_{\text{eff}}$ is a $4\times 4$ matrix, with matrix elements $\langle \rho^{0,\mu'\nu'}_{L}|\delta\Gamma|\rho^{0,\mu\nu}_{R}\rangle$. Eigenvalues of $\Gamma^{(1)}_{\text{eff}}$ correspond to the spectrum splitting caused by perturbation. 
For simplicity, we consider the square lattice with $n=L^{2}$ vertices. We find that the largest eigenvalue of $\Gamma^{(1)}_{\text{eff}}$ is exactly zero, consistent with the Liouvillian dynamics that there is at least one steady state with zero eigenvalue. 
The second largest eigenvalue is $-L^{2}h/\log(L)$, indicating that the spectrum splitting is proportional to $h$. 
Therefore, the degeneracy of the steady state is already broken at the first order of perturbation, with a unique steady state. Further details are provided in Appendix \ref{sec:appendix2dpertub}.

We note that $\Gamma^{0}$ is actually gapless, which will shown in Sec. \ref{sec:relaxation}, and therefore $\Gamma^{(1)}_{\text{eff}}$ does not really govern the long-time dynamics. Nevertheless, it is sufficient for understanding the breakdown of topological degeneracy.
 
\subsubsection{Fragility of topological degeneracy: an exact solution of the steady state with $h\neq 0$}\label{sec:2dexactform}
In the previous section, we observed that different from closed systems, the topological degeneracy of steady states in our open quantum model can be lifted by local perturbations at the first order. Therefore, we expect a unique steady state when the spin-flip perturbation is turned on. In this section, we non-perturbatively solve the exact steady state when adding the uniform spin-flip quantum jump operators.

Here, we define a larger stabilizer group to cover all four classes of topologically distinct loops:
\begin{equation}
\begin{aligned}
G' =\{g|g\in G,\ W_{x}G,\ W_{y}G,\ W_{x}W_{y}G\}.
 % W_{x}&=\prod_{i\in\gamma_x}\sigma_i^{z}, \  W_{y}=\prod_{j\in\gamma_y}\sigma_j^{z}.
\end{aligned}
\end{equation}
$W_{x}$ and $W_{y}$, defined in Eq. \eqref{eq:2dgs}, are 't Hooft loop operators along the non-contractible dual loops $\tilde{\gamma}_{x}$ and $\tilde{\gamma}_{y}$ on 2-torus. $G$, defined in Eq. \eqref{eq:groupG}, is the stabilizer group of $A_{v}$.  Due to the periodic boundary condition $\prod_{v}A_{v}=I$, the size of group $G'$ is $|G'|=4|G|=2^{n+1}$. We directly give the form of the steady state as follows, and put the detailed derivation in Appendix \ref{sec:appendix2dss}:
\begin{eqnarray}
\rho_{ss}= \frac{1}{T'}\sum_{k=0}^{\lfloor\frac{n}{2}\rfloor}\beta^{2k}\sum_{g\in G'}\sum_{\{r\}}g|m_{2k}(\{r\})\rangle\langle m_{2k}(\{r\})|g.
\label{eq:2dss1}
\end{eqnarray}
Here, the parameter $\beta=\sqrt{h/(1+h)}$. $\lfloor\frac{n}{2}\rfloor$ denotes the largest integer not exceeding $\frac{n}{2}$, which corresponds to the maximum number of $m$ particle pairs that can be created on $n$ plaquettes. $|m_{2k}(\{r\})\rangle\langle m_{2k}(\{ r\})|$ is a specific spin configuration given the distribution of $m$ particles and $\{r\}=\{r_{1},r_{2}\ldots r_{2k}\}$ are the coordinates of $m$ particles where $B_{r_i}=-1$. $T'$ is the overall normalization constant given by
\begin{equation}
T'=|G'|\sum_{k=0}^{\lfloor\frac{n}{2}\rfloor}\beta^{2k}{n\choose 2k}=4|G|\frac{(1+\beta)^n+(1-\beta)^n}{2}.
\label{eq:normalT}
\end{equation}

In Eq. \eqref{eq:2dss1}, we perform a triple summation. First, for given $k$ pairs of m particles, we need to sum over all possible particle distributions $\{r\}$, and there are ${n \choose 2k}$ different distributions in total. When the positions of particles are fixed, we choose one specific spin configuration to realize this $m$ particle distribution. Any particle is connected to another by a string operator $S_{\tilde{t}}^{x}=\prod_{l\in \tilde{t}}\sigma^{x}_{l}$ and $\tilde{t}$ is an open path connecting two particles [See Fig. \ref{fig:2dexcitation1}.(b)]. Second, by acting $ g\in G'$ on  $|m_{2k}(\{r\})\rangle\langle m_{2k}(\{r\})|$, we can get all other possible spin configurations with the same distribution of $m$ particles. Because $g$ only flips spins around a vertex, this action does not change the value of $B_{p}$. Finally, we sum over all possible number $2k$ of $m$ particle pairs. 

%From the expression of the steady state, we can see that the weight coefficient of the configuration $\beta^{2k}$ only depends on the number of $m$ particles, irrespective of their location or the specific spin configuration. $T'$ is the overall normalization constant and the coefficient $\beta$ depends on the spin-flip parameter,
%\begin{equation}
%\begin{aligned} T'=|G'|\sum_{k=0}^{\lfloor\frac{n}{2}\rfloor}\beta^{2k}{n\choose 2k}&=4|G|\frac{(1+\beta)^n+(1-\beta)^n}{2},\\
%\beta&=\sqrt{\frac{h}{1+h}}.\label{eq:normalT}
%\end{aligned}
%\end{equation}
On a $L_{x}\times L_{y}$ square lattice, there are $n=L_{x}L_{y}$ sites, which means that the number of vertex operator $A_{v}$ and plaquette operator $B_p$ are both $n$. The number of spins on the links is $N=2n$ and the dimension of the Hilbert space is $2^{2n}$. Here in the steady state, we add up all possible spin configurations with any distributions of $m$ particles, and then the number of independent states is $\sum_{k=0}^{\lfloor\frac{n}{2}\rfloor}{n \choose 2k}|G'|=2^{n-1}2^{n+1}=2^{2n}$. Therefore, when adding a spin-flip term, we find that the steady state is a mixture of all possible spin configurations. For $k=0$, $\sum_{g\in G'}g|m_{0}\rangle\langle m_{0}|g$ is just the maximally mixed state of the four degenerate steady states in Eq. \eqref{eq:2dss0}. We can check this solution in two limits. When $\beta=0$ $(h=0)$, without the single spin-flip term $L_{x,l}$, the states with $m$ particles all vanish, and only loop states remain. It is consistent with the former result. 
In the opposite limit $\beta\rightarrow 1$ $(h\rightarrow \infty)$, the steady state becomes the maximally mixed state (the identity $I$) when the perturbation term $L_{x,l}$ dominates. The steady state can also be written in a familiar form of Gibbs state: $\rho_{ss}=\text{exp}(-\sum_{s}B_{p}/T_{\text{eff}})$, with $T_{\text{eff}}=4/\ln\frac{h+1}{h}$ \cite{castelnovo2007finiteT}. We check that the steady state satisfies the detailed balance principle in Appendix \ref{sec:appendix2dss}.

\subsubsection{Topological entropy}\label{sec:2dentropy}
In the three preceding sections, we mainly focus on the topological degeneracy of steady states, and its fragility under perturbation.%degeneracy of our model originating from the topology of the underlying manifold. However, the degeneracy is fragile against local perturbations. 
Topological entanglement entropy (TEE) is another important diagnosis of topological order in closed systems. In this section we aim to answer the following question: Is there any counterpart of TEE for the (mixed) steady states? Despite the lack of quantum entanglement between different regions due to decoherence, we discover that the entropy of the subsystem still contains a topological part when the steady states have topological degeneracy.
Moreover, as already shown in Sec. \ref{sec:2dperturbation}, the topological degeneracy would be instantaneously lifted by perturbation, and we find that the topological entropy also immediately drops to zero. We compare two schemes for computing the topological entropy: the Levin-Wen scheme \cite{levin2006detecting} and the Kitaev-Preskill scheme \cite{kitaev2006topological}. Although the two definitions produce identical results (up to a factor of $2$) for ground states in closed systems, they give qualitatively different outcomes for the steady state of our dissipative model. In the Levin-Wen scheme, the topological entropy drops from $\log2$ to zero immediately when $h\neq0$, as expected. However, in the Kitaev-Preskill scheme, the topological entropy exhibits a non-universal behavior and even depends on the details of tripartition. These findings suggest that Levin-Wen's scheme is better suited for characterizing the topological order of steady states in open quantum systems.

\begin{figure}[htb]
 \centering
\includegraphics[width=0.40\linewidth]{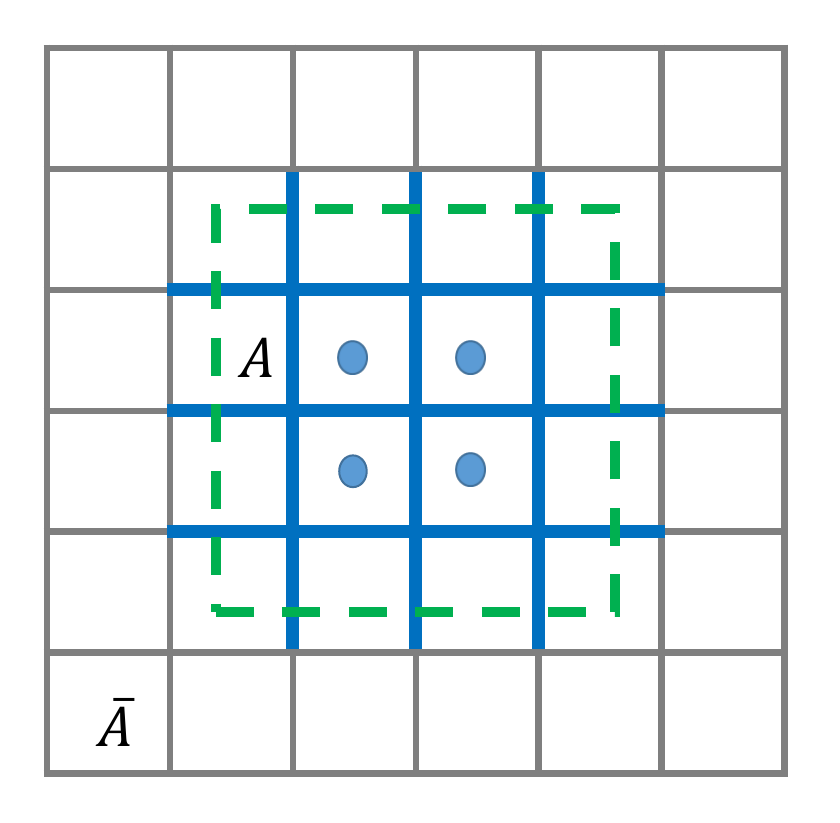}
\caption{
 An example of bipartition $A\cup \bar A$. The links crossed by the green dotted line are the boundary $|\partial A|$ and are included in $A$.  Here $|A|=9,|\partial A|=12, m_{A}=4$ (blue solid circle).} \label{fig:bipartition}
\end{figure}

In the following, we present a detailed calculation of the von Neumann entropy $S_{A}=-\text{Tr}(\rho_A\text{log}\rho_A)$ of subsystem $A$, following a similar approach to that in Ref. \cite{hamma2005ground}. To calculate $\rho_A$ and $S_A$, we need to make some preparations.
Consider a bipartition of the system $A\cup\bar{A}$. In this subsection, we focus on the case where both $A$ and $\bar A$ are path-connected (no disjointed parts). See Fig. \ref{fig:bipartition} for an example. Recall that group $G$ in Eq. \eqref{eq:groupG} is defined as the stabilizer group of vertex operators $A_{v}$. The group elements can be decomposed as $g=g_A\otimes g_{\bar{A} }$, where $g_{A}$ and $g_{\bar{A}}$ are actions of $g$ on $A$ and $\bar{A}$, respectively, and generally, they are not group elements of $G$. There exist two special subgroups $G_A$ and $G_{\bar{A}}$,
\begin{equation}
\begin{aligned}
G_{A}&=\{ g\in G|\ g=g_{A}\otimes I_{\bar{A}}\},\\
% \\ g_{A}&\in G\ \text{only acts non-trivially on $A$}\\
G_{\bar A}&=\{g\in G|\ g=I_{A}\otimes g_{{\bar A}}\}.
% g_{{\bar A}}&\in G\ 
%  \text{only acts  non-trivially on $\bar A$}.
\end{aligned}
\end{equation}
Since the generators $A_{v}$, $W_{x}$, and $W_{y}$ commute with each other, $G$ and $G'$ are Abelian groups, and $G_{A}$ and $G_{\bar A}$ are normal subgroups. Then we can further define two quotient groups $G/G_{\bar A}$ and $G/G_{A}$. 

% \paragraph{The Kitaev-Preskill scheme}
% \subparagraph{$h=0$}\label{sec:kph0}
First,  we analyze the simplest case without the spin-flip term ($h=0$), where the steady states are four-fold degenerate. Evidently, the reduced density matrix $\rho_A$ and $S_A$ do not depend on the topological sector. Without loss of generality, we choose $\rho^{00}=\frac{1}{|G|}\sum_{g\in G}g|\Uparrow\rangle\langle \Uparrow|g$ as an example. Here the loop configurations $g|\Uparrow\rangle\langle \Uparrow|g$ are all contractible and $|\Uparrow\rangle\langle\Uparrow|$  is the state with all spins up in $\sigma^z$ basis. The reduced density matrix of region $A$ is
\begin{equation}
\begin{aligned}
  \rho_{A} &=\text{Tr}_{\bar A}\rho^{00}=\text{Tr}_{\bar A}\frac{1}{|G|}\sum_{g\in G}g|\Uparrow\rangle\langle \Uparrow|g\\
           % &=\text{Tr}_{\bar A}\frac{1}{|G|}\sum_{g\in G}g_{A}\otimes g_{\bar A}|0\rangle\langle 0|g_{A}\otimes g_{\bar A}\\
           &=\frac{|G_{\bar A}|}{|G|}\sum_{g\in G/G_{\bar A}}g_{A}|\Uparrow_{A}\rangle\langle\Uparrow_{A}|g_{A}.
\end{aligned}
\end{equation}
Given that the reduced density matrix $\rho_{A}$ is diagonal, we can easily obtain the von Neumann entropy
\begin{equation}
\begin{aligned}
   S_{A}&=-\text{Tr}(\rho_{A}\log\rho_{A})=-\sum_{i}\lambda_i \log\lambda_i\\
     % &=-\frac{|G_{\bar A}|}{|G|}\frac{|G|}{|G_{\bar A}|}\log\frac{|G_{\bar A}|}{|G|}= \log\frac{|G|}{|G_{\bar A}|}\\
     % &={\log}2^{n-1-|\bar A|}= \log 2^{|A|+|\partial A|-1}\\
     &=(|A|+|\partial A|-1)\log2.\label{eq:SA0}
\end{aligned}
\end{equation}
$|A|$ denotes the number of vertices inside $A$ (here a vertex is defined to be inside $A$ if and only if its four adjacent links all belong to $A$), and $|\partial A|$ is the number of vertex operators crossed by the boundary. $\lambda_{i}$ is the eigenvalue of the reduced density matrix. 
% Whether the boundary spins are included in $A$ or $\bar{A}$ does not affect the entropy. %We can compare this result to the Hermitian case. 

We can compare the above result to the entanglement entropy of the toric-code ground state: $S_{A}^{\text{g}}= (|\partial A|-1) \log2=|\partial A|\log2-S_{\text{topo}}$, which obeys the area law, and has a subleading term $S_{\text{topo}}=\log 2$, which is the celebrated topological entanglement entropy (TEE) \cite{kitaev2006topological,levin2006detecting}. Compared to $S_{A}^{\text{g}}$, for our mixed steady state, $S_{A}$ has an additional term that is proportional to the volume of $A$ \cite{castelnovo2007classical}. It originates from the decoherence effect in the open system which makes our steady state a mixed state. The last $O(1)$ term can be identified as the topological entropy and We discuss this later.
% \subparagraph{$h\neq0$}\label{sec:kph}

Second, we investigate the behavior of topological entropy when topological degeneracy breaks down at nonzero $h$. %With the exact form of the steady state obtained in Eq. \eqref{eq:2dss1}, we can directly calculate $S_A$ with the spin-flip term turned on. 
The reduced density matrix is
\begin{equation}
\rho_{A}=\text{Tr}_{\bar A}\frac{1}{T'}\sum_{k=0}^{\lfloor\frac{n}{2}\rfloor}\beta^{2k}\sum_{g\in G'}\sum_{\{r\}}g|m_{2k}(\{r\})\rangle\langle m_{2k}(\{r\})|g.
% &= \text{Tr}_{\bar A}\frac{1}{T'}\sum_{k=0}^{\lfloor\frac{n}{2}\rfloor}\beta^{2k}\sum_{g,\{r\}}g_{A}\otimes g_{\bar A}|e_{2k}(\{r\})\rangle\langle e_{2k}(\{r\})|g_{A}\otimes g_{\bar A}\\
% \end{aligned}
\end{equation}
For states with $k$ pairs of $m$ particles, there are overall $|G'|{n \choose 2k}$ different spin configurations. For states with the same particle distribution but generated from different topological classes, they contribute the same to the reduced density matrix and we can replace $G'$, $T'$ (Eq. \eqref{eq:normalT}) by $G$ and $T=T'/4$. 

When the system is divided into $A$ and $\bar A$, these $2k$ particles are separately distributed in two parts. We denote the number of plaquettes inside $A$ as $m_A$. The value of plaquette operators $B_{p}$ on these $m_{A}$ plaquettes solely depends on spin configurations in $A$. See Fig. \ref{fig:bipartition} for an illustration. When tracing out $\bar{A}$, the reduced density matrix $\rho_A$ only keeps the information of $m$-particle distribution on these plaquettes. In other words, $m$-particle distribution inside $\bar{A}$ or dangling on the boundary of $A$ generates identical configurations in $A$. After summing over the action of stabilizers $g\in G$, we have    %Because $m$ particles are living on the plaquettes, one may naively classify the position of $m$ particles by whether they are in $A$ or $\bar{A}$.  However, this gives the wrong counting on the number of states.  Here, we define an important variable $m_{A}$.
%\begin{equation}
%m_{A}\equiv\#\text{plaquettes totally inside } A.\label{eq:mvertex}
%\end{equation}
%Here a plaquette is totally inside $A$ means that all four spins constituting the plaquette operator $B_{p}$ belong to $A$. on these plaquettes can generate independent configurations in the reduced density matrix $\rho_A$ states with the same $m$-particle configuration inside $A$ contribute to a specific reduced density matrix.
% Only $m$ particles on these plaquettes can generate independent configurations in the reduced density matrix $\rho_A$: 
\begin{widetext}
\begin{equation}
\begin{aligned}
\rho_{A}&= \text{Tr}_{\bar A}\frac{1}{T}\sum_{j=0}^{m_{A}}\sum_{k=\lceil \frac{j}{2}\rceil}^{\lfloor \frac{n-m_{A}+j}{2}\rfloor}\beta^{2k}\sum_{g\in G}\sum_{\{r_{A}\}}\sum_{\{r_{\bar A}\}}g|m_{{2k-j}}(\{r_{\bar A}\})m_{j}(\{r_{A}\})\rangle\langle m_{2k-j}(\{r_{\bar A}\})m_{j}(\{r_{A}\})|g \\
        % &= \frac{|G_{\bar A}|}{T}\sum_{j=0}^{m}\sum_{k=\lceil \frac{j}{2}\rceil}^{\lfloor \frac{n-m+j}{2}\rfloor}\beta^{2k} {n-m \choose 2k-j}\sum_{g\in G/G_{\bar A}}\sum_{\{r_{A}\}}g_{A}|e_{j}(r_{A})\ra\la e_{j}(r_{A})|g_{A}\\
        &= \frac{|G_{\bar A}|}{T}\sum_{j=0}^{m_{A}}\sum_{\{r_{A}\}}\sum_{g\in G/G_{\bar A}}\left(\sum_{k=\lceil \frac{j}{2}\rceil}^{\lfloor \frac{n-m_{A}+j}{2}\rfloor}\beta^{2k} {n-m_{A} \choose 2k-j}\right)g_{A}|m_{j}(\{r_{A}\})\rangle\langle m_{j}(\{r_{A}\})|g_{A}.
\end{aligned}    
\end{equation}
\end{widetext}
Here $\lfloor x \rfloor$ and $\lceil x \rceil$ are the floor and ceiling functions \footnote{$\lfloor x \rfloor=\text{max}\{z\in Z|z\leq x \}$ and $\lceil x \rceil=\text{min}\{z\in Z|z\geq x\}$}.
$\{r_{A}\}$ is the set of positions of $j$ particles that totally determined by spins in $A$ and $\{r_{\bar A}\}$ is the positions of the rest $2k-j$ particles, including those inside $\bar A$ and dangling on the boundary. When tracing out $\bar A$, there are $ {n-m_{A} \choose 2k-j}$ different $m$ particle distributions giving the same spin configuration of $A$, up to an action of $g_A$. 

To calculate the von Neumann entropy $S_{A}=- \text{Tr}\rho_{A}\log\rho_{A}$, we need to work out the spectrum of $\rho_A$. For each eigenstate $g_{A}|m_{j}(\{r_{A}\})\rangle$ of $\rho_A$, the eigenvalue only depend on $j$, and thus we denote it by $\lambda_j$. Each eigenvalue is $D_j$-fold degenerate, contributed by states with different $m$-particle positions $\{r_A\}$ and string configurations $g_A$. We list the eigenvalue and the corresponding degeneracy below:
\begin{equation}
\lambda_{j}=\frac{|G_{\bar A}|}{T}\sum_{k=\lceil \frac{j}{2}\rceil}^{\lfloor \frac{n-m_{A}+j}{2}\rfloor}\beta^{2k} {n-m_{A} \choose 2k-j},\quad 
D_{j}=\frac{|G|}{|G_{\bar A}|}{m_{A} \choose j}.\label{eq:kpeigenvalue}
\end{equation}
Then we can obtain the von Neumann entropy. The details of the calculation are in Appendix \ref{sec:appendixkp}. Here we directly give the result in the thermodynamic limit $n\to \infty$:
\begin{equation}
\begin{aligned}
     S_{A}=&-\sum_{j=0}^{m_{A}} D_j\lambda_{j} \log\lambda_{j}\\
     % &= S_{A}^{(1)}+S_{A}^{(2)}\\
     =& (|A|+|\partial A|-1)\log2
     +m_{A} f(\beta). \label{eq:SAh}
     % (\log(1+\beta)-\frac{\beta}{1+\beta}\log\beta)\\
     % &:= (|A|+|\partial A|-1)\log2+mf(\beta)
\end{aligned}
\end{equation}
The additional term proportional to $m_A$ comes from the nonzero density of $m$ particles due to the perturbation. The explicit form of $f(\beta)$ is
\begin{equation}
    f(\beta)=\log(1+\beta)-\frac{\beta}{1+\beta}\log\beta,
\end{equation}
 where $\beta=\sqrt{\frac{h}{1+h}}$. Here $f(\beta)$ is a monotone-increasing function with $f(0)=0$ and $f(1)=\log 2$. %The entropy has two parts $S_{A}=S_{A}^{(1)}+S_{A}^{(2)}$ where $S_{A}^{(1)}=(|A|+|\partial A|-1)\log2$ and $S_{A}^{(2)}=mf(\beta)$ . When $\beta=0$ $(h=0)$, we have $S^{(2)}_{A}=mf(0)=0$, and  $S_{A}=S^{(1)}_{A}$ is contributed by all loop states which is the same as the $h=0$ case in Eq. \eqref{eq:SA0}. $S^{(2)}_{A}$ is given by open loop states and depends on the perturbation strength.

\begin{figure}[htb]
\centering
\includegraphics[width=1.0\linewidth]{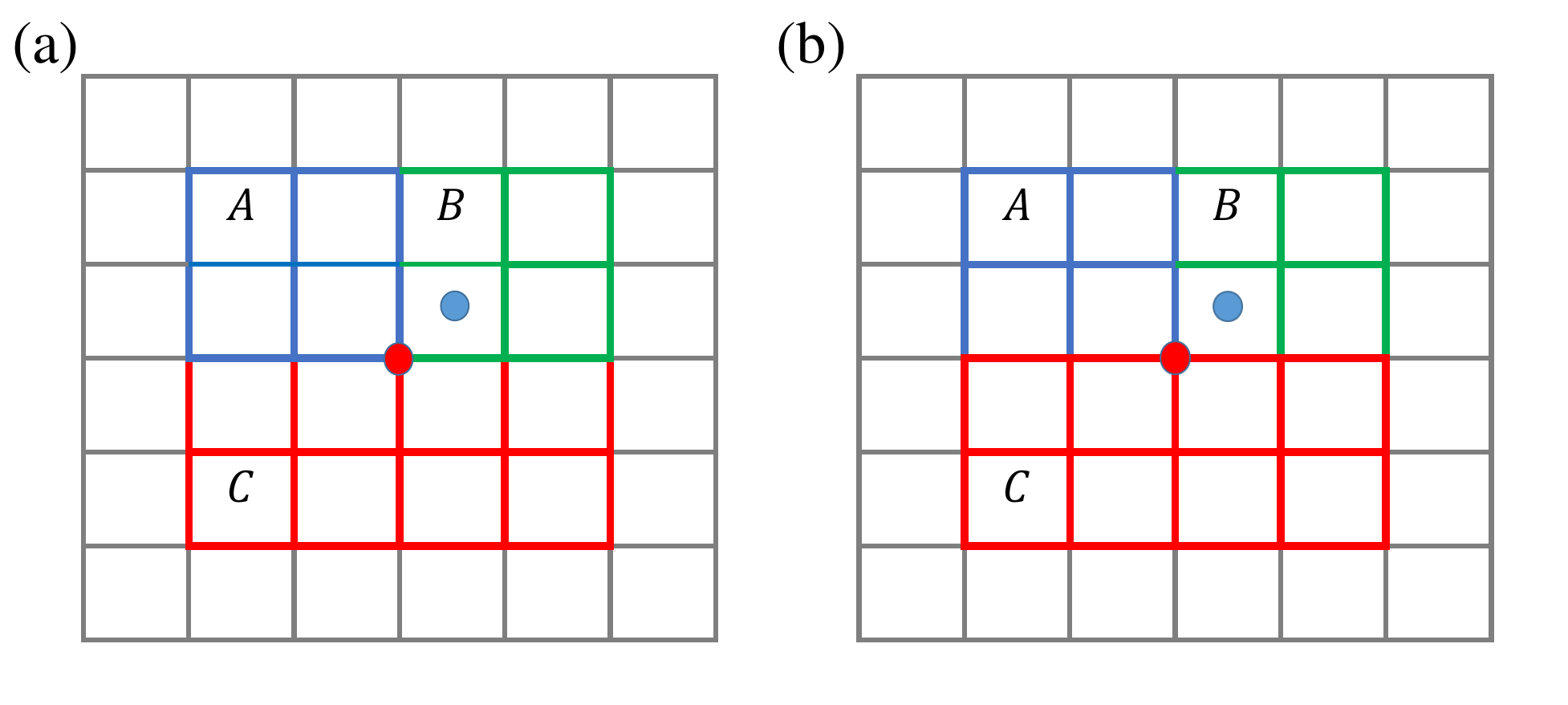}
\caption{
Two similar tripartitions. (a) $\delta |A|=1$, $\delta m=0$. The central vertex (red solid circle) only appears in $ABC$, while the plaquette nearby (blue solid circle) appears in both $AB$ and $ABC$ and finally cancels. (b) $\delta |A|=0$, $\delta m=1$. The central vertex (red solid circle) appears in both $AC$ and $ABC$ and cancels in the end. However, the plaquette (blue solid circle) only appears in $ABC$. }
\label{fig:kpentropy}
\end{figure}

\paragraph{The Kitaev-Preskill scheme}
With these preparations, we can follow Kitaev-Preskill's scheme to get the topological entropy. As in the closed system, we can extract that topological entropy by calculating tripartite mutual information $S_{\text{topo}}=-I(A:B:C)$,
\begin{equation}
    I(A:B:C)=S_{A}+S_{B}+S_{C}-S_{AB}-S_{BC}-S_{AC}+S_{ABC}.
\end{equation}
With the explicit expression in Eq. \eqref{eq:SAh}, we have
\begin{equation}
\begin{aligned}
S_{\text{topo}}=(1-\delta |A|)\log2-\delta mf(\beta).\label{eq:kpstopo1}
\end{aligned}
\end{equation}
 The terms proportional to the area $|\partial A|$ (boundary of $A$) cancel out. Here 
\begin{equation}
     \begin{aligned}
         &\delta m=m_{A}+m_{B}+m_{C}-m_{AB}-m_{BC}-m_{CA}+m_{ABC}\\
         &\delta |A|=|A|+|B|+|C|-|AB|-|BC|-|AC|+|ABC|.
     \end{aligned}
\end{equation}
$\delta m$ $(\delta |A|)$ is the difference of independent plaquettes (vertices) in $I(A:B:C)$. Now, we can check whether $S_{\text{topo}}$ is universal and non-vanishing in different tripartitions (See Fig. \ref{fig:kpentropy}).
% Notice that $m+|\partial A|=V$ is the total number of vertices associated with $A$. We can relate $m$ to $|A|$ and $|\partial A|$ through the following equation: $m= L_{A}-|A|-|\partial A|+\chi$ where $L_{A}$ is the number of spins in $A$ and $\chi=V-L_{A}+|A|$ is the Euler characteristic number. The expression of $S_A$ can be rewritten as:
% \begin{equation}
% S_A=L_Af(\beta)+(|A|+|\partial A|-1)(\log2-f(\beta)).
% \end{equation}
% $L_{A}$ and $|\partial A|$ will be canceled when considering tripartite mutual information $I(A:B:C)$, so we have the final expression of topological entropy:
% \begin{equation}
% \begin{aligned}
% S_{\text{topo}}&=(1-\Delta|A|)(\log2-f(\beta))\\
% &=\Delta m(\log2-f(\beta)).
% \end{aligned}
% \end{equation}
% Here $\Delta m=m_{A}+m_{B}+m_{C}-m_{AB}-m_{BC}-m_{CA}+m_{ABC}$.
% $\Delta m$($\Delta |A|$)  is the difference of independent vertices(plaquettes) in $I(A:B:C)$. We can check whether $S_{\text{topo}}$ is universal and non-vanishing in two situations.

Here in Fig. \ref{fig:kpentropy}(a), we observe that $S_{\text{topo}}=0$ for any dissipation strength. This is because the central vertex of $ABC$ is only included in $S_{ABC}$ which gives $\delta |A|=1$, and all plaquettes are canceled, resulting in $\delta m=0$. Therefore we have $S_{\text{topo}}=(1-\delta |A|)\log2-\delta mf(\beta)=0$. %There are no long-range correlations between different parts and the topological entropy is always zero. 

In Fig. \ref{fig:kpentropy}(b), if we move part $C$ to include the boundary of $AC$ and $BC$, then the central vertex appears in both $AC$ and $ABC$. As a result, we obtain $\delta |A|=0$.  The central plaquette only appears in $ABC$, giving $\delta m =1$. In this case, $S_{\text{topo}}=(1-\delta |A|)\log2-\delta mf(\beta)=\log 2-f(\beta)$. The mutual information of the three parts is non-vanishing and there exists non-zero entropy. Recall that $\beta=\sqrt{\frac{h}{1+h}}\in [0,1)$, then $S_{\text{topo}}=\log 2-f(\beta)$ decreases from $\log 2$ to $0$ smoothly when increasing the perturbation strength $h$. It is non-vanishing except in the limit $h\rightarrow \infty$, indicating the existence of non-local correlations between different parts of the system for the entire parameter regime, although weakened by the spin-flip term. 

For open quantum systems,  we find that the Kitaev-Preskill scheme is not universal and the topological entropy depends on the specific partition. The correlation between different parts is affected by spins along the boundary. 

\begin{figure}[htb]
 \centering
\includegraphics[width=0.90\linewidth]{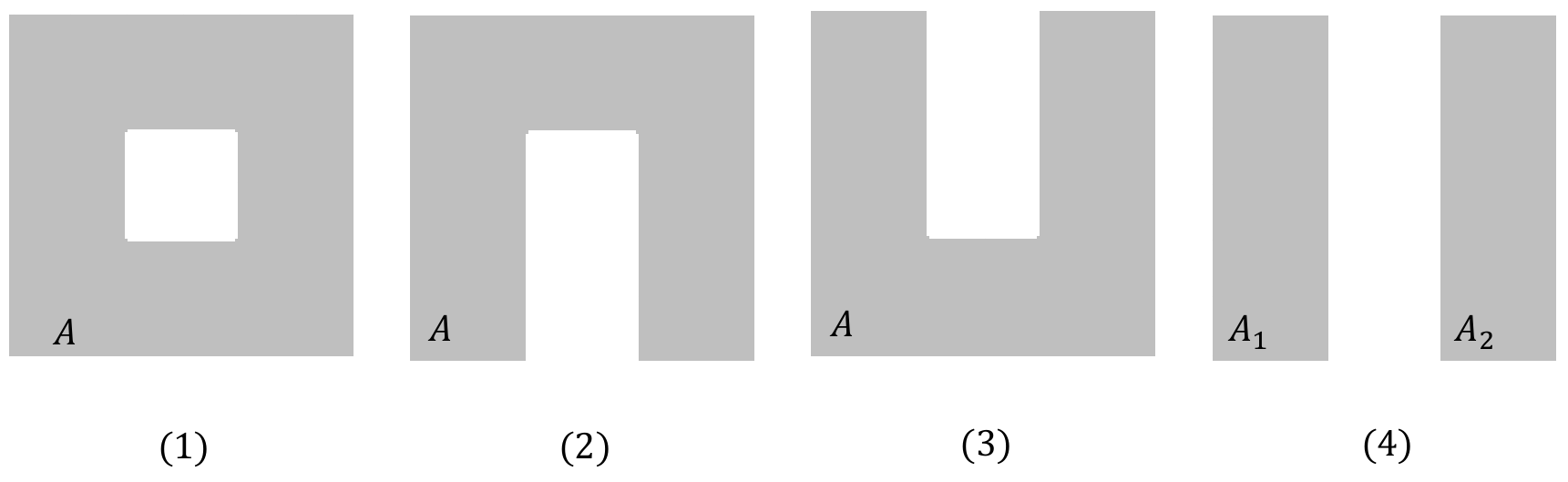}

\caption{
An illustration of the four bipartitions used in Levin-Wen's definition of topological entanglement entropy. Especially in partition-4, $A=A_{1}\cup A_{2}$}
\label{Levin-Wen}
\end{figure}

\paragraph{The Levin-Wen scheme}
In this part, we adopt Levin-Wen's definition of topological entropy:
\begin{equation}
S_{\text{topo}}=-S_{1A}+S_{2A}+S_{3A}-S_{4A}.
\label{eq:lwtopo}
\end{equation}
Here $1,2,3,4$ refer to different choices of bipartition as illustrated in Fig. \ref{Levin-Wen}. Although results given by the two definitions only differ by a factor of $2$ for the ground states in closed systems, we find that they are qualitatively different for the steady states in dissipative systems.
\subparagraph{$h=0$}
First, we consider the unperturbed case. Following exactly the same procedure in Eq. \eqref{eq:SA0}, we get
\begin{equation}
   S_{A}=-\text{Tr}(\rho_{A}\log\rho_{A})=\log\frac{|G|}{|G_{\bar A}|}.
     %-\sum_{i}\lambda_i \log\lambda_i\\
     % &=-\frac{|G_{\bar A}|}{|G|}\frac{|G|}{|G_{\bar A}|}\log\frac{|G_{\bar A}|}{|G|}\\
     %&= \log\frac{|G|}{|G_{\bar A}|}%={log}2^{n-1-|\bar A|}\\
     %&= \log 2^{|A|+|\partial A|-1}=(|A|+|\partial A|-1)\log2
\label{eq:SA}
\end{equation}
That works for all four partitions. When encountering the case where subsystem $A$ has more than one connected piece, as is the case for partition-$4$, we need to be more careful about counting the elements in $G_{\bar A}$. There are extra elements apart from the products of $A_{v}$ operators that solely act on $\bar A$. In the example of partition-$4$ in Fig. \ref{Levin-Wen}, $A=A_1\cup A_2$,  we have one more independent element in $G_{\bar A}$, which is the product of all $A_{v}$ operators acting non-trivially on $A_1$ ($A_2$). In general, we have
\begin{equation}
|G_{\bar A}|=2^{|\bar A|+p_A-1},
\label{eq:number}
\end{equation}
where $p_A$ is the number of connected parts of subsystem $A$. Substitute Eq. \eqref{eq:number} and Eq. \eqref{eq:SA} into Eq. \eqref{eq:lwtopo}, and we can get the following results:
\begin{equation}
\begin{aligned}
S_A&=(|A|+|\partial A|-p_A)\log 2,\\
S_{\text{topo}}&=(p_{1A}-p_{2A}-p_{3A}+p_{4A})\log 2=\log 2.\label{eq:lwtopo1}
\end{aligned}
\end{equation}
There indeed exists a term that is reminiscent of TEE in the unitary case, which indicates nonlocal correlations between different parts of the system. We term it the topological entropy \cite{castelnovo2007classical}.

\subparagraph{$h\neq0$}
Now we turn on a finite $h$. For partition-$2,3,4$, the calculation of $S_A$ is identical to Eq. \eqref{eq:SAh} in the preparation part. The result is
\begin{equation}
S_{iA}=\log \frac{|G|}{|G_{\bar A}|}+m_Af(\beta), \quad i=2,3,4.
\end{equation}
Here $m_{A}$ is the number of plaquettes that are totally inside $A$.

\begin{figure}[htb]
 \centering
\includegraphics[width=0.75\linewidth]{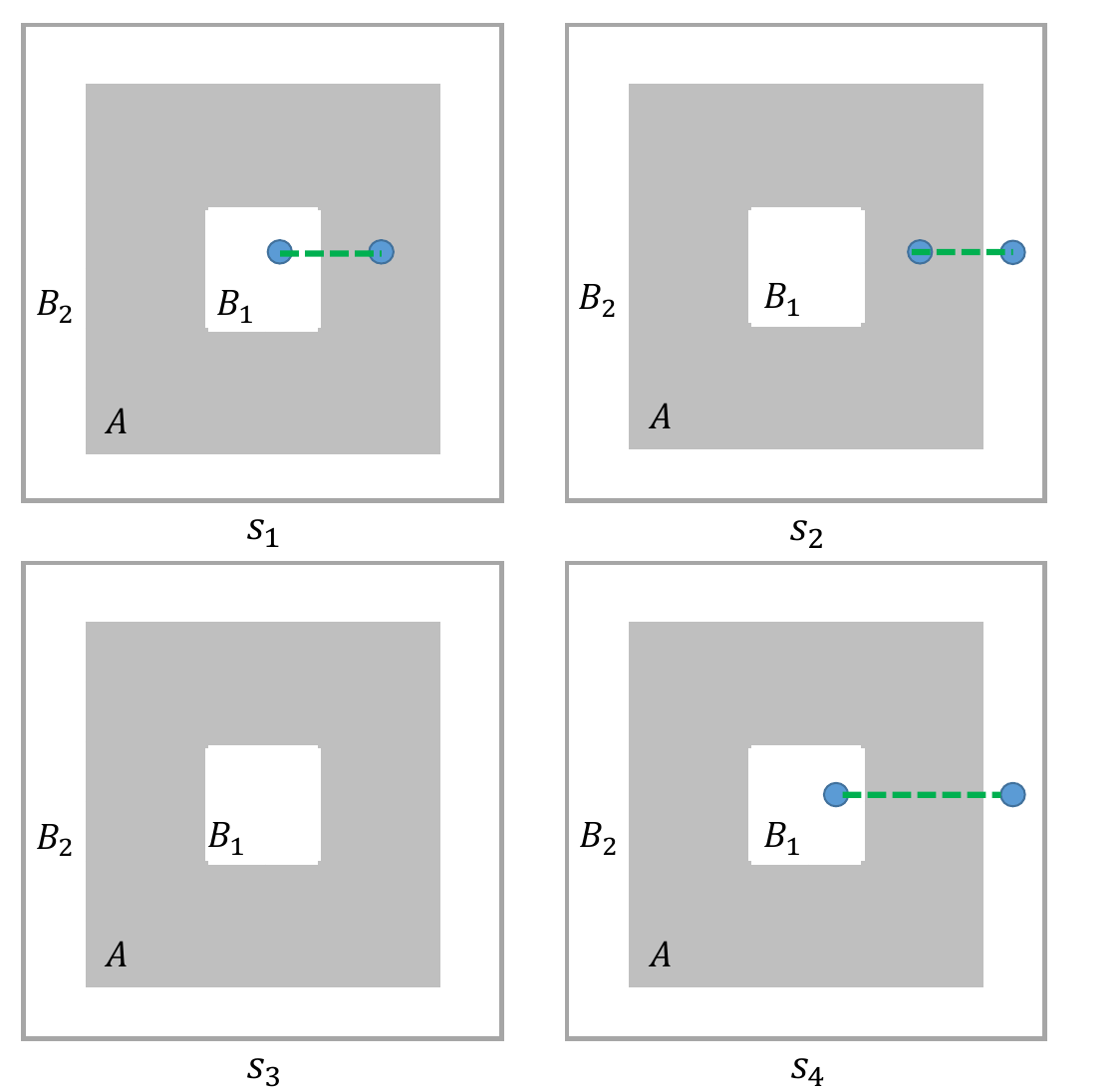}
\caption{
Representative examples of the four topologically inequivalent scenarios. The blue solid circle represents the $m$ particle and the green dashed line represents a string operator.}
\label{fig:fourclasses}
\end{figure}

In partition-1, however, the situation becomes a bit more complicated due to the fact that subsystem $\bar A$ has two disjointed pieces: $\bar A=B_1\cup B_2$ (each is path connected). When tracing out $\bar{A}$, there are four classes of topologically different states $s_1, s_2, s_3$, and $s_4$ (See Fig. \ref{fig:fourclasses}). First, according to particle number, the states in $\rho_A$ can be classified into two classes: odd $s_1, s_2$, and even $s_3, s_4$. Then, $s1$ and $s2$  are distinguished by a string operator $S^{x}_{\tilde{t}}$ which crosses $A$ once, and this is also the difference between $s_{3}$ and $s_{4}$.

We directly give the result of $\rho_A$ below and the details can be found in Appendix
\ref{sec:appendixkp}. In the thermodynamic limit $n\to\infty$ with the linear sizes of $A$, $B_{1}$, and $B_{2}$ all going to infinity, we have
\begin{widetext}
\begin{equation}
\begin{aligned}
\rho_A=\frac{G_{\bar A}}{2|G|(1+\beta)^{m_A}}\sum_{g\in G/G_{\bar A}}\bigg[ &\sum_{j\text{ odd}}\sum_{\{r_A\}}\beta^j\Big(  g_A|m_{j}(\{r_A\}),s_1\rangle\langle m_{j}(\{r_A\}),s_1|g_A
+g_A|m_{j}(\{r_A\}),s_2\rangle\langle m_{j}(\{r_A\}),s_2|g_A\Big)\\
+&\sum_{j\text{ even}}\sum_{\{r_A\}}\beta^j\Big(  g_A|m_{j}(\{r_A\}),s_3\rangle\langle m_{j}(\{r_A\}),s_3|g_A
+g_A|m_{j}(\{r_A\}),s_4\rangle\langle m_{j}(\{r_A\}),s_4|g_A\Big)\bigg].\label{eq:rhoA1}
\end{aligned}
\end{equation}
\end{widetext}
Its eigenvalues and corresponding degeneracies are
\begin{equation}
\lambda_j=\frac{|G_{\bar A}|\beta^j}{2|G|(1+\beta)^{m_A}},\quad D_j=2{m_A\choose j}\frac{|G|}{|G_{\bar A}|}.
\end{equation}
We can calculate the entropy of system A for partition-$1$:
\begin{equation}
\begin{aligned}
S_{1A}&=\sum_{j=0}^{m_A}D_j\lambda_j\log \lambda_j\\
% &=\sum_{j=0}^{m_A}\frac{{m_A\choose j}\beta^j}{(1+\beta)^{m_A}}\log \frac{|G_{\bar A}|\beta^j}{2|G|(1+\beta)^{m_A}}\\
% &=\log{\frac{|G|}{G_{\bar A}}}+\log 2+m_Af(\beta)\\
&=(|A|+|\partial A|)\log 2+m_Af(\beta).
\end{aligned}
\end{equation}
Finally, we can get the topological entropy
\begin{equation}
\begin{aligned}
S_{\text{topo}}=&-S_{1A}+S_{2A}+S_{3A}-S_{4A}\\
=&(p_{1A}-p_{2A}-p_{3A}+p_{4A}-1)\log 2\\ 
&+(-m_{1A}+m_{2A}+m_{3A}-m_{4A})f(\beta)\\
=&0.\label{eq:lwstopo}
\end{aligned}
\end{equation}
The topological entropy immediately drops from $\log2$ to $0$ under arbitrarily small perturbation. That indicates the absence of a topologically ordered phase when turning on the perturbation. In Sec. \ref{sec:2dperturbation} and Sec. \ref{sec:2dexactform}, we have seen that the topological degeneracy of steady states is fragile under any perturbation, which is consistent with the result of topological entropy. 
It is important to note that our analysis of topological entropy using the Levin-Wen and Kitaev-Preskill schemes reveals that Levin-Wen's definition is universal and physically meaningful. It establishes a correspondence between topological entropy and topological degeneracy of steady states in our dissipative models. As we have shown, the quantized topological entropy drops to zero when the degeneracy of steady states is broken by perturbation.

\subsubsection{Dissipative $Z_2$ gauge theory}\label{sec:2dgauge}
\begin{figure}[htb]
\centering
\includegraphics[width=0.60\linewidth]{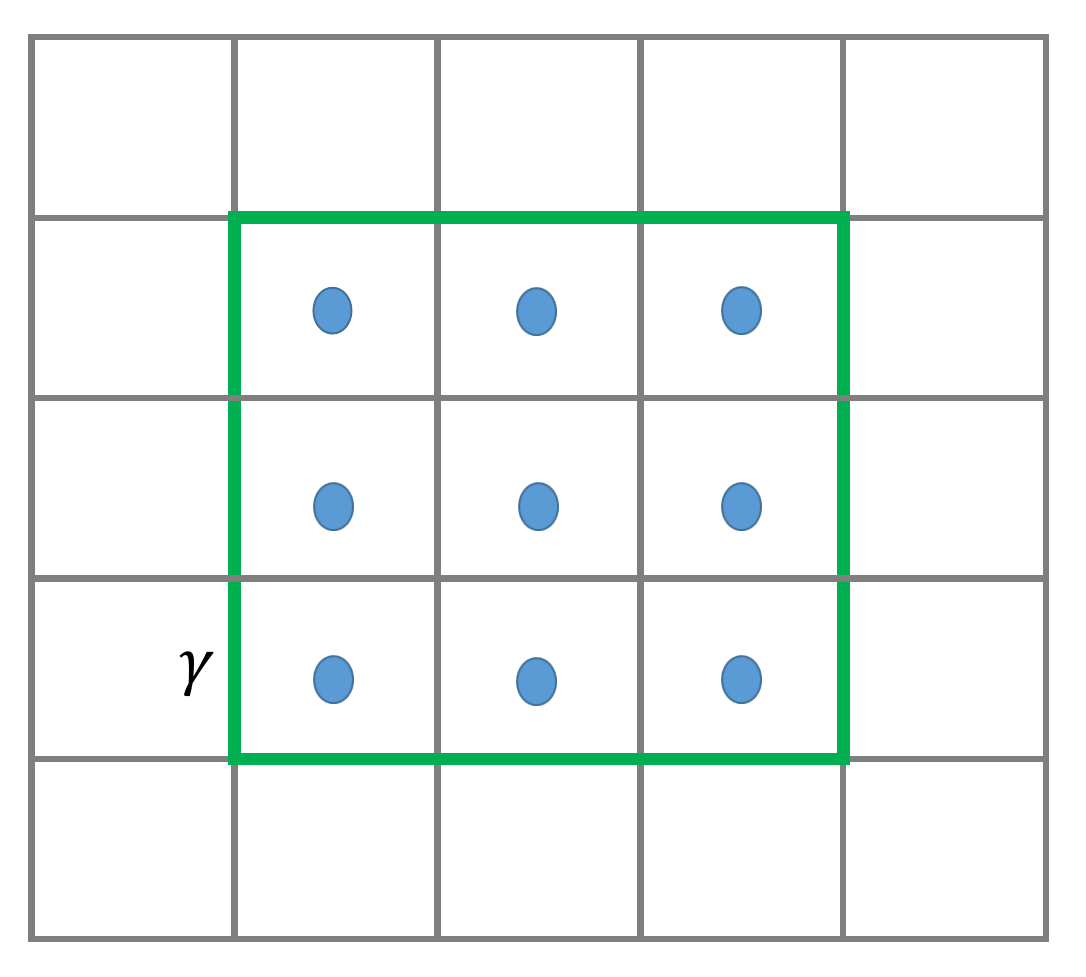}
\caption{
Wilson loop $\gamma$ is represented by the green line. $m_{\gamma}=9$ in this example}\label{fig:Wilson loop}
\end{figure}

As already mentioned in Eq. \eqref{eq:map1} and Eq. \eqref{eq:map2}, with the dephasing term and all the other quantum jump operators that are compatible with the diagonal structure of the density matrix, we can restrict the density matrix to the diagonal subspace. Correspondingly, with the perturbation turned on, the Liouvillian superoperator is mapped to some operator in this vector space $\mathcal{L}=\mathcal{L}^{0}+\delta \mathcal{L}\rightarrow \Gamma=\Gamma^{0}+\delta \Gamma$:
\begin{equation}
\begin{aligned}
\Gamma=\sum_l(\sigma^x_l-1)P^{2}(\sum_{p|l\in\partial p}B_{p})+\kappa_{v}\sum_v(A_{v}-1)&\\
+h\sum_i(\sigma_i^x-1)&.
\label{eq:2deff}
\end{aligned}
\end{equation}
Then the steady state is mapped to the ``ground state" of the non-Hermitian Hamiltonian $-\Gamma$, with eigenvalue $\lambda=0$. In this representation, one can immediately notice a local symmetry: $[A_{v}, \Gamma]=0$. Starting from an arbitrary initial state, the system would eventually relax to the subspace with $A_v=1$, controlled by the second term in $\Gamma$. Thus the long-time ($t\gg {\kappa_v}^{-1}$) relaxation dynamics can be described by a pure $Z_2$ gauge theory, with the gauge invariance condition $A_{v}=1$. %If ignoring the non-Hermitian part $\sum_{l}(\sigma^x_l-1)P^{2}(\sum_{p|l\in\partial p}B_{p})$ in $\Gamma$, we will get an Ising gauge theory which will be in the deconfined phase for small $h$ and undergo a deconfinement-confinement transition at some critical field $h_c$ \cite{kogut1979introduction}. 
It is known that the conventional $Z_2$ gauge theory undergoes a deconfinement-confinement transition in 2d. However, we show below that, due to non-Hermiticity, this theory is always confined in 2d for finite $h$. To see this, we calculate the expectation value of the Wilson loop operator:
\begin{equation}
\begin{aligned}
W_{\gamma}=\prod_{l\in\gamma}\sigma_{l}^{z},\quad
\langle W_{\gamma}\rangle =\frac{\langle I|  W_{\gamma}|\rho_{ss}\rangle}{\langle I|\rho_{ss}\rangle}  = \frac{\text{Tr}(  W_{\gamma}\rho_{ss})}{\text{Tr}(\rho_{ss})},
\end{aligned}
\end{equation}
here $\gamma$ is a closed loop as shown in Fig. \ref{fig:Wilson loop} and $|I\rangle=\sum_{\{s\}}|\{s\}\rangle$ is the left steady state $I$ (the identity) written in the diagonal space. The rightmost equality follows from the fact that both $W_{\gamma}$ and $\rho_{ss}$ in Eq. \eqref{eq:2dss1} are diagonal in the $\sigma^{z}$ basis. This is crucial because only then do Wilson loop operators in this effective gauge theory really correspond to physical observables in the original model. Since any $m$ particle is connected to another one by a string operator $S^{x}_{\tilde{t}}=\prod_{l\in \tilde{t}}\sigma^{x}_{l}$, we have $W_{\gamma}=\pm 1$ if there is an even/odd number of $m$ particles inside $\gamma$. 
Therefore, we can easily calculate its expectation value:
\begin{equation}
\begin{aligned}
 \langle W_{\gamma}\rangle
                         &= \frac{1}{T'}\sum_{k=0}^{\lfloor\frac{n}{2}\rfloor}\beta^{2k}\sum_{g\in G'}\sum_{\{r\}}\langle m_{2k}(\{r\})|g\prod_{j\in \gamma}\sigma_{j}^{z}g|m_{2k}(\{r\})\rangle\\
                         % &=\frac{1}{t^{+}(\beta,n)}\sum_{k=0}^{\lfloor\frac{n}{2}\rfloor}\beta^{2k}\sum_{\{r\}}\la e_{2k}(\{r\})|\prod_{j\in \gamma}\sigma_{j}^{z}|e_{2k}(\{r\})\ra\\
                         % &=\frac{1}{t^{+}(\beta,n)}\sum_{j=0}^{m_{\gamma}}{m_{\gamma} \choose j}(-1)^{j}\sum_{k=\lceil \frac{j}{2}\rceil}^{\lfloor \frac{n-m_{\gamma}+j}{2}\rfloor}\beta^{2k}{n-m_{\gamma} \choose 2k-j}\\
                         % &=\frac{1}{t^{+}(\beta^{2},n)}\sum_{j=0}^{m_{\gamma}}{m_{\gamma} \choose j}(-\beta)^{j}\sum_{k=\lceil \frac{j}{2}\rceil}^{\lfloor\frac{n-m_{\gamma}+j}{2}\rfloor}\beta^{2k-j}{n-m_{\gamma} \choose 2k-j}\\
                         &=\frac{1}{t^{+}(\beta,n)}\sum_{j=0}^{m_{\gamma}}{m_{\gamma} \choose j}(-\beta)^{j}\sum_{k'}\beta^{k'}{n-m_{\gamma} \choose k'},
\end{aligned}
\end{equation}
where $t^{\pm}=\frac{(1+\beta^{n})\pm (1-\beta^{n})}{2}$ and $m_{\gamma}$ is the number of plaquettes inside $\gamma$ (See Fig. \ref{fig:Wilson loop}). 
This is also an exact result, and like the entropy calculation in Appendix \ref{sec:appendixkp}, it is greatly simplified in the thermodynamic limit:
\begin{equation}
\begin{aligned}
\lim_{n\to\infty}\langle W_{\gamma}\rangle&= \lim_{n\to\infty}\sum_{j=0}^{m_{\gamma}}(-\beta)^{j}{m_{\gamma} \choose j}\frac{t^{(-1)^{j}}(\beta,n-m_{\gamma})}{t^{+}(\beta,n)}\\
&= \frac{(1-\beta)^{m_{\gamma}}}{(1+\beta)^{m_{\gamma}}}= \exp[-m_{\gamma}\ln \frac{1+\beta}{1-\beta}].\label{eq:2dwilson}
\end{aligned}
\end{equation}
When $\beta=0$ $(h=0)$, we can get $W_{\gamma}=1$. This is obvious because only loop states contribute. However, at any finite $h$, the Wilson loop obeys an area law, signifying a confined phase. This is also consistent with the broken degeneracy and the zero topological entropy.

%Going back to Eq. \eqref{eq:2deff}, we find that for $h=0$, there are three types of anyon excitations $e, m, f=e\times m$ as in the toric code model. However, adding a $\sigma^x$ term immediately leads to the condensation of $m$ particles, the $e$ particles will be confined, which is consistent with the above calculation.

\subsection{Model-2}\label{sec:2dmodelquamtum}
\subsubsection{Topologically degenerate steady states}
In the above model, the steady states are always diagonal, which reveals a lack of quantum coherence. Moreover, the steady state can be written in another form $\rho_{ss}=\exp(-\sum_p B_{p}/T_{\text{eff}}-0\cdot \sum_{v} A_{v})/Z$. We can see that the fluctuation of $m$ particles is suppressed by the $L_{m,l}$ which acts like an effective temperature $T_{\text{eff}}$, while the $e$ particles are heated up to an infinite temperature configuration. We can restore quantum coherence by suppressing the fluctuations of $e$ particles as well as $m$ particles by the following quantum jump operators:
\begin{equation}
    \begin{aligned}
        L_{m,l}=&\sigma^x_lP(\sum_{p|l\in \partial p}B_{p});\\
        L_{e,l}=&\sigma^z_lP(\sum_{v|v\in\partial l }A_{v}).
    \end{aligned}
\end{equation}
Like the definition of $\sum_{p|l\in \partial p}B_{p}$, $\sum_{v|v\in\partial l }A_{v}$ gives the sum of two vertex operators connected by link $l$. Then the action on $e$ and $m$ particles are entirely symmetric. These quantum jump operators drive the system to their common dark space $\{|\psi \rangle|\ A_{v}|\ \psi\rangle=|\psi\rangle, B_p|\psi\rangle =|\psi\rangle\}$. Such methods of preparing steady states, that is, by projecting to the dark space of the quantum jump operators, are first proposed in the two seminal papers \cite{kraus2008preparation,verstraete2009quantum}. In our case, the steady states are the combinations of ground states of the original toric code model. The bases of the steady-state subspace are
\begin{equation}
\rho^{\mu\nu,\mu'\nu'}=|\psi^{\mu\nu}\rangle\langle\psi^{\mu'\nu'}|.
\end{equation}
Here $\mu,\nu,\mu', \nu'=0,1$ and $|\psi^{\mu\nu}\rangle$ is one of the four topological degenerate ground states in the toric code model (Eq. \eqref{eq:2dgs}). The physical steady state is a linear combination of diagonal parts and coherent parts. If we are not concerned about the trace, there are $D^{2}=4^{2\mathcal{G}}$ degenerate steady states, where $D=4^\mathcal{G}$ ($\mathcal{G}$ is the genus of the manifold, here $\mathcal{G}=1$ on a 2-torus) is the ground state degeneracy in the original toric code model. 

\subsubsection{Fragility of topological degeneracy:\\an exact solution of the steady state under perturbation}
Parallel to the discussion in Model-${1}$, we consider the effect of the following perturbation:
\begin{equation}
\begin{aligned}
L_{x,l}&=\sqrt{h_{x}}\sigma^{x}_{l},\\
L_{z,l}&=\sqrt{h_{z}}\sigma^{z}_{l}.
\end{aligned} 
\end{equation}
The steady state can also be exactly solved based on the following observations:

1. Here $L_{m,l}$ and $L_{x,l}$ are the same with Model-1. $L_{e,l}$ and $L_{z,l}$ do not change $m$ particles.

2. $L_{e,l}$ and $L_{z,l}$ act on $e$ particles. The action on $e$ and $m$ particles are symmetric.

The above discussion reveals that the steady state is diagonal with respect to the $e$ and $m$ particle configurations respectively.  Although the $e,m$ particles have nontrivial mutual statistics in the unitary case, i.e., when we drag one $e/m$ particle around another $m/e$ particle, the state acquires a minus sign, and the situation in the steady state is quite different. For steady states, the minus sign from braiding cancels out, and the dynamics of the $e/m$ particles become completely independent. In this sense, the model can be regarded as a doubled version of Model-${1}$, and the exact form of the steady state can be obtained by generalizing the results from the previous section.

\paragraph{$h_{z}=0, h_{x}\neq0$}
First, when $h_{z}=0$ and $h_{x}\neq0$, only $m$ particles proliferate, and one of the steady states is
\begin{equation}
\begin{aligned}
\rho_{m} &= \frac{1}{T'_{m}}\sum_{g,g'\in G'}g(\sum_{k,\{r\}}\beta_{m}^{2k}|m_{2k}(\{r\})\rangle \langle m_{2k}(\{r\})|)g' \\
&=\frac{1}{T'_{m}}\sum_{g,g'\in G'}g(\sum_{k,\{\partial \tilde{t}_{i}\}}\beta_{m}^{2k}\prod_{i=1}^{k}S^{x}_{\tilde{t}_{i}}|\Uparrow\rangle \langle \Uparrow|\prod_{i=1}^{k}S^{x}_{\tilde{t}_{i}})g'.
\end{aligned}
\end{equation}
Here, $S^{x}_{\tilde{t}_{i}}=\prod_{l\in \tilde{t}_{i}}\sigma^{x}_{l}$ is the string operator that creates two $m$ particles living on the endpoints of the string $\tilde{t}_{i}$, where $B_{\partial \tilde{t}_{i}}=-1$, and $2k$ gives the number of $m$ particles. $\beta_{m}=\sqrt{\frac{h_{x}}{1+h_{x}}}$ is the decaying coefficient and $T'_{m}$ is the normalization constant,
\begin{equation}
\begin{aligned}
 T_{m}'=|G'|\sum_{k=0}^{\lfloor\frac{n}{2}\rfloor}\beta_{m}^{2k}{n\choose 2k}&=4|G|\frac{(1+\beta_{m})^n+(1-\beta_{m})^n}{2}.
\end{aligned}
\end{equation}
In the summation, we require that no strings share any common endpoints, and then, the distribution of $m$ particles in $\rho_m$ is the same as in Eq. \eqref{eq:2dss1}. The difference here is the absence of $e$ particles, which follows the fact $A_{s}\rho=\rho A_{s}=\rho$. Indeed, there are four degenerate steady states on a torus. We explicitly give their expression: 
\begin{equation}
\begin{aligned}
\rho_{m}^{\mu\nu}&= \tilde{W}^{\mu}_{x}\tilde{W}^{\nu}_{y}\rho_{m}\tilde{W}^{\nu}_{y}\tilde{W}^{\mu}_{x},\quad  \mu,\nu = {0,1}\\
\tilde{W}_{x}&=\prod_{i\in \gamma_{x}} \sigma^{x}_{i}, \ \tilde{W}_{y}=\prod_{i\in \gamma_{y}} \sigma^{x}_{i}.
\end{aligned}
\end{equation}
Here $\tilde{W}_{x}$ and $\tilde{W}_{y}$ are non-contractible 't Hooft loop operators. $\gamma_{x}$ and $\gamma_{y}$ are non-contractible loops on the lattice (See Fig. \ref{fig:2dexcitation2}). In the special case where $h_{x}=0$ and $h_z=0$ (the unperturbed case), the 16-fold steady states can also be rewritten as $\rho_m^{\mu\nu,\mu'\nu'}=\tilde{W}^{\mu}_{x}\tilde{W}^{\nu}_{y}\rho_{m}\tilde{W}^{\nu'}_{y}\tilde{W}^{\mu'}_{x}$ and $\mu,\nu,\mu',\nu' = {0,1}$. By turning on $h_x$, the degeneracy is broken down to 4 by requiring $\mu=\mu',\nu=\nu'$, the non-contractible 't Hooft loops have no effect on the dynamics of the spin-flip term. 

\paragraph{$h_{x}=0, h_{z}\neq0$}
Similarly, when $h_{x}=0$, $h_{z}\neq 0$, only $e$ particles proliferate. The steady  state is
\begin{equation}
\begin{aligned}
\rho_{e} = \frac{1}{T'_{e}}\sum_{k,\partial t_{i}}\beta_e^{2k}\prod^{k}_{i=1}S^{z}_{t_{i}}\left[\sum_{g\in G}g|\Uparrow\rangle\langle \Uparrow|\sum_{g'\in G}g'\right]\prod^{k}_{i=1}S^{z}_{t_{i}}.
\end{aligned}
\end{equation}
$\beta_{e}=\sqrt{\frac{h_{z}}{1+h_{z}}}$ is the decaying coefficient and $T'_{e}$ is the normalization constant,
\begin{equation}
 T_{e}'=|G'|\sum_{k=0}^{\lfloor\frac{n}{2}\rfloor}\beta_{e}^{2k}{n\choose 2k}=4|G|\frac{(1+\beta_{e})^n+(1-\beta_{e})^n}{2}.
\end{equation}
Here, $S^{z}_{t_{i}}=\prod_{l\in t_{i}}\sigma^{z}_{l}$ is a string operator and $t$ is the path on the lattice (See Fig. \ref{fig:2dexcitation1}.(b)). $S^{z}_{t}$ can create two $e$ particles at the endpoints of $t_{i}$, where $A_{\partial t_{i}}=-1$.   In the summation, we require that no strings share a common endpoint. The steady states are 4-fold degenerate:
\begin{equation}
\begin{aligned}
\rho_{e}^{\mu\nu}&= W^{\mu}_{x}W^{\nu}_{y}\rho_{e}W^{\nu}_{y}W^{\mu}_{x},\quad  \mu,\nu = {0,1}\\
W_{x}&=\prod_{i\in \tilde{\gamma}_{x}} \sigma^{x}_{i}, \ W_{y}=\prod_{i\in \tilde{\gamma}_{y}} \sigma^{x}_{i}.
\end{aligned}
\end{equation}
$W_{x}$ and $W_{y}$ are non-contractible Wilson loop operators. $\tilde{\gamma}_{x}$ and $\tilde{\gamma}_{y}$ are non-contractible loops on the dual lattice [See Fig. \ref{fig:2dexcitation1}(a)].

\paragraph{$h_{x}\neq 0$, $h_{z}\neq 0$}
Finally, when $h_{x}\neq 0$, $h_{z}\neq 0$, both kinds of particles proliferate, and since they are completely independent, we can simply combine the solution in the above two cases. The steady state is
\begin{widetext}
\begin{equation}
\begin{aligned}
\rho_{em} &= \frac{1}{T'_{em}}\sum_{\mu,\nu}\sum_{k_e,\{r_{e}\}}\beta_{e}^{2k_e} \prod_{j=1}^{k_e}S^{z}_{t_{j}}\sum_{g,g'\in G}g\left(\sum_{k_m,\{p_m\}}\beta_{m}^{2k_m}   W_x^\mu {W_y}^\nu\prod_{i=1}^{k_m}S^{x}_{\tilde{t}_{i}}|\Uparrow\rangle \langle \Uparrow|\prod_{i=1}^{k_m}S^{x}_{\tilde{t}_{i}}W_x^\mu W_y^\nu\right)g'\prod_{j=1}^{k_e}S^{z}_{t_{j}}, \\
% &= \frac{1}{T'_{em}}\sum_{k_m,\{p_m\}}\beta_{m}^{2k_m}\sum_{g,g'\in G}(-1)^{C(g,\{p_m\})}g\sum_{k_e,\{r_e\}}\beta_{e}^{2k_e}\sum_{\mu,\nu}W_x^\mu W_y^\nu|e_{2k_e}(\{r_e\})\rangle\langle e_{2k_e}(\{r_e\})|{W_x}^\mu {W_y}^\nu g'(-1)^{C(g',\{p_m\})}
\end{aligned}
\end{equation}
\end{widetext} 
where $p_m=\partial\tilde{t}_{i}$ and $r_e=\partial t_{i}$ are positions of $m$ and $e$ particles. $\beta_{e}$ and $\beta_{m}$ are given in the above cases. $T_{em}'$ is the total normalization constant,
\begin{equation}
T_{em}'=|G'|\sum_{k_{e}=0}^{\lfloor\frac{n}{2}\rfloor}\beta_{e}^{2k}{n\choose 2k_{e}}\sum_{k_{m}=0}^{\lfloor\frac{n}{2}\rfloor}\beta_{m}^{2k}{n\choose 2k_{m}}=\frac{T'_{e}T'_{m}}{|G'|}.
 % =&|G|\left((1+\beta_{e})^n+(1-\beta_{e})^n\right)\left((1+\beta_{m})^n+(1-\beta_{m})^n\right)
\end{equation}
It is clear that  $\rho_{em}$ has all possible distributions of  $e$ and $m$ excitations created by string operators $S^{z}_{t_{i}}$ and $S^{x}_{\tilde{t}_{j}}$. Though it is diagonal in the  $e$ and $m$ distribution space ($\{r_{e}\}$ and $\{p_{m}\}$),  there are quantum coherences between different spin configurations generated by $g$ and $g'$ which actually contribute to the same particle distribution.

In conclusion, topological degeneracy is sensitive to perturbations. In the absence of perturbation, the 16-fold degeneracy arises from the contributions of both gauge structures,  with $e$ and $m$ being the corresponding charges. However, once we introduce fluctuations in either type of charge, the degeneracy is reduced to 4-fold, and it is completely lifted when both $e$ and $m$ particles are allowed to fluctuate.

\subsubsection{Topological entropy}
As shown in Sec. \ref{sec:2dentropy}, Levin-Wen's definition of topological entropy is more suitable for characterizing topological order in open systems, and we use their definition for our calculation. It is worth noting that in dissipative systems, the topological entropy defined by Levin and Wen is closely related to the topological degeneracy of steady states.

\paragraph{$h_x=h_z=0$}
Firstly, we calculate the topological entropy in the unperturbed case $h_x=h_z=0$. As discussed before, the steady state degeneracy is 16 on the 2-torus. Since they are locally indistinguishable, the subsystem entropy of any region $A$ should be identical for all the steady states. We can simply take one of them
\begin{equation}
\rho_{ss}=\frac{1}{|G|}\sum_{g,g'\in G}g|\Uparrow\rangle\langle \Uparrow|g',
\end{equation}
for calculation, which is just the ground state of the toric code model. The calculation follows that in Ref. \cite{hamma2005ground} and the result is simple:
\begin{equation}
    S_{A}=\log\frac{|G|}{|G_A||G_{\bar A}|}.
\end{equation}
Using $|G|=2^{n-1},|G_A|=2^{|A|+p_{\bar A}-1},|G_{\bar A}|=2^{|\bar A|+p_A-1}$, we finally get
\begin{equation}
    S_A=\left(|\partial A|+1-p_A-p_{\bar A}\right)\log 2.
\end{equation}
Here the entanglement entropy follows the area law and the TEE is obviously the same as the Hermitian case:
\begin{equation}
\begin{aligned}
S_{\text{topo}}=&(p_{A_{1}}+p_{\bar A_{1}}-p_{A_2}-p_{\bar{A}_2}\\
& -p_{A_3}-p_{\bar A_3}+p_{A_4}+p_{\bar A_4})\log 2\\
=&2\log 2.
\end{aligned}
\end{equation}
That is twice the topological entropy of the unperturbed Model-${1}$ (Eq. \eqref{eq:lwtopo1}) since both gauge structures with charges $e$ and $m$ contribute a $\log2$ in this model. That is also reflected in the fact that the steady-state topological degeneracy is the square of that of Model-${1}$, where only the gauge structure with charge $m$ is in the game. 

\paragraph{$h_z=0, h_x\neq0$}
Next, we allow $m$ $(e)$ to fluctuate by turning on $h_x$ $(h_z)$. Since they are equivalent to each other, without loss of generality, we only discuss the case $h_z=0, h_x\neq0$. After tedious calculation in Appendix \ref{sec:appendixqm}, we get the topological entropy
\begin{equation}
    S_{\text{topo}}=\log 2,
\end{equation}
which is half of the topological entropy of the unperturbed Model-2, and identical to that of unperturbed Model-1. This is closely related to the reduction of the steady state degeneracy from $16$ to $4$. This result again shows the rationality of generalizing the topological entanglement entropy defined by Levin and Wen for steady states in open systems. 

\paragraph{$h_x\neq0, h_z\neq0$}
We learn that the topological entropy is vulnerable to the fluctuation of $m$ particles. The remaining topological entropy as well as the topological degeneracy is due to the fact that the $e$ particles remain unaffected. Therefore we expect that the topological entropy completely vanishes once the $e$ particles also begin to fluctuate by turning on $h_{z}$. The calculation in Appendix \ref{sec:appendixqm}  finally confirms our prediction:
\begin{equation}
    S_{\text{topo}}=0.
\end{equation}
That is, in the thermodynamic limit the topological entropy immediately drops to zero under perturbations on both gauge sectors, as expected from the fact that the topological degeneracy also completely lifted at the same time.

\subsection{Discussion about the topological degeneracy in 2d}\label{sec:2dfragile}
In two dimensions, we have constructed Model-1 and Model-2 to demonstrate the degeneracy of steady states. We establish the correspondence between topological degeneracy and topological entropy of steady states and also identify the long-time relaxation dynamics as a lattice gauge theory.  

For our two specific models in two dimensions, we show that the steady-state topological degeneracy is fragile. Nevertheless, this is expected to be a generic phenomenon in two dimensions. Below, we present our argument.

Previous studies of topological order have shown that topologically ordered phases are characterized by topological excitations that cannot be created or annihilated by local operators (excluding the case of invertible topological order). Our primary objective is to stabilize such topologically ordered phases through local dissipation. In this case, these topological excitations would manifest as defects such as the $e$ and $m$ particles in our dissipative toric model. However, a proliferation of such topological defects would lead to the destruction of the topological degeneracy, prompting us to seek ways for the dissipators to eliminate these defects. As local annihilation of topological defects is infeasible, our feasible local operations involve moving the defects and annihilating them in pairs when two defects come into contact. However, if we only allow local dissipators and there is no long-range interaction between defects, a defect would not be aware of the presence of other defects except its neighbors. In other words, the defects are deconfined, making it impossible to directly bring two defects together. Therefore, it seems these defects can only wander around randomly, which is exactly the case in the two models constructed above. Consider a situation where we acquire a state in a specific topological sector; it is inevitable that random noise would create pairs of defects somewhere. Given the aforementioned discussion, there are no particular constraints preventing the defects from separating and winding around, and eventually, the initial state evolves into another topological sector.

%One possibility that was not considered in the previous discussion is that the defects can be localized in a specific region, where they can then be annihilated, rather than moving randomly. For example, under open boundary conditions, we can restrict the defects to only move rightwards or upwards, they eventually gather at the upper-right corner. This could potentially make the topological degeneracy more robust. However, this approach cannot be directly applied to a torus with no boundary or corner. To gather the defects at a specific point on a torus, we would need to introduce two branch cuts where the topological defects can move in either direction. In this case, the topological degeneracy would be robust against noise in most areas, but perturbations near the two branch cuts would likely lift the topological degeneracy.

\section{Open systems with topologically degenerate steady states on 3-torus}\label{sec:3dmodel}
% In the previous section, we argued that achieving robust topological degeneracy of steady states under local dissipative dynamics in two dimensions is challenging, due to deconfined point-like topological defects. However, in higher dimensions, loop-like topological excitations, such as loop excitations of $m$-defects in 3d toric code model or vortex loops in 3d BCS superconductors with dynamic gauge fields, have an extensive energy cost with increasing linear size, resulting in confinement to small lengths at low energy. Thus, this essential feature may be used to stabilize dissipative topological order against perturbation in higher dimensions. In this section, we will demonstrate that robust topological degeneracy of steady states can  be achieved in Lindblad systems. Additionally, analysis of the emergent gauge field reveals a deconfined phase.
In the last section, we argued that achieving robust topological degeneracy of steady states under local dissipative dynamics in two dimensions is particularly challenging due to the presence of point-like topological defects. However, in higher dimensions, there also exist extended topological defects, exemplified by loop excitations in the 3d toric code model and vortex loops in 3d BCS superconductors with dynamic gauge fields. Such topological defects cost an extensive amount of energy with increasing linear size and thus are confined to small lengths at low energy. This feature is essential for the stability of classical topological memory at low temperature. For non-equilibrium dynamics in generic open quantum systems, such extended topological defects can also be dynamically suppressed using engineered dissipation, thus stabilizing dissipative topological order against perturbation in higher dimensions. In this section, we demonstrate the achievement of robust topological degeneracy of steady states in three-dimensional open quantum systems. Moreover, our analysis of the emergent gauge theory reveals a stable deconfined phase.

\subsection{Model-1\label{sec:3dclassical}}

\subsubsection{Topologically degenerate steady states}

To realize steady states with robust topological degeneracy, we design the following set of quantum jump operators:
\begin{equation}
\begin{aligned}
%L_{l}&=\sigma^{x}_l\left(1-\frac{1}{2}\sum_{p:l\in \partial p}{B_p}\right)\left(1-\frac{1}{4}\sum_{p:l\in \partial p}{B_p}\right), \ {B_p}=\prod_{l\in\partial p }\sigma^{z}_{l}\\
L_{m,l}&=\sigma^{x}_lP(\sum_{p|l\in\partial p }B_{p}),\\
L_{z,l}&=\sqrt{\kappa_{z}}\sigma^{z}_{l},\\
 L_{v}&=\sqrt{\kappa_{v}}A_{v},\label{3Dclassical}
\end{aligned}
\end{equation}
where $\kappa_{z}$ ($\kappa_{v}$) is the dissipative strength and it is the same for all links (vertices). These three sets of quantum jump operators are of the same form as Eq. \eqref{eq:2dclassical} in the 2d case. In 3d, $\sum_{p|l\in p}B_p$ gives the sum of four plaquette operators around link $l$ and each vertex connects to six links. $P(x)$ is the same projection operator in Eq. \eqref{eq:projection}: 
\begin{equation}
P(x)\equiv\left\{
\begin{aligned} 
0,\quad &x>0;\\
1,\quad &x<0;\\ 
1/\sqrt{2},\quad &x=0. 
\end{aligned}
\right.\nonumber
\end{equation}
\begin{figure}[htb]
\centering
\includegraphics[width=0.95\linewidth]{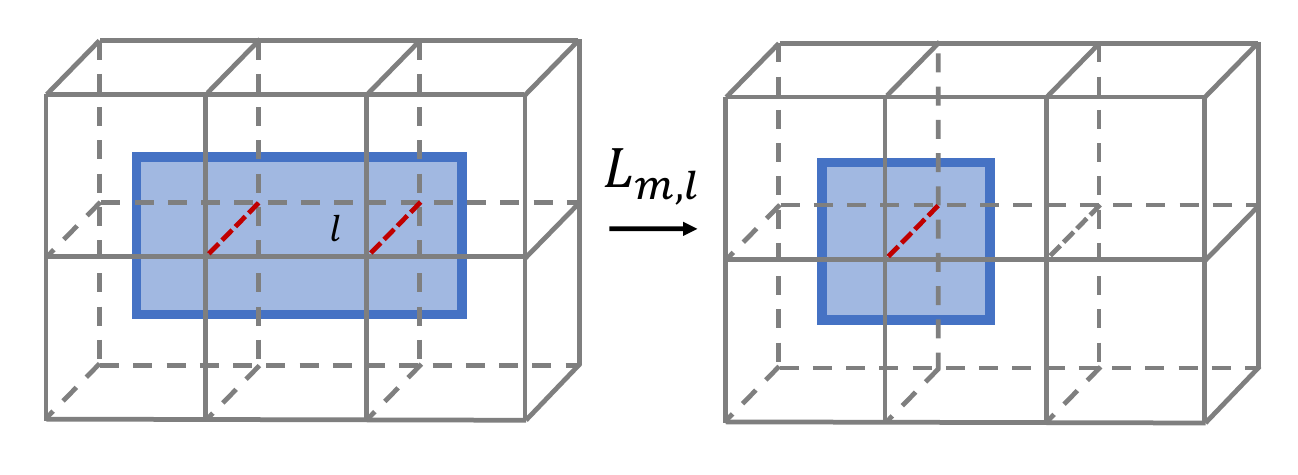}
\caption{The action of operator $L_{m,l}$ on loop defects. $L_{m,l}$ can move and annihilate $m$ excitations, which leaves the length of the dual loop non-increasing.}
\label{fig:3dexcitation2}
\end{figure}

In Fig. \ref{fig:3dexcitation2}, we can see that for any link $l$ there are four plaquettes around it and the sum of $B_p$ around $l$ has five possible values $4,2,0,-2,-4$. The form of the quantum jump operator $L_{m,l}$ is to ensure that only the spins on link $l$ with $\sum_{p|l\in\partial p }B_{p}\leq 0$ can be flipped. Consequently, $L_{m,l}$ has a dark space spanned by all closed membrane states, i.e., $L_{m,l}|C\rangle=0$ where $|C\rangle=g|\Uparrow\rangle$, since $B_{p}|C\rangle=|C\rangle,\forall p$. Then, with the effect of dephasing term $L_{z,l}$, all off-diagonal elements are damped, which makes all diagonal closed membrane states $|C\rangle\langle C|$ become steady states. Finally, $L_{v}=\sqrt{\kappa_{v}}A_{v}$ mixes all possible closed membrane states. In this way, we get steady states in the trivial sector with all contractible closed membranes: $\rho_{ss}=\frac{1}{2^{N}}\sum_{C}|C\rangle\langle C|$. Other degenerate steady states with non-contractible closed membranes can be generated as follows:
\begin{widetext}
\begin{equation}
    \rho_{ss}^{\{\mu_{i}\}}=V_{xy}^{\mu_{1}} V_{yz}^{\mu_{2}}V_{zx}^{\mu_{3}}\frac{1}{2^{N}}\sum_{\{v\}}\left(\prod_{v}A_{v}|\Uparrow\rangle\langle \Uparrow|\prod_{v}A_{v}\right)V_{xy}^{\mu_{1}} V_{yz}^{\mu_{2}}V_{zx}^{\mu_{3}},\label{eq:3dss}
\end{equation}
\end{widetext}
where $\mu_{i}=0,1$ and $N$ is the number of vertices on the 3-torus. $V_{ij}=\prod_{l\perp ij}\sigma^x_l$ creates a non-contractible membrane in the $ij$-plane. There are in total 8 topological sectors distinguished by $\{\mu_{i}\}$ which is the parity of non-contractible membranes.

\begin{figure}[htb]
 \centering
\includegraphics[width=1\linewidth]{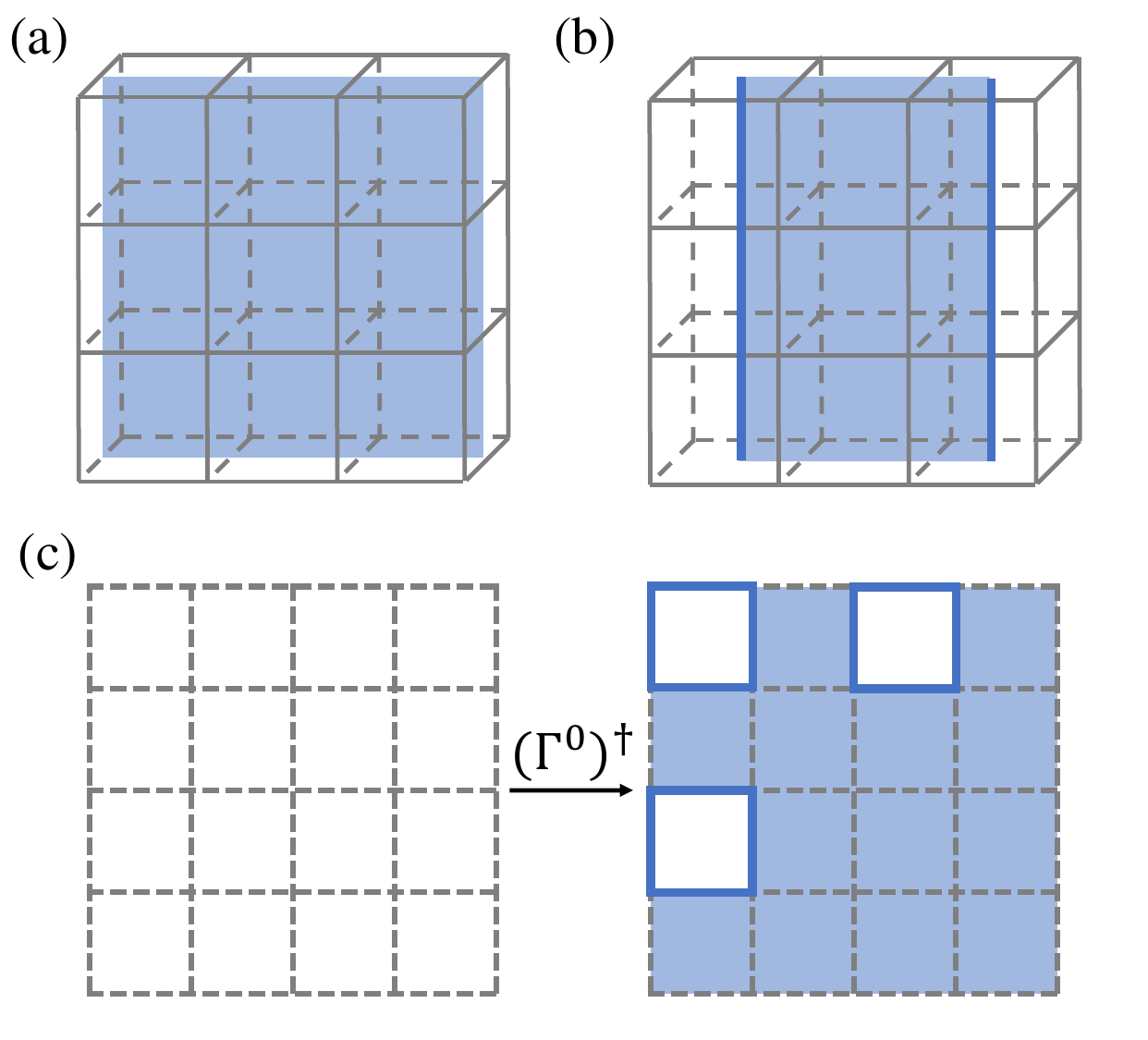}
\caption{
 (a) Type-A states with closed membranes on 3-torus. Here the blue surface gives a type-A state with a non-contractible closed membrane. (b) Type-B states with open membranes. The blue strip is a non-contractible open membrane with $m$ defects (blue solid line) along the boundary. (c) Left steady state configurations on the dual lattice. Left panel: a right steady-state configuration in the trivial sector $|ss^{0}_{1}\rangle_{R}$. Right panel: a configuration in $|ss^{0}_{1}\rangle_{L}$ }
\label{fig:typeAB}
\end{figure}

However, the dark space of $L_{m,l}$ contains more than just closed membrane states. It also includes some open membrane states, where the boundary must form a non-contractible loop, resulting in an exponentially large number of additional steady states. For later convenience, we refer to the closed membrane steady states as type-A states, and the open membrane steady states as type-B states (Fig. \ref{fig:typeAB}). However, we show in the following section that the degeneracy of type-B steady states is instantly lifted under perturbation. In contrast, the 8-fold topological degeneracy of type-A states is robust against any weak local perturbations, indicating the presence of dissipative topological order (DTO).

\subsubsection{Robustness of topological degeneracy\label{sec:3drobust}}
In this section, we aim to study the influence of local perturbations on steady states. First, we note that this model, similar to the discussion of Model-1 in Sec. \ref{sec:model1}, can be reduced to a classical Markovian dynamics, with the generator
\begin{equation}
\Gamma^{0}=\sum_l(\sigma^x_l-1)P^{2}(\sum_{p|l\in\partial p }B_{p})+\kappa_{v}\sum_{v}(A_{v}-1).\label{eq:gamma0}
%+\sum_{l}\kappa_{z}(\sigma^{z}-1).
\end{equation}
Since the steady states are all reduced to the diagonal space, $L_{z,l}$ can be ignored.

We already know that the first term is given by $L_{m,l}$ and it only acts on the links with $\sum_{p|l\in\partial p }B_{p}\leq 0$. This means that starting from any states with $m$ excitations, the loop length (boundary of open membranes on the dual lattice) is non-increasing under this dissipative process. Now, we add some local perturbation
\begin{equation}
    L_{x,l}=\sqrt{h}\sigma^x_l,
\end{equation}
and this leads to $\delta \Gamma=h\sum_l(\sigma^x_l-1)$, which can create open membranes on the dual lattice by flipping spins. Consider that we start from the steady state configuration and let the system evolve under the perturbed Markovian generator $\Gamma^{0}+\delta \Gamma$. First, $\delta \Gamma$ flips spins with equal probability, which creates open membranes. Second, $\Gamma^{0}$ keeps the loop length (boundary of open membrane) non-increasing and prefers to make it decrease. When the perturbation $h$ is weak, $h\ll 1$, one can imagine that the boundaries of configurations in steady states are still restricted to small lengths. 

In contrast to our model in 2d discussed in Sec. \ref{sec:model1} where the topological degeneracy is vulnerable to perturbations, in 3d, the steady-state topological order persists under weak local perturbations.  
% We demonstrate that the transition amplitude for states in two topological sectors $\{\mu_{i}\}$ and $\{\mu'_{i}\}$ is exponentially small and the 8-fold type-A topological degeneracy is robust against weak local perturbations in our 3d model. 
The robust topological degeneracy of steady states also means that for two states from different topological sectors, it takes an exponentially long time to become indistinguishable. On the other hand, the additional degeneracy of type-B states is broken under perturbation. These states arise in the unperturbed model because the loop length of the boundaries of open membranes is restricted to be non-increasing. This restriction causes the evolution to get stuck when these boundaries form non-contractible loops, resulting in metastable configurations. However, these loops are extensive, and the restriction is released by perturbation, causing the non-contractible loops to shrink further and leading to the emergence of robust steady states.     
The hand-waving argument above can be justified by degenerate perturbation theory in the degenerate steady-state subspace $S_{0}$ spanned by type-A and type-B states. Our goal is to obtain the effective Markovian generator $\Gamma_{\text{eff}}$ in the subspace $S_{0}$. Following the standard perturbation theory, we need to calculate matrix elements of  $\delta \Gamma$ to $n^{\text{th}}$ order:
\begin{equation}
\begin{aligned}
_L\langle ss^{0}_i|\delta \Gamma|ss^{0}_j\rangle_R,\ \cdots,\ _L\langle ss^{0}_i|\delta \Gamma (V\delta \Gamma)^{n-1} |ss^{0}_j\rangle_R, 
\label{eq:perturbation}
\end{aligned}
\end{equation}
where $V=\left[(1-P)\Gamma^0(1-P)\right]^{-1}$ and $P$ is the projection operator of the steady-state subspace $S_{0}$, $P=\sum_i |ss^{0}_i\rangle_R\langle ss^{0}_i|_L$. Here $|ss^{0}_i\rangle_{R(L)}$ denotes the $i^{th}$ right (left) steady state of $\Gamma^{0}$:
\begin{equation}
\begin{aligned}
    \Gamma^{0}|ss^{0}_i\rangle_{R}&=0,\quad P_{A}=\sum_{i\in A} |ss^{0}_i\rangle_R\langle ss^{0}_i|_L,\\
    _L\langle ss^{0}_i|\Gamma^{0}&=0,\quad P_{B}=\sum_{i\in B} |ss^{0}_i\rangle_R\langle ss^{0}_i|_L.
    \end{aligned}
\end{equation}
We assume $|ss^{0}_{1\leq i\leq 8}\rangle_{R}$ stands for the eight type-A states and $|ss^{0}_{i>8}\rangle_{R}$ represents those type-B states. $P_{A(B)}$ is the projection operator of type-A (B) steady states and we have $P=P_{A}+P_{B}$.  The general form of the effective generator is
\begin{equation}
\Gamma_{\text{eff}}=\left(\begin{array}{ccc}
h_A & h_{AB} \\
 h_{BA} & h_B
\end{array}\right),
\end{equation}
where the block $h_A=P_A\Gamma_{\text{eff}}P_A$ and  $(h_{A})_{ij}$ are matrix elements within the subspace of type-A states. $h_{BA}=P_B\Gamma_{\text{eff}}P_A$ gives the transition amplitude from the right type-A states to the left type-B states, and similar for $h_B$ and $h_{AB}$. We should note that similar to the 2d situation in Sec.\ref{sec:2dperturbation}, $\Gamma_{\text{eff}}$ may not really govern the long-time dynamics of the system, since we find that there is no finite Liouvillian gap to separate the degenerate steady-state subspace $S_{0}$ and other eigenstates (the difference between degenerate states in $S_{0}$ and other gapless states to be discussed in Sec. \ref{sec:lifetime}). However, this analysis helps us understand why the type-A configurations are stable, exhibiting robust topological degeneracy, whereas type-B configurations are metastable and will eventually relax to type-A configurations under perturbations. 

Since the Markovian generator $\Gamma^{0}$ is non-Hermitian, the left steady state and right steady state are not conjugate in general. Therefore, we need to figure out the left steady state $|ss^{0}_i\rangle_L$. Although it is difficult to get the exact form of left steady states, we are able to obtain sufficient information for our analysis. Assuming the right steady states $|ss^{0}_i\rangle_R$ are chosen to be bi-orthonormal, we have the following identities: 
\begin{equation}
\begin{aligned}
&|ss^{0}_i\rangle_L=\lim_{t\rightarrow \infty}e^{(\Gamma^{0})^\dagger t}|ss^{0}_i\rangle_R,\quad
_L\langle ss^{0}_i|ss^{0}_j\rangle_R=\delta_{ij}.\\
&(\Gamma^{0})^{\dagger}=\sum_lP^{2}(\sum_{p|l\in\partial p }B_{p})(\sigma^x_l-1)+\kappa_{v}\sum_{v}(A_{v}-1).
\end{aligned}
\end{equation}
Starting from the right steady state, evolving the state under  $(\Gamma^{0})^{\dagger}$, we can get the corresponding left steady state. In contrast to the dynamics generated by $\Gamma^{0}$, the length of loop excitation is non-decreasing under the evolution generated by $(\Gamma^{0\dagger})$. For example, start from a trivial type-A state $|ss^{0}_1\rangle_R$, where there are no loop excitations and no non-contractible membranes. $(\Gamma^{0})^{\dagger}$ flip spins on link $l$ with $\sum_{p|l\in \partial p}B_{p}\geq 0$. Loops can be created and they expand to achieve a larger perimeter. Due to this dynamics, we can see that all possible configurations contained in $|ss^{0}_1\rangle_L$ are ``far from" another type-A configuration in $|ss^{0}_{2\leq i\leq 8}\rangle_R$. It means that starting from any configuration contained in $|ss^{0}_1\rangle_L$, it takes at least $O(L)$ local operations to form a non-contractible membrane configuration which appears in $|ss^{0}_{2\leq i\leq 8}\rangle_R$. Here $L$ is the linear size of the cubic. 

For example, starting from a configuration in $|ss^{0}_{1}\rangle_R$ [left panel in Fig. \ref{fig:typeAB}(c)] and evolving under $(\Gamma^{0})^{\dagger}$ which makes the loop-length non-decreasing,  we can get all possible configurations in the left steady state and among them, the right panel in Fig. \ref{fig:typeAB}(c) is the closest one to configurations in $|ss^{0}_{2}\rangle_R$ which have non-contractible closed membrane [Fig. \ref{fig:typeAB}(a)]. We can see that it requires about $L/2+L/2=L\sim O(L)$ local operations to connect these two configurations from different topological sectors. Therefore, the off-diagonal terms of $h_A$ are exponentially small: $(h_{A})_{i\neq j}\sim O(e^{-\alpha_{1}L}) $. By similar approaches, one can show all the elements in $h_{BA}$ are also exponentially small: $(h_{BA})_{ij}\sim O(e^{-\alpha_{2}L})$. $\alpha_{1}$ and $\alpha_{2}$ are finite constants depending on the perturbation strength $h$. Also, the diagonal terms in $h_A$ are finite and identical. This can be easily seen by noting that different type-A steady states are related by applying $V_{xy}$, $V_{yz}$, or $V_{zx}$ introduced in Eq. \eqref{eq:groundstate}, which commutes with $V$ in Eq. \eqref{eq:perturbation} as well as $\delta \Gamma$. This tells us that the 8-fold topological degeneracy of type-A states is robust against perturbation. 

However, for the other two blocks $h_{B}$ and $h_{AB}$, the perturbation affects them in the first order $(h_B)_{ij}\sim O(hN)$ and $(h_{AB})_{ij}\sim O(hL)$. This implies that type-B states can easily evolve into either type-A or other type-B states under perturbation. Therefore, the degeneracy of type-B states is easily broken under perturbation, and these states are merely metastable. On the contrary, type-A steady states are robust against perturbation and remain in their original sectors. Finally, only the 8-fold topological degeneracy of type-A steady states is robust.

The previous analysis raises an important question: is the steady-state degeneracy precisely 8 or are there any nearly degenerate steady states left under perturbation, such as some superposition of type-B states? Here degenerate states refer to those states with exponentially small eigenvalues relative to the linear size of the system, as is the case for the 8-fold type-A steady states. This leads us to inquire whether the type-B metastable states can exhibit an exponentially long lifetime. We will explore this problem in Sec. \ref{sec:lifetime}.  Our findings reveal that type-B states possess algebraically long lifetimes with respect to the linear size, indicating the Liouvillian gap decays algebraically with system size. This is qualitatively different from the exponentially small splitting of the 8-fold degenerate steady states. Hence, it is appropriate to refer to type-B states as metastable states.  The topological degeneracy of steady states comes with an algebraically vanishing gap which is distinct from the topological order in the Hermitian system where there is a finite gap. This intriguing characteristic in open quantum systems will be further elucidated in Sec. \ref{sec:relaxation}

\subsubsection{Dissipative deconfined $Z_2$ gauge field}
In closed systems, one of the characteristics of topologically ordered phases is the emergence of a deconfined gauge field \cite{kogut1979introduction,savary2016quantum, gregor2011diagnosing}. In this section, we demonstrate the emergence of $Z_2$ gauge field in the long-time dynamics of our dissipative model. %Following the procedure in Ref. \cite{kogut1979introduction}. 
By studying the steady-state property of the gauge field using perturbation theory as well as numerical simulation, we reveal the presence of a stable deconfined phase, consistent with the robust topological degeneracy. Furthermore, we show that increasing the perturbation strength $h$ can drive a phase transition to the confined phase, which corresponds to the breaking of the topological degeneracy.

First, following the same procedure in Sec. \ref{sec:2dgauge} we give the full perturbed Markovian generator $\Gamma=\Gamma^0+\delta \Gamma$:
\begin{equation}
\begin{aligned}
 \Gamma=\sum_l(\sigma^{x}_l-1)P^{2}(\sum_{p|l\in \partial p}B_{p})+\kappa_{v}\sum_{v}(A_{v}-1)&\\
 +h\sum_l(\sigma^x_l-1)&.
 \label{eq:eff}
 \end{aligned}
\end{equation}
Like the analysis in 2d in Sec. \ref{sec:2dgauge}, we find the model has a local $Z_2$ symmetry: $[\Gamma, A_{v}]=0$. Then the steady state must lie in the $A_{v}=1$ subspace and $\Gamma$ can be reduced to
\begin{equation}
    \Gamma|_{A_{v}=1} = \sum_l(\sigma^{x}_l-1)P^{2}(\sum_{p|l\in \partial p}B_{p})+h\sum_l(\sigma^x_l-1).\label{eq:eff2}
\end{equation}
With the gauge invariance condition, we obtain a (non-Hermitian) $Z_2$ gauge theory in the dissipative system. As it is known in the research of closed systems, the topologically ordered phase usually corresponds to the deconfined phase in the lattice gauge theory formulation. From the robustness of topological degeneracy in 3d, we expect a deconfinement-confinement transition in our 3d dissipative model at some finite $h$, in contrast to the 2d case where there is no stable deconfined phase. %In Sec. \ref{sec:2dgauge}, we have learned that our 2d model is always in the confined phase when the perturbation is turned on.

Similar to the 2d case, the Wilson loop operator is defined as follows:
 \begin{equation}
 \begin{aligned}
W_{\gamma}=\prod_{l\in\gamma}\sigma_{l}^{z},\quad
\langle W_{\gamma}\rangle =\frac{\langle I|  W_{\gamma}|ss\rangle}{\langle I|ss\rangle}  = \frac{\text{Tr}(  W_{\gamma}\rho_{ss})}{\text{Tr}(\rho_{ss})},
 \end{aligned}
 \label{eq:Wilsonloop}
 \end{equation}
where $\gamma$ is a closed loop (See Fig. \ref{fig:Wilson loop}).

 \paragraph{$h\ll1$}
In this limit, we can treat $\delta \Gamma$ as a small perturbation. Although the exact form is hard to get, the steady state can be written in perturbative expansion $|ss\rangle=\sum_n|ss^{(n)}\rangle$ . 
We choose the $0^{\text{th}}$ order $|ss^{(0)}\rangle$ from the trivial sector:
\begin{equation}
|ss^{(0)}\rangle=\prod_{v}\frac{1+A_{v}}{2}|\uparrow\uparrow\cdots\uparrow\rangle.
\end{equation}
With the steady state equation $(\Gamma^0+\delta\Gamma)\sum_n|ss^{(n)}\rangle=0$, the higher orders are as follows:
\begin{equation}
\begin{aligned}
% &1^{st} \ \text{order}: \Gamma^0|ss^{(1)}\rangle+\delta \Gamma|ss^{(0)}\rangle =0,\\
% &2^{nd} \text{ order}: \Gamma^0|ss^{(2)}\rangle+\delta \Gamma|ss^{(1)}\rangle=0,\\
% &\cdots\\
n^{\text{th}} \text{ order}: \Gamma^0|ss^{(n)}\rangle +\delta\Gamma |ss^{(n-1)}\rangle=0.\label{eq:perturbexpand}
\end{aligned}
\end{equation}
These states can be labeled by the number of the loop configuration (boundary of the open membrane on the dual lattice) and the loop length. For example,
\begin{equation}
|ss^{(1)}\rangle=h\sum_O |\text{one 4-loop}\rangle,
\end{equation}
where $n$-loop means there is a loop excitation with its length equal to $n$ and $O$ represents a specific configuration with open membranes whose boundary is the $n$-loop.
The form of the higher-order solution is complicated and even in $2^{\text{nd}}$, the expression is cumbersome:
\begin{equation}
\begin{aligned}
|ss^{(2)}\rangle=&\alpha^{(2)}\ \sum_O|\text{two independent 4-loops}\rangle\\
&\quad +\beta^{(2)}\sum_O|\text{two adjacent 4-loops}\rangle+\cdots.
\end{aligned}
\end{equation}
% \begin{equation}
% \begin{aligned}
% |ss^{(2)}\rangle=&\alpha^{(2)}\ \sum_O|\text{two independent 4-loops}\rangle\\
% &\quad +\sum_O\text{weight}(O)|\text{two adjacent 4-loops}\rangle\\
% &\quad +\sum_O\text{weight}(O)|\text{one 8-loop}\rangle\\
% &\quad +\sum_O\text{weight}(O)|\text{one 6-loop}\rangle\\
% &\quad +\beta^{(2)}\sum_O|\text{one 4-loop}\rangle.
% \end{aligned}
% \end{equation}
Here, we denote loops that do not merge into a larger loop under the action of $\Gamma^{0}$ as independent loops. The ellipses represent 6-loop and 8-loop terms. By the perturbation expansion $\Gamma^0|ss^{(2)}\rangle+\delta \Gamma|ss^{(1)}\rangle =0$, we get the coefficient of independent loops $\alpha^{(2)}=h^{2}$. For higher orders:
\begin{equation}
|ss^{(n)}\rangle=\alpha^{(n)}\sum_{O}|n \text{ independent 4-loops}\rangle +\cdots.
\label{eq:expansion}
\end{equation}
By similar analysis in Eq. \eqref{eq:perturbexpand}, the independent loops cancel and we have  $\alpha^{(n)}=h^n$. The ellipsis represents terms with other loop configurations and the weight coefficients of these configurations are $O(h^{n})$. However, we demonstrate that only independent 4-loops contribute to the leading term in $W_{\gamma}$ and we can focus on independent loops. For a specific configuration, $W_{\gamma}$ gives the parity of the number that the loop $\gamma$ crosses the membrane. In Fig. \ref{fig:3dexcitation2}, we can see that the link that crosses the membrane is flipped. The loop $\gamma$  goes through a closed membrane for even times and odd times when it meets an open membrane. Therefore, $W_{\gamma}=1$ for even times and $W_{\gamma}=-1$ for odd times. Following the procedure in Ref. \cite{kogut1979introduction}, we have
\begin{equation}
\langle I|W_{\gamma}|ss\rangle\sim\sum_n\frac{1}{n!}(N_{l}-P_{\gamma}-P_{\gamma})^nh^n=\text{exp}[h(N_{l}-2P_{\gamma})],
\end{equation}
and the denominator in Eq. \eqref{eq:Wilsonloop} is
\begin{equation}
\langle I|ss\rangle=\sum_n\frac{1}{n!}N_{l}^nh^n=\text{exp}[hN_{l}],
\end{equation}
where $N_{l}$ is the number of all spins and $P_{\gamma}$ is the number of spins on the loop $\gamma$ which is also the length of $\gamma$. Finally, the expectation of the Wilson loop operator,
\begin{equation}
\langle W_{\gamma}\rangle=\frac{\langle I|W_{\gamma}|ss\rangle}{\langle I|ss\rangle}\sim\text{exp}(-2hP_{\gamma}),
\end{equation}
satisfies a perimeter law in the small $h$ regime, which reveals that the model is in a deconfined phase.

\paragraph{$h\gg1$}
When in the large $h$ limit, we take $\Gamma^{0}/h$ as the perturbation and $|ss^{(0)}\rangle$ is the zero energy state of $\delta \Gamma/h=\sum_{l}(\sigma^{x}-1)$:
\begin{equation}
|ss^{(0)}\rangle=|\rightarrow\rightarrow\cdots\rightarrow\rangle=|I\rangle.
\end{equation}
Since $\langle I|\sigma_{z}|ss^{0}\rangle=0$, the first nonzero contribution which is also the leading term in $\langle W_{\gamma}\rangle$ is given by flipping all the spins on the minimal membrane enclosed by the Wilson loop $\gamma$. $A_{\gamma}$ is the number of plaquettes surrounded by $\gamma$  which is also the area of the minimal membrane. In each order of perturbation, there are at most four plaquettes created and two of them can appear in the membrane circled by $\gamma$. Therefore, the leading term is in the $[A_{\gamma}/2]^{\text{th}}$ order:
\begin{equation}
\langle W_{\gamma}\rangle\sim h^{-A_{\gamma}/2}=\text{exp}(-\frac{1}{2}A_{\gamma}\ln h),
\end{equation}
which is obviously the area law and the system is in the confined phase.

In the regimes of two limits, we find that $\langle W_{\gamma}\rangle\sim \text{exp}(-2hP_{\gamma})$ when $h\ll 1$ and $\langle W_{\gamma}\rangle\sim \text{exp}(-\frac{1}{2}A_{\gamma}\ln h)$ when $h\gg 1$. We expect that the system would undergo a deconfinement-confinement transition at some critical $h_c$, and that is when the topological degeneracy of steady states is broken. 

\begin{figure}[htb]
 \centering
\includegraphics[width=1\linewidth]{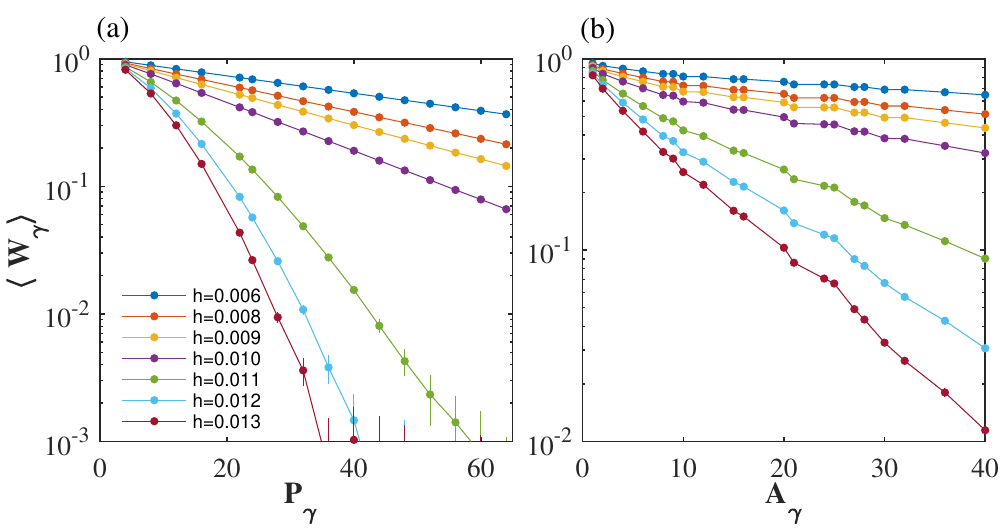}

\caption{
Semi-log plot of  the expectation value of Wilson loop $W_{\gamma}$ (on $32\times32\times32$ cubic lattice) versus the perimeter $P_{\gamma}$ (a) and the area $A_{\gamma}$ (b).  There is a phase transition driven by perturbation $h$ and the critical point is about $0.010<h_c<0.011$. For $h<h_c$, the system is in the deconfined phase, with $\langle W_{\gamma}\rangle$ satisfying a perimeter law; for $h>h_c$, the system is in the confined phase, with $\langle W_{\gamma}\rangle$ satisfying an area law. }
\label{fig:deconfine-confine}
\end{figure}

To examine whether the perturbation expansion indeed gives the right prediction, we also give a numerical simulation of the corresponding Markovian dynamics using the Monte Carlo method. For each step, the link $l$ is chosen randomly and the spin $\sigma_{l}^{z}$ is flipped with probability given by the $L_{m,l}$ and $L_{x,l}$:
\begin{equation}
w_{l}=\left\{
\begin{array}{lll}
\frac{h}{1+h}              & \text{if the flip decreases }  \sum_{p|l\in \partial p}B_{p}; \\
\frac{1/2+h}{1+h}       & \text{if the flip keeps } \sum_{p|l\in \partial p}B_{p} \text{ invariant}; \\
1                                   &  \text{if the flip increases }\sum_{p|l\in \partial p}B_{p}.  
\end{array}
\right.
\end{equation}
The numerical results are shown in Fig. \ref{fig:deconfine-confine}, which confirms our expectation.
In Eq. \eqref{eq:projection}, we give the projection operator with $q_{1}$ and $q_{2}$. Then analyze the Markovian dynamics in Eq. \eqref{eq:eff}, the amplitudes for $\sum_{p|l\in \partial p}B_{p}$ decreasing, invariant, and increasing are $h$, $q^{2}_{2}+h$, and $q^{2}_{1}+h$. The corresponding probability in the Metropolis algorithm is $\frac{h}{q_{1}^{2}+h}$, $\frac{q^{2}_{2}+h}{q_{1}^{2}+h}$, and $1$. As long as $q_{1}>q_{2}$, the loop-shrinking process dominates when $h$ is relatively small and the loop excitations would not proliferate.

The existence of a deconfined phase when $h<h_{c}$ supports our statement that the topological degeneracy is robust in 3d. As a comparison, recall that in 2d the expectation of the Wilson loop operator always obeys an area law for any finite $h$ (See Eq. \eqref{eq:2dwilson}), which means the system is always in a confined phase.

\subsection{Model-2}
Like the discussion in two-dimension (Sec. \ref{sec:2dmodelquamtum}), we can generalize the classical Model-1 to a quantum version with phase coherence, by dropping the dephasing operator $L_{z,l}$ and suppressing the fluctuation of both $e$ and $m$ defects. We construct the following dissipators:
\begin{equation}
\begin{aligned}
%L_{l}&=\sigma^{x}_l\left(1-\frac{1}{2}\sum_{p:l\in \partial p}{B_p}\right)\left(1-\frac{1}{4}\sum_{p:l\in \partial p}{B_p}\right), \ {B_p}=\prod_{l\in\partial p }\sigma^{z}_{l}\\
L_{m,l}&=\sigma^{x}_lP(\sum_{p|l\in \partial p}B_{p}),\\
L_{e,l}&=\sigma^{z}_lP(\sum_{v|v\in \partial l}A_{v}). \label{eq:3dquantum}
\end{aligned}
\end{equation}
Similarly, for link $l$, it connects two vertices, and $\sum_{v|v\in \partial l}A_{v}$ gives the sum of two vertex operators attached to link $l$. The quantum jump operators $L_{m,l}$, and $L_{e,l}$  are symmetric. However, it is crucial to keep in mind that in $d=3$, these two kinds of defects are inequivalent, because the $e$ particle is point-like while the $m$ excitation in 3d is loop-like (See Fig. \ref{fig:3dexcitation1}). 
% \begin{figure}[t]
%   \centering
%   \subfigure{\includegraphics[width=3.5cm, height=4cm]{loopfig.pdf}}     
%   \subfigure{\includegraphics[width=4cm,height=4.2cm]{looprelaxation.pdf}}
%   \subfigure{\includegraphics[width=8.5cm,height=4.2cm]{gaptime.pdf}}
%   \caption{fig}
% \end{figure}
Without perturbation, the zero-eigenvalue subspace (whose dimension corresponds to the steady-state degeneracy) of the Liouvillian contains $8^2=64$ type-A states and an exponentially large number of type-B states. Type-A states are dark states of operators $L_{m,l}$ and $L_{e,l}$, then they correspond to the ground states of the 3d toric code model. The independent bases of the steady-state subspace are
\begin{equation}
 \rho^{\{\mu_{i}\},\{\mu'_{i}\}}_{A}=|\psi^{\{\mu_{i}\}}\rangle\langle \psi^{\{\mu'_{i}\}}|,\label{eq:3dquantum2}
\end{equation}
where $|\psi^{\{\mu_{i}\}}\rangle$ is one of the 8 ground states of the 3d toric code and the explicit form is in Eq. \eqref{eq:groundstate}. 
 
Next, we add the following perturbation:
\begin{equation}
\begin{aligned}
 L_{x,l}&=\sqrt{h_{x}}\sigma^{x}_{l},\\
 L_{z,l}&=\sqrt{h_{z}}\sigma^{z}_{l}.
\end{aligned}
\end{equation}
As in the 2d model, since $e$ and $m$ defects have completely independent dynamics, we can treat them separately. For the $e$ defects in 3d which is similar to that in 2d, the $e$ defects are point-like particles that can hop, and be created/annihilated in pairs. Even the amplitudes associated with these processes are the same as those in 2d. Based on our previous discussion, $e$ defects would proliferate for any finite $h_z$. Therefore, the steady state is similar to the solution in Eq. \eqref{eq:2dss1} :
\begin{equation}
\rho_{e} \sim \sum_{k,\{r\}}\beta_{e}^{2k}|e_{2k}(\{r\})\rangle \langle e_{2k}(\{r\})|. 
\end{equation}
Here for simplicity, we omit the additional information describing the configuration of $m$ defects. The above expression represents the reduced density matrices of $e$ defects by tracing out the $m$ defects. For $m$ defects, the situation is identical to that in Model-1. The perturbation $L_{x,l}$ would make the type-B states metastable, and the $m$ defects would not proliferate until $h_x>h_c$. For small but finite $h_z$ and $h_x$, the steady state degeneracy is 8, which again is characterized by the non-contractible membrane on the dual lattice. Indeed, for $h_z\rightarrow \infty$, this model reduces to model-1.   

In conclusion, we have shown that in the 3d quantum model, the 64-fold degeneracy of type-A steady states is easily broken down to 8-fold by perturbation, while the remaining 8-fold topological degeneracy is robust. This is qualitatively similar to the behavior observed in the steady states discussed in Sec. \ref{sec:3dclassical}.

% It is noteworthy that utilizing the exact expressions of steady states as outlined in Eq. \eqref{eq:3dss} and Eq. \eqref{eq:3dquantum2}, we are equipped to compute the topological entropy for both Model-1 and Model-2 in three dimensions \cite{castelnovo2008topological}. Analogous to the two-dimensional scenario, the topological entropy in three dimensions will exhibit quantization. However, it is important to acknowledge that the satisfaction of detailed balance for Model-1 and Model-2 in two dimensions no longer holds in the three-dimensional perturbed case ($h>0$), and obtaining the corresponding steady states becomes challenging. We posit that the quantization of topological entropy endures within the ordered phase ($h<h_{c}$) for both Model-1 and Model-2.

\section{Topological degeneracy implies slow relaxation}\label{sec:relaxation}
In this section, we discuss the relaxation dynamics, which provides information about the low-lying Liouvillian spectrum. We demonstrate that when the steady states have topological degeneracy, the relaxation time always diverges in the thermodynamic limit, most likely algebraically with the system size. This implies that the Liouvillian is gapless \footnote{Here we implicitly make the assumption that that the system has translation symmetry, which excludes special cases such as systems with boundary dissipation or skin effect.}.

First, we note the Liouvillian gap is defined as
\begin{equation}
    \Delta=\min_{\beta\notin S_{0}}\{\text{Re}(-\lambda_\beta)\},
\end{equation}
here $S_{0}$ is the subspace of topologically degenerate steady states where the splitting of eigenvalues is of $O(e^{-cL})$ ($L$ is the linear system size and $c>0$.) order and $\lambda_{\beta}$ is the eigenvalue of Liouvillian. $\Delta$ is a measure of the gap between the topologically degenerate steady states and the rest of the spectrum. It determines the relaxation time for a random initial state to decay into the degenerate steady-state subspace. Our result shows that in the topologically ordered phase, as the system size $L$ goes to infinity, $\Delta$ approaches zero. Moreover, it is likely that $\Delta$ decays algebraically, i.e., $\Delta\sim L^{-\alpha}$ with a finite exponent $\alpha$ (See Fig. \ref{fig:degenerate_gapless} for an illustration).

This is in sharp contrast to the topologically ordered phase in closed systems, where there is a finite bulk gap and the gap only closes at the critical point of a phase transition. In open systems, however, the Liouvillian gap also closes at the transition from the trivial phase to the topologically ordered phase but then remains closed within the topologically ordered phase. To support this picture, we first demonstrate this behavior in the models of the present paper, and then provide a general argument for its universality.
% \begin{figure}[htb]
%  \centering
% \includegraphics[width=1.0\linewidth]{degenerate_gapless.pdf}

% \caption{Gap in the topologically ordered phase. Left panel: In the closed systems, the energy splitting of the degenerate ground states (blue lines) is $O(e^{-L})$ and there is a finite gap between the degenerate ground states (blue lines) and excited states (dark lines). Right panel: In open quantum systems, the splitting of degenerate steady states (blue lines) is also $O(e^{-L})$. There are slow-relaxation modes (red lines) that contribute to the Liouvilian gap $\Delta\sim L^{-a}$. Here $L$ is the system size and $a$ is a finite power. }
% \label{fig:degenerate_gapless}
% \end{figure}

Note that in Model-2, the dynamics of $e,m$ defects are completely independent and can be treated individually, we only need to figure out whether one of the defects exhibits slow relaxation. Then, the problem can be reduced to the same as in Model-1. Therefore we only discuss Model-1 in the following sections.

\subsection{Model-1 in 2d}
First, we discuss the relaxation dynamics of Model-1 in 2d (Eq. \eqref{eq:2dclassical}). We use the effective Markovian generator description:
\begin{equation}
\begin{aligned}
\Gamma_{2d}=\sum_l(\sigma^x_l-1)P^{2}(\sum_{p|l\in \partial p}B_{p})+\kappa_{v}\sum_v(A_{v}-1)&\\
+h\sum_i(\sigma_i^x-1)&.
 \label{eq:2deff2}
\end{aligned}
\end{equation}
\subsubsection{$h=0$}
As analyzed in Sec. \ref{sec:model1}, this model can only have the steady-state degeneracy at $h=0$. Though the $m$ particles appear in pairs, two particles can be separated far away from each other and in the limit thermodynamic $L\to\infty$,  the finite time evolution of one particle can be identified as a random walk process. Therefore, the dynamics of one particle can be well approximated by a tight-binding Hamiltonian. The dispersion on the 2d square lattice is $\lambda_k\sim2- \text{sin}(k_x)-\text{cos}(k_y)$ which shows that the Liouvillian is gapless in the thermodynamic limit $\Delta\sim L^{-2}$. Starting from any excited state, all $m$ particles would finally annihilate in pairs under the random walk process and the relaxation time $\tau$ follows a power law: $\tau\propto L^2$.

\subsubsection{$h\neq0$}
Then, when the perturbation is nonzero, the degeneracy is broken immediately. In the long time limit $t\gg \kappa_{v}^{-1}$, all $m$ particle configurations are mixed by $A_{v}$ and $\Gamma_{2d}$ can be mapped to a spin model on the dual lattice in the $A_{v}=1$ subspace, as long as we care about the low-lying spectrum around the steady state. With the mapping: $B_{p}=-1\to \tau^{z}=-1$ and $B_{p}=+1\to\tau^{z}=+1$, we have $\Gamma_{2d}\rightarrow H$:
\begin{equation}
\begin{aligned}
H =-&\sum_{\langle i,j\rangle}\left(\tau^{-}_{i}\tau^{- }_{j}+\frac{\tau^{-}_{i}\tau^{+}_{j}}{2}+\frac{\tau^{+}_{i}\tau^{-}_{j}}{2}\right)+2\sum_i n_{i}\\
-h&\sum_{\langle i,j\rangle}(\tau^{x}_{i}\tau^{x}_{j}-1),
\end{aligned}
\end{equation}
here $n_{i}=\frac{1+\tau^{z}_{i}}{2}$ and the set of new spin variables $\tau^{z}$ are living on the vertices of the dual lattice corresponding to the lattice operator $B_{p}$. In the subspace of $A_{v}=1$, since $m$ particles (plaquettes with $B_{p}=-1$) are created in pairs, then for $H$, only states with an even number of spins $\tau^{z}$ flipped are considered. The ground state of $H$ is 
\begin{equation}
|\rho\rangle  = \otimes_{i}(|\downarrow\rangle_{i}+\beta|\uparrow\rangle_{i}),
\end{equation}
where $\beta = \sqrt{\frac{h}{h+1}}$. Though $H$ is non-Hermitian, it can be transformed into a Hermitian Hamiltonian $H_{s}$ with a similarity transformation:
\begin{equation}
H_{s} = SHS^{-1},\quad S=\beta^{-\sum_in_i/2}. 
\end{equation}
After ignoring a constant term, we have
\begin{equation}
H_{s}=-\sum_{\langle i,j\rangle}(\beta_{1}\tau^{x}_{i}\tau^{x}_{j}+\beta_{2}\tau^{y}_{i}\tau^{y}_{j})+2\sum_{i}\tau^{z}_{i}.
\end{equation}
$\beta_{1}=(h+\frac{1}{2})+\sqrt{h(h+1)}$, $\beta_{2}=(h+\frac{1}{2})-\sqrt{h(h+1)}$ and $\beta_{1}\beta_{2}=\frac{1}{4}$.  This is an $XY$ model with anisotropic interaction and a $z$-direction magnetic field.
$H_{s}$ can be rewritten into a more illuminating form:
\begin{equation}
\begin{aligned}
\frac{H_s}{(2h+1)} =& -\sum_{\langle i,j\rangle}\left[\frac{1}{2}(1+\eta)\tau^{x}_{i}\tau^{x}_{j}+\frac{1}{2}(1-\eta)\tau^{y}_{i}\tau^{y}_{j}\right]\\
&+h_{z}\sum_{i}\tau_{i}^{z}.
\end{aligned}
\end{equation}
where $\eta=\frac{2\sqrt{h(h+1)}}{2h+1}$ and $h_{z}=\frac{2}{2h+1}$.  This model has been analyzed in Ref. \cite{henkel1984statistical} and we find that our model is just in a special parameter regime $\eta^{2}+(\frac{h_{z}}{2})^{2}=1$. When $h_{z}<2$ ($h>0$), $H_{s}$ is in the Ising ordered phase which is obviously gapped. While $h_{z}=2$ ($h=0$) is a critical point where a ferromagnetic phase transition happens, then $H_{s}$ is gapless. Since the similarity transformation keeps the spectrum invariant, $H$ is also in the gapless phase when $h=0$ and this is consistent with the discussion in the last part. $H$ would be in the gapped phase when $h>0$. Therefore, the Liouvillian is gapped as long as $h$ is nonzero and any initial states would decay into the steady state in a finite time. Finally, we find that in 2d, the breaking of topological degeneracy and the gap opening happen at the same time when the perturbation is turning on.

\subsection{Model-1 in 3d}
At the end of Sec. \ref{sec:3drobust}, we mention that in the 3d model, the type-B metastable states have power law decay $\tau\sim L^{a}$ and $\Delta\sim L^{-a}$ when the system is in the topologically ordered phase ($h<h_{c}$) and exponential decay when in the topologically trivial phase ($h>h_{c}$). Next, we confirm that topological order in dissipative systems comes with slow relaxation (power law decay).

\subsubsection{$h=0$}
When $h=0$, type-A and type-B states are all steady states. In this part, we study the relaxation process of those states with contractible loops which finally evolve into steady states under $\Gamma^{0}$ [Eq. \eqref{eq:gamma0}].

For any states with loop excitation, because the loop length is non-increasing and prefers to decrease under $\Gamma^{0}$, the loop excitation gradually shrinks and finally vanishes. When $L\to\infty$, It seems that the relaxation time will diverge if the loop length is proportional to the system size.  

\begin{figure}[htb]
 \centering
\includegraphics[width=0.60\linewidth]{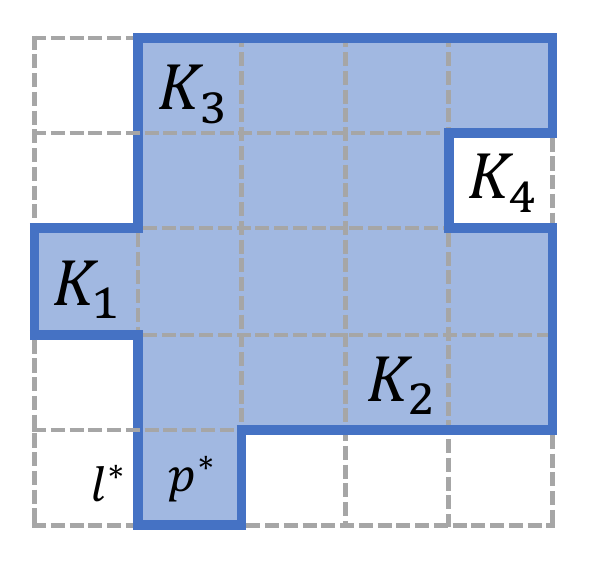}

\caption{A specific configuration with concave, convex, and flat parts. On the dual lattice, the solid link represents $B_{p}=-1$ and the dotted link represents $B_{p}=+1$.  }
% (b) An initial type-B state with non-contractible loop defects (solid blue circle).}
\label{fig:cloop}
\end{figure}

The dynamics of the loop evolution under $\Gamma^{0}$ is similar to the phase-ordering kinetics \cite{bray2002theory,spirin2001freezing}. In the following, we give an estimation of the relaxation process. The operator $L_{m,l}$ in Eq. \eqref{3Dclassical} only acts on the dual plaquettes which are concave or convex or at the corner along the loop. In Fig. \ref{fig:cloop}, we give an example of the loop defects on the dual lattice where the solid links $l^*$ represent original plaquettes $p=l^*$ with $B_{p}=-1$. The dual plaquettes $p^*=l$ with $\sum_{l^*\in\partial p^*}B_{p=l^*}=-2$ labeled by the curvature $K_{1}>0$ (convex) and $K_{4}<0$ (concave) can be flipped by $L_{m,l}$, and in this way, the loop defect is flattened. The corner plaquette with $\sum_{l^*\in\partial p^*}B_{p=l^*}=0$ labeled by $K_{3}>0$ would also be flipped but with a smaller probability. The one with $\sum_{l^*\in\partial p^*}B_{p=l^*}=2$ giving curvature $K_{2}=0$ stays unchanged, which is exactly the reason the type-B states with non-contractible loops can be steady states. Like the shrinking process of a surface with tension, the shrink of the loop depends on curvature. With coarse-graining, we can approximate these loop defects with smooth curves and we can assume the shrinking rate $\sigma$ of the loop defect to be a smooth function of local curvature, $\sigma=\sum_{n}a_{n}K^{2n}$, with coefficients $a_{n}$. Here we consider the relaxation of large defects with length $R$, so that the curvature $K$ is about $1/R$. To the leading order, we have $\sigma\sim a_{2}K^{2}\sim a_2/R^2$. Then the time evolution can be approximated by
\begin{equation}
\begin{aligned}
\frac{dR(t)}{dt}\approx -2\pi R(t)\sigma.
% \Rightarrow & R(t)\frac{dR(t)}{dt}\approx-2\pi a_2,\\
% \Rightarrow  &R(t)\approx\sqrt{R^2_0-2\pi a_2 t}.\ \ \ (t<\frac{R^2_0}{2\pi a_2})
\end{aligned}
\end{equation}
We can straightforwardly get the solution
\begin{equation}
    \begin{aligned}
         % &R(t)\frac{dR(t)}{dt}\approx-2\pi a_2,\\
        R(t)\approx\sqrt{R^2_0-2\pi a_2 t},
    \end{aligned}
\end{equation}
where $R_0$ is the initial length of the loop defects. Thus the relaxation time $\tau$ is proportional to $R_{0}^{2}$. %For other configurations that are not circular, we have similar results: $\tau\sim P^{2}$ where the total loop length $P$ is the perimeter of the loop and it can be a linear approximation of radius $R$. 
The numerical verification of this result can be found in a related paper \cite{wang2023topologically}, where the result fits the above estimation really well.
% We give a numerical check of this result. Take a $R_0\times R_0$ loop defect (boundary of membrane) as the initial state and check the evolution under $\Gamma^{0}$. The results in Fig. \ref{fig:cloop}(b) and (c) confirm our derivation. 
% \begin{figure}[htb]
%  \centering
% \includegraphics[width=0.99\linewidth]{cloop.pdf}

% \caption{
% Relaxation dynamics at $h=0$ with a $R_0\times R_0$ square membrane in the initial state. (a) A specific configuration with concave, convex, and flat parts.
% (b) When $h=0$, start form an open membrane ($R_0\times R_0$) and then evolve under $\Gamma^{0}$. The square of the loop length $P^{2}$ decreases linearly with time $t$ for a wide range. 
% (c) Relaxation time $\tau$ verse the initial size $R_{0}$. $\tau\propto R_0^2$ fits well.
% To calculate $\tau$, We record the time when the loop length drops to zero for each of the 10000 trajectories and take the average.}
% \label{fig:cloop}
% \end{figure}
When $h=0$, type-A and type-B states are steady states, and those contractible loops (contractible open membranes on the dual lattice) contribute to the low-lying spectrums: $\Delta\sim L^{-2}$. 

\subsubsection{$h\neq0$}\label{sec:lifetime}
For the trivial contractible loop states, we expect that a similar relaxation process also happens in other regimes of the topologically ordered phase ($0<h<h_c$), where the loop defects do not proliferate in the steady state. The shrinking process of a large open membrane would still happen slowly (diffusively) and dominate the long-time dynamics.  The relaxation spectrum in this phase should be gapless. However, for $h>h_c$, the loop defects proliferate, and any large open membrane in the initial configuration will be quickly separated into pieces by the strong local fluctuation of defects. 

Then, we deal with the special type-B metastable states with non-contractible loops.
In this part, we solve the problem left at the end of Sec. \ref{sec:3drobust}: the evolution of the type-B metastable states for small but finite $h$. We find that the topological degeneracy of steady states is robust under perturbation and the Liouvillian is gapless. This is quite different compared with the case in closed systems. As already mentioned, it is crucial to prove that these states do not have an exponentially long lifetime, because only then do these states detach from $S_{0}$ the subspace of the 8-fold topologically degenerate steady states. Otherwise, the topological degeneracy would be ill-defined. 

\begin{figure}[htb]
 \centering
\includegraphics[width=1.0\linewidth]{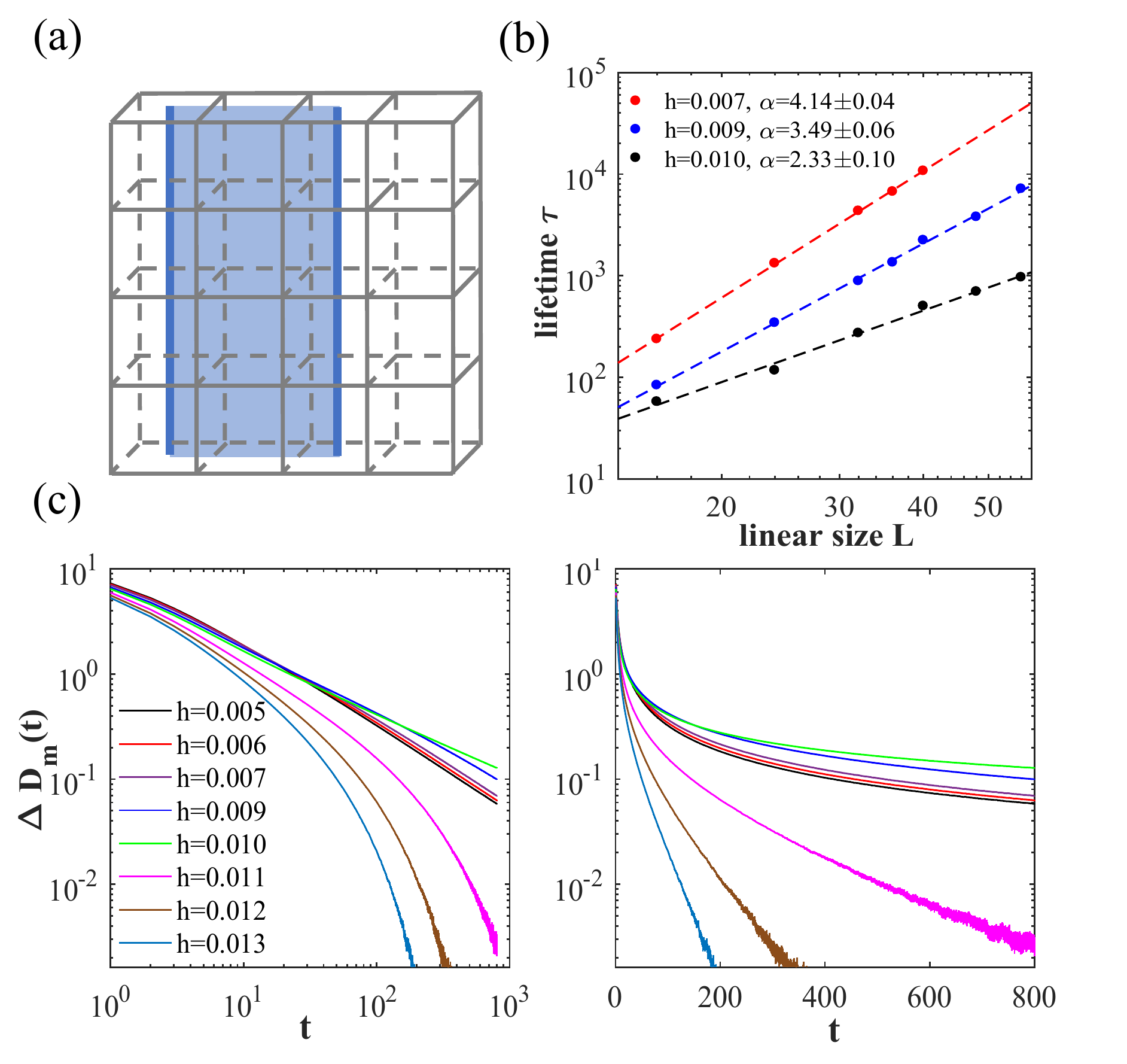}
\caption{
The lifetime of metastable states. (a) An initial type-B state on the dual lattice. (b) log-log plot of relaxation time $\tau$ versus the linear size $L$. For different $h<h_{c}$, $\tau\sim L^{\alpha}$ fits well with power $\alpha$ depending on perturbation strength. 
(c) Relaxation dynamics for various $h$ and random initial states. The left/right panel shows a log-log/semi-log plot of the evolution of the loop defect density $\Delta D_{m}(t)$. 
The results are obtained by averaging over 96000 trajectories and the transition point is about $0.010<h_{c}<0.011$.}
\label{fig:relax}
\end{figure}

We perform the numerical simulation of the loop evolution using the Monte Carlo method. We set the initial condition in the following way: a large membrane on the $xy$ plane, whose boundaries form two non-contractible loops circling around the $x$-axis, as shown in Fig. \ref{fig:relax}(a). For simplicity, we assume there is no other defect. 
To determine the lifetime of the type-B states, we examine whether there are still any remaining non-contractible loops every $N\tau_0$ steps, with properly chosen time resolution $\tau_0$ for different sizes, and the lifetime is defined by the time when no non-contractible loops are found. We did the simulation for several parameters in the deconfined phase $h<h_{c}$ and found similar results in these cases. As shown in Fig. \ref{fig:relax}(b),  the lifetime scales as a power law with the system size: $\tau\propto L^\alpha$ with $\alpha\approx 4$ when $h$ is well below $h_{c}$. The lifetime of these type-B states diverges algebraically for large systems, which is consistent with our expectation that they are metastable states. These states contribute to the lowest-lying Liouvillian spectrum above the 8-fold degenerate steady states, with an algebraically small Liouvillian gap. Moreover, we simulate the relaxation process of all kinds of loops with the density of loop defects $\Delta D_{m}$ versus time. The defect density is defined as  $\Delta D_{m}(t)=(P_{\text{tot}}(t)-P_{\text{tot}}(\infty))/L^{3}$ and $P_{\text{tot}}(t)$ is the length of all loops.
We choose configurations randomly and evolve under $\Gamma^{0}+\delta \Gamma$ (Eq. \eqref{eq:eff}). In Fig. \ref{fig:relax}(c), it is clear that there is a 
transition point at $0.010<h_{c}<0.011$, which is consistent with the deconfinement-confinement phase transition (See Fig. \ref{fig:deconfine-confine}). 
When $h<h_{c}$, the loop states decay algebraically $\Delta D_{m}(t)\sim t^{-a}$, indicating a gapless Liouvillian spectrum \cite{cai2013algebraic}. When $h>h_{c}$, the defect density decays exponentially $\Delta D_{m}(t)\sim e^{-bt}$, and there is a finite Liouvillian gap. Here $a$ and $b$ are nonuniversal coefficients depending on the perturbation strength $h$.

\subsection{General Discussion}
\begin{figure}[htb]
 \centering
\includegraphics[width=0.90\linewidth]{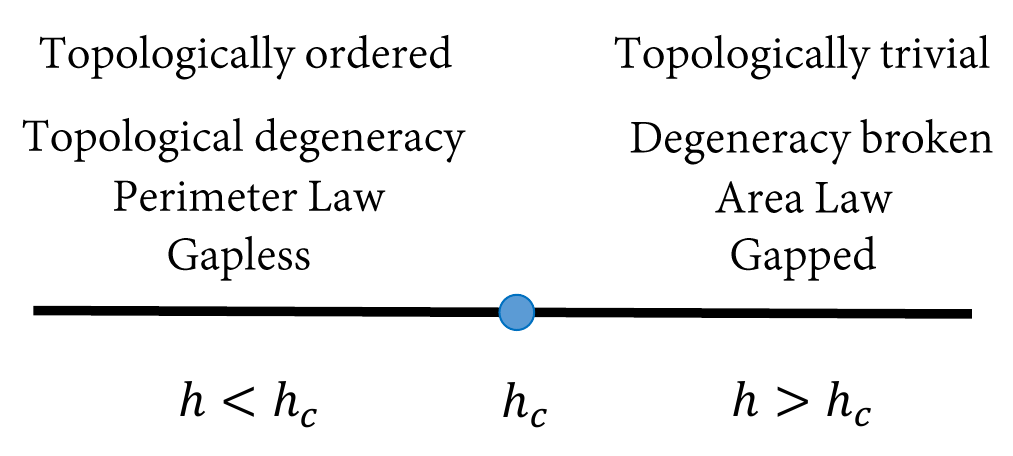}

\caption{
Phase diagram as a function of the perturbation strength $h$. The topologically ordered phase is characterized by the topologically degenerate steady states, quantized topological entropy, the perimeter law of the Wilson loop, and gapless modes.}
\label{fig:phase}
\end{figure}

We observe that in all the models considered, the Liouvillian gap vanishes in the topologically ordered phase where there are topologically degenerate steady states. Furthermore, the Wilson loop operator follows the perimeter law and the topological entanglement entropy is quantized.  In 2d, the Liouvillian gap closes at $h_{c}=0$ where the system has topological degeneracy. In 3d, we have a robust topologically ordered phase where the Liouvillian is gapless and the gap opens at finite $h_{c}$ (See Fig. \ref{fig:phase}). 

In this section, we present a heuristic argument that this behavior is a general feature of a local Liouvillian with topological degeneracy.
Based on the previous study of topological order in closed systems, we know that if the ground state exhibits topological degeneracy, there would be topological excitations such as $e$ and $m$ excitations that cannot be singly created or annihilated by local operators. The low-lying energy spectrum is determined by the excitations and interactions. In the case of open quantum systems, we expect similar behavior to happen: if the steady states have topological degeneracy, then the low-lying Liouvillian spectrum is dominated by the relaxation dynamics of such topological defects. These defects are akin to point-like particles in 2d and can also be extensive objects in 3d and higher dimensions.  A proliferation of such topological defects would lead to the destruction of the topological degeneracy and the emergence of a trivial phase.
%Therefore, it is more appropriate to refer to these excitations as topological defects rather than topological particles.

First, we assume the steady states have topological degeneracy. Then, starting from some random initial state full of topological defects, the system will relax into the steady state by annihilating the defects. However, they can not be annihilated locally. If they are point-like particles, then they can only travel to get close to another point-like defect and annihilate in pairs. Obviously, the time of this process diverges as the distance between the defects goes to infinity. If they are extended objects, like the case in Model-1 in 3d, then a large defect must shrink to a small size to reach a steady state. The time of this process also diverges as the initial size of the defect goes to infinity. In both cases, the relaxation time diverges in the thermodynamic limit, and this is usually equivalent to the vanishing (algebraically) of the Liouvillian gap. 

The above argument about the relaxation dynamics of the ordered phase appears to be quite universal in open quantum systems. In fact, it may not only apply to topological degeneracy but also to degeneracy resulting from symmetry breaking (discrete or continuous), in which case the slow relaxation dynamics is due to the presence of defects associated with the broken symmetries, such as domain walls and vortices. This leads us to conjecture that there might be a counterpart of the Goldstone theorem in open systems that has a wider range of applicability: not just for continuous symmetry breaking phase but also for discrete symmetry, and topologically ordered phase as well. This question is left for future research to explore.

\section{Conclusion}
In summary, we have studied topological order in open quantum systems, for which the steady-state topology takes the place of ground-state topology. This dissipative topological order is characterized by the topological degeneracy of steady states, quantized topological entropy, and dynamically deconfined emergent gauge field. In contrast to topological order in closed systems, dissipative topological order exhibits a unique feature: The topological degeneracy is typically accompanied by gapless modes that lead to the slow relaxation of the system. Notably, topological degeneracy remains well defined in spite of the gapless modes, because the gap decays algebraically with the system size while the spectral range of the steady states is exponentially small. %This finding highlights the unique behavior of open quantum systems. 

Several questions remain open for further investigation. For instance, it is desirable to conduct a more comprehensive study on the dimensional dependence of DTO. Specifically, a rigorous proof regarding the fragility of DTO in generic 2d open quantum systems is called for. Furthermore, the two models we construct can be generalized to four and higher dimensions, where both types of topological defects are extended objects, and thus the topological degeneracy of Model-2 also becomes robust. In that case, the steady states exhibit long-range entanglement and robust quantum memory can be realized, and thus can potentially be useful for robust dissipative quantum computation. It would also be intriguing to study the effect of gapless modes on dissipative quantum computation.
%Moreover, our construction of DTO can be generalized to four and higher dimensions where robust long-range entangled steady states may be realized, and thus can potentially be used for robust dissipative quantum computation. The effect of gapless modes on robust dissipative quantum computation needs further exploration.
%analogous to the Hermitian topological order which can only exist in two and higher dimensions \cite{chen2011complete}, DTO has a subtle dependence on spatial dimensions and we need rigorous proof of whether robust DTO can manifest within a two-dimensional framework. 

\begin{acknowledgments}
We thank Xuan Zou, Zhengzhi Wu, and Hao-Xin Wang for useful discussions. This work is supported by NSFC under Grant No. 12125405.
\end{acknowledgments}

\begin{widetext}

\appendix
\section{2d perturbation analysis}\label{sec:appendix2dpertub}
To calculate the first order perturbation effect $\langle \rho^{0,\mu'\nu'}_{L}|\delta\Gamma|\rho^{0,\mu\nu}_{R}\rangle$, we have to figure out the two $m$ particles configuration $\alpha(m_{2})|m_{2}\rangle$ in the left steady state:
 \begin{equation}
\begin{aligned}
|\rho^{0,\mu\nu}_{R}\rangle &= \frac{1}{|G|}W_{x}^{\mu}W_{y}^{\nu}\sum_{g\in G}g|\Uparrow\rangle,\\
|\rho^{0,\mu\nu}_{L}\rangle &= W_{x}^{\mu}W_{y}^{\nu}\sum_{g\in G}g|\Uparrow\rangle + \alpha(e_{2})|m_{2}\rangle +\alpha(m_{4}) |m_{4}\rangle\ldots.
\end{aligned}
\end{equation}
With the translation symmetry on the 2-torus and the stabilizer group $G$, starting from one specific configuration of the state $|m_{2}\rangle$, we can generate many other spin configurations with two $m$ particles at the same location and these configurations have the same weight. Thus, the position and configuration of open loops are not important; rather, it is the relative position of the two ends of the open loop, where particles are situated, that matters. We choose a representative configuration and denote it by the relative position of two $m$ particles $\{(x,y)|x=|x_1-x_2|, y=|y_1-y_2|\}$. The corresponding coefficient in the left steady state is written as $\alpha(x,y)$. There are some special points 
representing the steady states. $\alpha(0,0)$ is the coefficient of a steady state without any non-contractible loops, and $\alpha(L_{x},0)$ gives the coefficient of a steady state with a non-contractible loop in the $x$ direction. $\alpha(0,L_{y})$ and $\alpha(L_{x},L_{y})$ are similar. $L_{x}$ and $L_{y}$ are the linear sizes of the 2d lattice. All the other coefficients $\alpha(i,j)$ represent open loop states with two $m$ particles. With Eq. \eqref{eq:map2} and Eq. \eqref{leftss}, since the coefficients of all states $|m_{2}\rangle$ cancel, we can write down all linear equations about $\alpha(x,y)$. For simplicity, take $L_{x}=L_{y}=L$ and then the coefficient $\alpha(x,y)$ is symmetric $\alpha(x,y)=\alpha(y,x)$. We have 
\begin{equation}
\begin{split}
&-4\alpha(i,j)+\alpha(i-1,j)+\alpha(i+1,j)+\alpha(i,j+1)
+\alpha(i,j-1)=0, \ 1\leq i,j\leq L\\
&-4\alpha(i,L)+\alpha(i-1,L)+\alpha(i+1,L)
+2\alpha(i,L-1)=0,\ 1\leq i\leq L-1\\
&-4\alpha(i,0)+\alpha(i-1,0)+\alpha(i+1,0)
+2\alpha(i,1)=0.\ 1\leq i\leq L-1
\end{split}
\end{equation}
This is a set of simple linear equations which can be solved numerically. To get the solution, we can map the matrix $\alpha(i,j)$ with $i,j=0,1,\cdots,L$ to a vector $|\alpha(i,j)\rangle$, and these equations transform into a matrix equation $M|\alpha(i,j)\rangle=0$. Then, the coefficients $\alpha(i,j)$ are elements of the zero-eigenstate to the matrix $M$. We can straightforwardly get four degenerate zero-eigenstates $|\alpha(i,j)\rangle$ and these four sets of degenerate coefficients are consistent with the degenerate left steady states $|\rho^{0,\mu\nu}_{L}\rangle$ in Eq. \eqref{eq:leftss}. This can be understood in another way. The left steady state can be generated from the corresponding right steady state with the conjugate Liouvillian,
\begin{equation}
|\rho^{0,\mu\nu}_{L}\rangle=\lim_{t\to\infty}e^{(\Gamma^0)^{\dagger}t}|\rho^{0,\mu\nu}_{R}\rangle.
\end{equation}
The four degenerate zero eigenstate $|\alpha^{(k)}(i,j)\rangle$ $(k=1,2,3,4)$ are generated by $\Gamma^{0\dagger}$ with the initial condition $\alpha(0,0)=1$, $\alpha(0,L) = 1$, $\alpha(L,0)=1$ and $\alpha(L,L)=1$ respectively. It can be easily checked that $\sum_{k=1}^{4}\alpha^{(k)}(i,j)=1$ for all $i$ and $j$, which is consistent with the requirement that the identity is always a left steady state $(\Gamma^{0\dagger})|I\rangle=0$. Now, we can calculate the first-order effect when the perturbation is turned on. 
With $\delta\Gamma = h\sum_{i}(\sigma^{x}_{i}-1)$, the first-order effective Hamiltonian is
\begin{equation}
\Gamma^{(1)}_{\text{eff}}=\langle \rho^{0,\mu'\nu'}_{L}|\delta\Gamma|\rho^{0,\mu\nu}_{R}\rangle=nh
\left(
\begin{array}{cccc}
a & b & c & d\\
b&a&d&c\\
c&d&a&b\\
d&c&b&a
\end{array}
\right),
\end{equation}
where $n=L^2$ is the number of vertices. Here the matrix elements are
\begin{equation}
    \begin{aligned}
    a = \alpha(1,0)+\alpha(0,1)-2,\ b = \alpha(L-1,0)+\alpha(L,1),\
    c = \alpha(1,L)+\alpha(0,L-1),\ d = \alpha(L-1,L)+\alpha(L,L-1).
    \end{aligned}
\end{equation}
Then we can get four eigenvalues $\delta_1, \delta_2, \delta_3$, and $\delta_4$:  
\begin{equation}
\begin{aligned}
\delta_1 = a-b-c+d,\quad \delta_2 = a+b-c-d,\quad \delta_3 = a-b+c-d,\quad \delta_4 = a+b+c+d,\\
\end{aligned}
\end{equation}
Particularly, $\delta_4 = 0$, which is consistent with the Liouvillian dynamics that there is always a zero eigenvalue. We numerically find that $\delta_2=\delta_3 \sim -1/\text{log}(L)$ and the splitting is proportional to $nh/\text{log}(n)$. Therefore, the degeneracy of the steady states is already broken at the first order and there is only one steady state when the perturbation is added.

\section{2d steady state}\label{sec:appendix2dss}
In this part, we are going to figure out the detailed Markovian process and find the exact solution of the steady state under perturbation. First, recall that the dephasing term $L_{z,l}=\sqrt{\kappa_{z}}\sigma^{z}_{l}$ does not change the spin configuration in $\sigma_{z}$ basis, but it will kill all the off-diagonal elements. Notice that all the other dissipators are also compatible with the diagonal structure of the density matrix (A diagonal matrix would still be diagonal after the action), therefore the steady state must be diagonal. Second, in the diagonal subspace, $A_{v}$ operators commute with other dissipators. The steady state must satisfy $A_{v}\rho_{ss}A_{v}=\rho_{ss}$, which suggests $\rho_{ss}$ is in the following form:
\begin{equation}
\rho_{ss}=  \sum_{\rm{conf}_m}\text{weight}(\text{conf}_m)\sum_{g\in G'}g|\text{conf}_m\rangle\langle\text{conf}_m|g,
\end{equation}
\begin{figure}[htbp]
\begin{minipage}[t]{0.5\linewidth}
    \centering
\includegraphics[width=\textwidth]{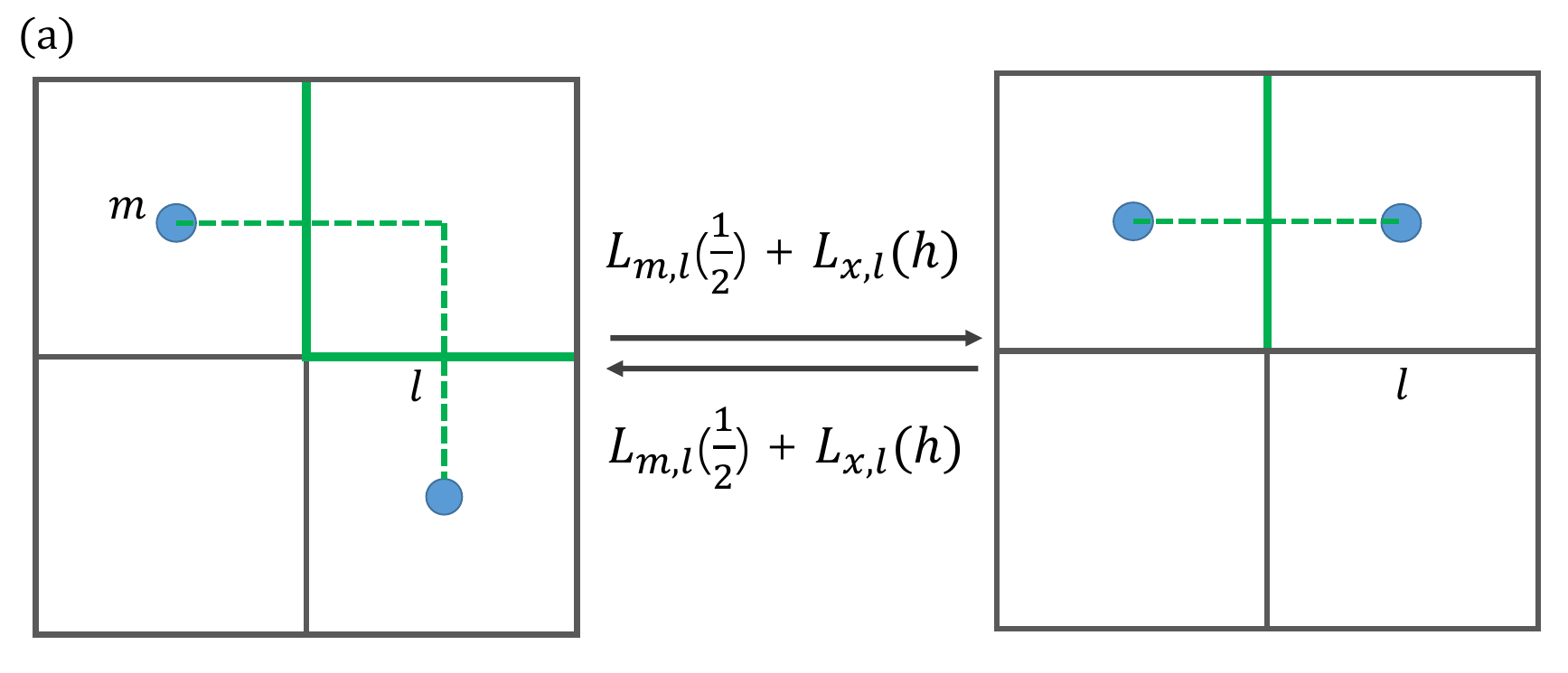}
    \end{minipage}
    \begin{minipage}[t]{0.5\linewidth}
    \centering
    \includegraphics[width=\textwidth]{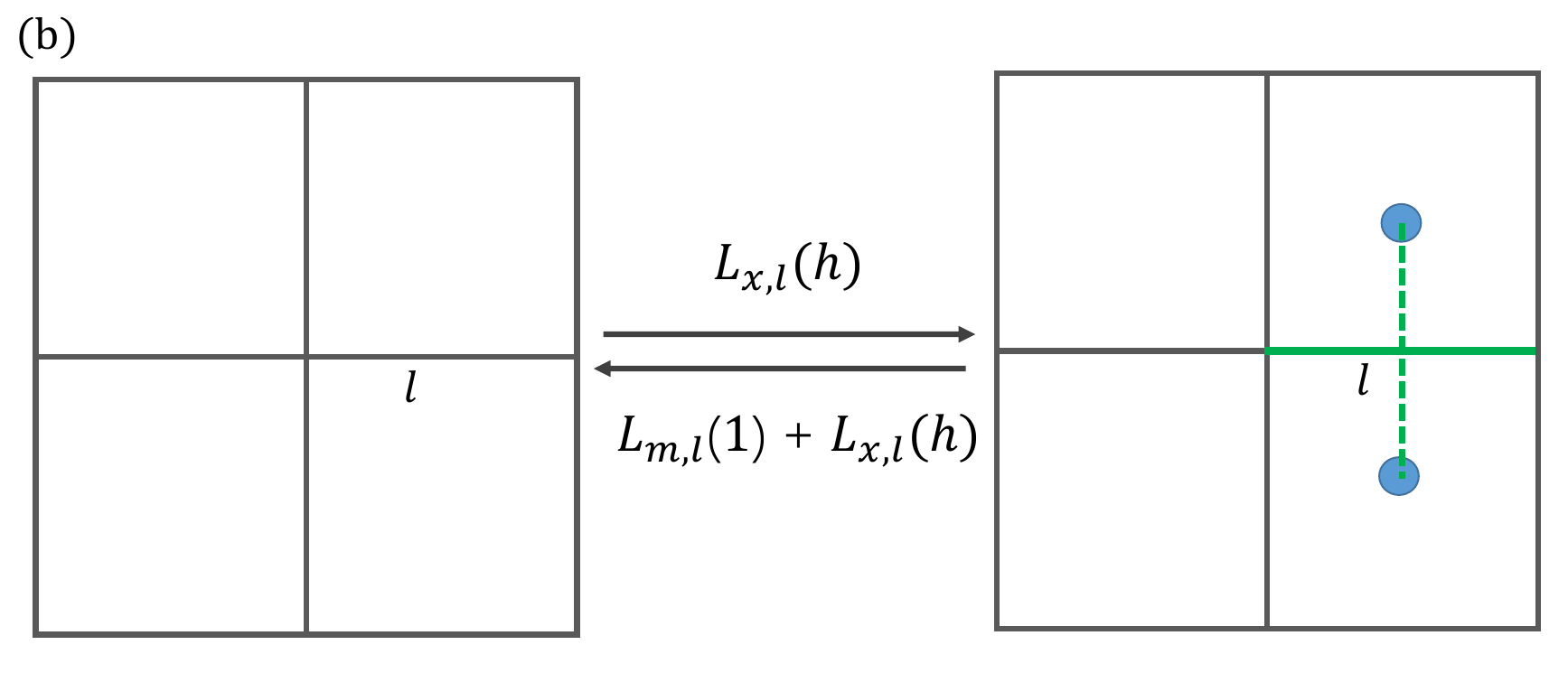}
    \end{minipage}
    \caption{The action of $L_{m,l}$ and $L_{x,l}$ on $m$ particles (represented by blue solid circles). 
 Links with $\sigma^{z}=-1$ are labeled by green solid lines and other spins are $\sigma^{z}=1$. (a) Both $L_{m,l}$ and $L_{x,l}$ can move one $m$ particle to a neighbor plaquette with no $m$ particle. (b) Both $L_{m,l}$ and $L_{x,l}$  can annihilate a pair of $m$ particles. Only $L_{x,l}$ can create a pair of $m$ particles. } \label{fig:action on e}
\end{figure}
where $\text{conf}_{m}$ gives one specific spin configuration of  $m$ particles distribution.
From this illustration, we see that the $m$ particle would move freely once created until it meets another $m$ particle when they might annihilate in pairs. With this intuition, we guess that the configuration weight only depends on the number of $m$ particles, irrespective of their detailed position (In the end, we will see that the assumption turns out to be correct.).  Now, we assume the steady state has the form $\rho\sim\sum_{g\in G'} \sum_{k}\beta_{2k}g|m_{2k}\rangle\langle m_{2k}|g$.  $L_{m,l}=\sigma^{x}_lP(\sum_{p|l\in p} B_{p})$ and $L_{x,l}=\sqrt{h}\sigma^{x}_{l}$ can mix $|m_{2k-2}\rangle\langle m_{2k-2}|$, $|m_{2k}\rangle\langle m_{2k}|$ and $|m_{2k+2}\rangle\langle m_{2k+2}|$ through moving, creating and annihilating $m$ particles. In Fig. \ref{fig:transition}, we give the transition rates associated with different terms. Here we define three characteristic values of a specific state:
\begin{equation}
\begin{aligned}
a &=\#\text{dual links connected to $m$ particles},\\
b &=\#\text{dual links with $m$ particles at both ends},\\
c &=\# \text{dual links with no $m$ particles at ends}.
\end{aligned}
\end{equation}
Here $a$, $b$, and $c$ have simple relations $a+b=4\cdot 2k$ and $a+c=2n$. $b$ is the number of pairs that can be annihilated by just flipping one link at a time and $c$ gives the number of pairs that can be created by flipping one link at a time. Now we can give the equation of coefficient of $|m_{2k}\rangle\langle m_{2k}|$ after the action of $L_{m,l}$ and $L_{x,l}$:
 \begin{figure}[htb]
 \centering
\includegraphics[width=0.8\linewidth]{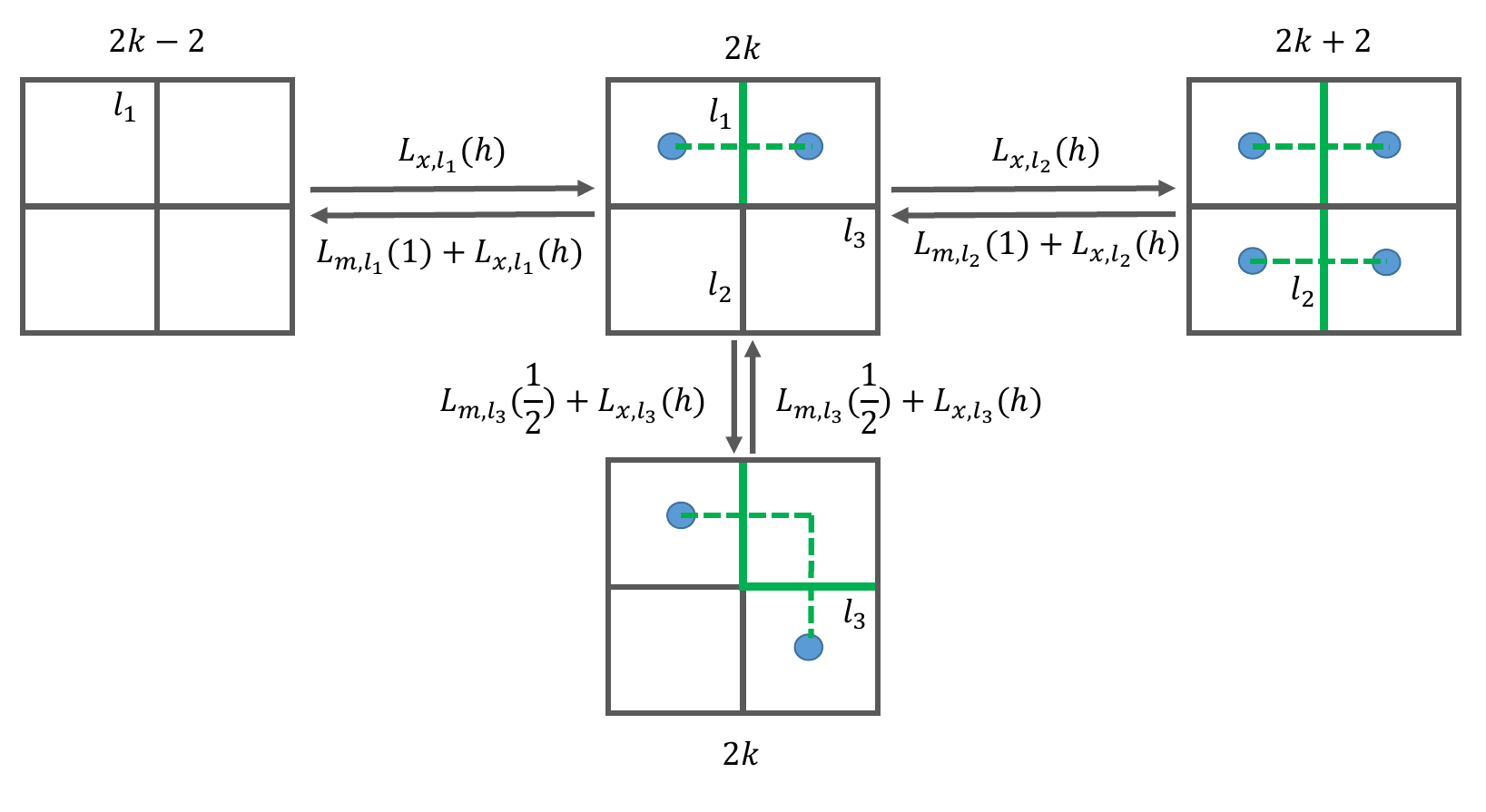}
\caption{
Transition from $|m_{2k-2}\rangle\langle m_{2k-2}|$, $|m_{2k}\rangle\langle m_{2k}|$ and $|m_{2k+2}\rangle\langle m_{2k+2}|$ to $|m_{2k}\rangle\langle m_{2k}|$. Inside the bracket, we mark the transition rate associated with the corresponding process. } \label{fig:transition}
\end{figure}
\begin{equation}
c(h+1)\beta_{2k+2}+((a-b-2n)h-b)\beta_{2k}+bh\beta_{2k-2}=0
\end{equation}
 Set $\beta_{2k}=\beta^{2k}$, then this equation can be simplified as follows:
\begin{equation}
c(h+1)\beta^4-(ch+b(h+1))\beta^{2}+bh=0
\end{equation}
where we have replaced $a$ with $2n-c$. We straightforwardly get two solutions $\beta^{2}=h/(h+1)$ and $\beta^{2}=b/c$. In the beginning, we assume the state coefficients only depend on the number of particles, and the solution $\beta^{2}=h/(h+1)$ is indeed consistent with the assumption. Finally, we get the exact form of the steady state:
\begin{equation}
    \rho_{ss}=\frac{1}{T'}\sum_{k}\beta^{2k}\sum_{g\in G'}\sum_{\{r\}}g|m_{2k}(\{ r \})\rangle \langle m_{2k}(\{r\})|g,
\end{equation}
which turns out to be the unique steady state by perturbation analysis in Sec. \ref{sec:2dperturbation} and numerical check. By the exact form of steady state, we can check that for any two configurations $\alpha, \beta$, the transition between these two satisfies the detailed balance
$\Gamma_{\alpha\beta}p(\beta)=\Gamma_{\beta\alpha}p(\alpha)$. Here $\Gamma$ is the Markovian generator and $p(\alpha)$ is the probability of configuration $\alpha$ (Eq. \eqref{eq:markov}). Therefore, the steady state is a Gibbs State.

\section{Calculation of $S_{A}$ in Model-1}\label{sec:appendixkp}
\subsection{$\bar{A}$ is path-connected}
In this part, we give the details of the calculation of entanglement entropy $S_{A}$ when $\bar{A}$ is path-connected ($\bar{A}$ has only one piece of connected area. See Fig. \ref{fig:bipartition}). For the eigenvalues $\lambda_{j}$ and degeneracy $D_{j}$ have already been listed in Eq. \eqref{eq:kpeigenvalue}, we have
\begin{equation}
\begin{aligned}
S_{A}&=- \text{Tr}\rho_{A}\log\rho_{A}=-\sum_{j} D_j\lambda_{j} \log\lambda_{j}\\
     &=- \sum_{j=0}^{m_{A}}{m_{A} \choose j}\frac{|G|}{T}\sum_{k=\lceil \frac{j}{2}\rceil}^{\lfloor \frac{n-m_{A}+j}{2}\rfloor}\beta^{2k} {n-m_{A} \choose 2k-j}\log\left[\frac{|G_{\bar A}|}{T}\sum_{k=\lceil \frac{j}{2}\rceil}^{\lfloor \frac{n-m_{A}+j}{2}\rfloor}\beta^{2k} {n-m_{A} \choose 2k-j}\right]\\
     &=-\frac{1}{t^{+}(\beta,n)}\sum_{j=0}^{m_{A}}\sum_{k=\lceil \frac{j}{2}\rceil}^{\lfloor \frac{n-m_{A}+j}{2}\rfloor}\beta^{2k} {n-m_{A} \choose 2k-j}{m_{A} \choose j}\left\{\log\frac{|G_{\bar A}|}{|G|}+\log\left[\frac{1}{t^{+}(\beta,n)}\sum_{k=\lceil \frac{j}{2}\rceil}^{\lfloor \frac{n-m_{A}+j}{2}\rfloor}\beta^{2k} {n-m_{A} \choose 2k-j}\right]\right\},
\end{aligned}    
\end{equation}
where we define two functions $t^{\pm}(\beta,n)$,
\begin{equation}
\begin{aligned}
t^{+}(\beta,n)&=\sum_{k}\beta^{2k}{n \choose 2k}=\frac{(1+\beta)^n+(1-\beta)^n}{2},\\
t^{-}(\beta,n)&=\sum_{k}\beta^{2k-1}{n \choose 2k-1}=\frac{(1+\beta)^n-(1-\beta)^n}{2},\\
\end{aligned}
\end{equation}
and the normalization constant can be written as $T=|G|t^{+}(\beta,n)$. With the combinatorial formula $\sum_{j}{n-m_{A} \choose 2k-j}{m_{A} \choose j}={n \choose 2k}$ and $j{m_{A} \choose j}=m_{A}{m_{A}-1 \choose j-1}$, we can  analyze these two terms separately. The first term in the parentheses reduces to the same result with the $h=0$ case, 
\begin{equation}
\begin{aligned}
S_{A}^{(1)}&=\frac{1}{t^{+}(\beta,n)}\sum_{k=0}^{\lfloor\frac{n}{2}\rfloor}\sum_{j=0}^{\text{min}(m_{A},2k)}\beta^{2k} {n-m_{A} \choose 2k-j}{m_{A} \choose j}\log\frac{|G|}{|G_{B}|}\\
         &=\frac{1}{t^{+}(\beta,n)}\sum_{k=0}^{\lfloor\frac{n}{2}\rfloor}\beta^{2k}{n \choose 2k}\log\frac{|G|}{|G_{\bar A}|}= \log\frac{|G|}{|G_{\bar A}|}\\
         &= (|A|+|\partial A|-1)\log2.
\end{aligned}
\end{equation}
The second term is more complicated,
\begin{equation}
\begin{aligned}
S_{A}^{(2)}&=-\frac{1}{t^{+}(\beta,n)}\sum_{j=0}^{m_{A}}\sum_{k=\lceil \frac{j}{2}\rceil}^{\lfloor \frac{n-m_{A}+j}{2}\rfloor}\beta^{2k} {n-m_{A} \choose 2k-j}{m_{A} \choose j}\log\left[\frac{1}{t^{+}(\beta,n)}\sum_{k=\lceil \frac{j}{2}\rceil}^{\lfloor \frac{n-m_{A}+j}{2}\rfloor}\beta^{2k} {n-m_{A} \choose 2k-j}\right]\\
         &=-\frac{1}{t^{+}(\beta,n)}\sum_{j=0}^{m_{A}}\beta^{j}{m_{A} \choose j}\sum_{k=\lceil \frac{j}{2}\rceil}^{\lfloor \frac{n-m_{A}+j}{2}\rfloor}\beta^{2k-j}{n-m_{A} \choose 2k-j}\log\left[\frac{\beta^{j}}{t^{+}(\beta,n)}\sum_{k=\lceil \frac{j}{2}\rceil}^{\lfloor \frac{n-m_{A}+j}{2}\rfloor}\beta^{2k-j} {n-m_{A} \choose 2k-j}\right]\\
         &=-\frac{1}{t^{+}(\beta,n)}\sum_{j=0}^{m_{A}}\beta^{j}{m_{A} \choose j}\sum_{k'=\text{even/odd}}\beta^{k'}{n-m_{A} \choose k'}\log\left[\frac{\beta^{j}}{t^{+}(\beta,n)}\sum_{k'=\text{even/odd}}\beta^{k'}{n-m_{A} \choose k'}\right]\\
         &=-\frac{1}{t^{+}(\beta,n)}\sum_{j=0}^{m_{A}}\beta^{j}{m_{A} \choose j}t^{\pm}(\beta,n-m_{A})\log\left[\frac{\beta^{j}}{t^{+}(\beta,n)}t^{\pm}(\beta,n-m_{A})\right].
\end{aligned}    
\end{equation}
In the third line, the summation is over even/odd $k'$ for j odd/even,  getting $t^{\pm}$  in the last line. This is the exact form of the second part of $S_{A}$. With the condition $\beta^{2}=h/(1+h)<1$ , we have
\begin{equation}
\begin{aligned}
\lim_{n\to \infty}\frac{t^{\pm}(\beta,n-m)}{t^{\pm}(\beta,n)}&=\lim_{n\to \infty}\frac{(1+\beta)^{n-m}\pm(1-\beta)^{n-m}}{(1+\beta)^{n}\pm(1-\beta)^{n}}=\frac{1}{(1+\beta)^{m}}.
\end{aligned}
\end{equation}
Therefore, when we consider the thermodynamic limit $n\rightarrow +\infty$, a simple result shows up,
\begin{equation}
\begin{aligned}
\lim_{n\to \infty}S_{A}^{(2)}&=-\sum_{j=0}^{m_{A}}\frac{\beta^{j}}{(1+\beta)^{m_{A}}}{m_{A} \choose j}\log\frac{\beta^{j}}{(1+\beta)^{m_{A}}}\\
&=-\sum_{j=0}^{m_{A}}\frac{\beta^{j}}{(1+\beta)^{m_{A}}}j{m_{A} \choose j}\log\beta+\sum_{j=0}^{m_{A}}\frac{m_{A}}{(1+\beta)^{m_{A}}}\beta^{j}{m_{A} \choose j}\log(1+\beta)\\
&=-\frac{\beta \log\beta}{(1+\beta)^{m_{A}}}\sum_{j}m_{A}{m_{A}-1 \choose j-1}\beta^{j-1}+m_{A}\log(1+\beta)\\
%&=- m_{A}\frac{\beta \log\beta}{1+\beta}+m_{A}\log(1+\beta)\\
&= m_{A}(\log(1+\beta)-\frac{\beta}{1+\beta}\log\beta).
\end{aligned}
\end{equation}
Finally, we get the exact form of $S_A$:
\begin{equation}
\begin{aligned}
S_{A}&= S_{A}^{(1)}+S_{A}^{(2)}\\
     &= (|A|+|\partial A|-1)\log2+m_{A}(\log(1+\beta)-\frac{\beta}{1+\beta}\log\beta)\\
     &= (|A|+|\partial A|-1)\log2+m_{A}f(\beta),
\end{aligned}
\end{equation}
where $f(\beta)=\log(1+\beta)-\frac{\beta}{1+\beta}\log\beta$.

\subsection{$\bar{A}$ is not path-connected}
In this part, we give the details of the calculation of $\rho_{A}$ and $S_{A}$ in partition-1 where $\bar{A}$ has two disjointed areas [See Fig. \ref{Levin-Wen}(1)]. There are four different states in $\rho_{A}$ and we can label these states with the particle number in different regions.  For convenience, we give the following definition:
\begin{equation}
\begin{aligned}
j\equiv&\# \text{ of $m$ particles totally inside $A$} ,\\
k_1+j\equiv&\# \text{ of $m$ particles totally inside $B_1\cup  A$},\\
k_2+j\equiv&\# \text{ of $m$ particles totally inside $B_2\cup A$},\\
m_{A}\equiv&\# \text{ of plaquettes totally inside $A$} ,\\
M_1+m_{A}\equiv&\# \text{ of plaquettes totally inside $B_{1}\cup A$},\\
M_2+m_{A}\equiv&\# \text{ of plaquettes totally inside $B_{2}\cup A$}.\label{eq:variable1}
\end{aligned}
\end{equation}
Then we have some constraints on these variables:
\begin{equation}
\begin{aligned}
0\leq j\leq m_A,\quad 0\leq k_1\leq M_1,\quad 0\leq k_2\leq M_2,\quad j+k_1+k_2=\text{even}.
\end{aligned}
\end{equation}
Unlike partition-2, 3, and 4 where $\bar A$ is path-connected, here the reduced configuration in $A$ does not only depend on $e$ particle configurations inside $A$, but also on the parity of $k_1$ and $k_2$. Since $j+k_1+k_2$ is even, there are in total $4$ classes. For $s_1$ states, $j=$ odd, $k_1=$ odd, and $k_2=$ even; for $s_2$ states, $j=$ odd, $k_1=$ even, and $k_2=$ odd; for $s_3$ states, $j=$ even, $k_1=$ even, and $k_2=$ even; for $s_4$ states, $j=$ even, $k_1=$ odd, and $k_2=$ odd. See Fig. \ref{fig:fourclasses} for an illustration. For each $m$ configuration in $A$ (with $j$ fixed), there are two topologically inequivalent configurations of $A$, which cannot be transformed to each other via the action of any $g_A$. The explicit form of $\rho_{A}$ is
\begin{equation}
\begin{aligned}
\rho_{A}=&\text{Tr}_{\bar A}\frac{1}{T} \sum_{j=0}^{m_{A}}\sum_{k=\lceil \frac{j}{2}\rceil}^{\lfloor \frac{n-m_{A}+j}{2}\rfloor}\beta^{2k}\sum_{g\in G}\sum_{\{r_{A}\}}\sum_{\{r_{\bar A}\}}g|m_{{2k-j}}(\{r_{\bar A}\})m_{j}(\{r_{A}\})\rangle\langle m_{2k-j}(\{r_{\bar A}\})m_{j}(\{r_{A}\})|g \\
=&\frac{|G_{\bar A}|}{T} \bigg( \sum_{j\text{ odd}}\sum_{\{r_A\}}   \sum_{\substack{k_1\text{ odd}\\k_2\text{ even}} }\beta^{j+k_1+k_2}{M_1 \choose k_1}{M_2\choose k_2} g_A|m_j(\{r_{A}\}),s_1\rangle\langle m_j(\{r_{A}\}),s_1|g_A\\
&\qquad +\sum_{j\text{ odd}} \sum_{\{r_A\}}   \sum_{\substack{k_1 \text{ even}\\k_2\text{ odd}} }\beta^{j+k_1+k_2}{M_1 \choose k_1}{M_2\choose k_2} g_A|m_j(\{r_{A}\}),s_2\rangle\langle m_j(\{r_{A}\}),s_2|g_A\\
&\qquad +\sum_{j\text{ even}} \sum_{\{r_A\}}   \sum_{\substack{k_1 \text{ even}\\k_2\text{ even}} }\beta^{j+k_1+k_2}{M_1 \choose k_1}{M_2\choose k_2} g_A|m_j(\{r_{A}\}),s_3\rangle\langle m_j(\{r_{A}\}),s_3|g_A\\
&\qquad+\sum_{j\text{ even}} \sum_{\{r_A\}}   \sum_{\substack{k_1 \text{ odd}\\k_2\text{ odd}} }\beta^{j+k_1+k_2}{M_1 \choose k_1}{M_2\choose k_2} g_A|m_j(\{r_{A}\}),s_4\rangle\langle m_j(\{r_{A}\}),s_4|g_A\bigg).\\
%=\lceil \frac{j}{2}\rceil}^{\lfloor \frac{n-m+j}{2}\rfloor}\beta^{2k} %{n-m \choose 2k-j}\sum_{g\in G/G_{\bar A}}x_{A}|e_{j}(r_{A})\ra\la e_{j}(r_{A})|x_{A}\\
\end{aligned}
\end{equation}
In the summation, the restrictions $0\leq k_{1,2}\leq M_{1,2}$ and $0\leq j \leq m_{A}$ are implied. Using the identity 
\begin{equation}
    t^{\pm}(\beta,M_{i})=\sum_{k_i\ \text{even/odd}}\beta^{k_i}{M_i\choose k_i}=\frac{(1+\beta)^{M_i}\pm (1-\beta)^{M_i}}{2},\quad i=1,2.
\end{equation}

and $T=|G|t^{+}(\beta,n)$, in the thermodynamic limit, $n\to\infty$ and $M_{i}\to\infty$, we can simplify the above expression which is given in Eq. \eqref{eq:rhoA1}:
\begin{equation}
\begin{aligned}
\rho_A=\frac{G_{\bar A}}{2|G|(1+\beta)^{m_A}}\sum_{g\in G/G_{\bar A}}\bigg[ &\sum_{j\text{ odd}}\sum_{\{r_A\}}\beta^j\Big(  g_A|m_j(\{r_A\}),s_1\rangle\langle m_j(\{r_A\}),s_1|g_A
+g_A|m_j(\{r_A\}),s_2\rangle\langle m_j(\{r_A\}),s_2|g_A\Big)\\
+&\sum_{j\text{ even}}\sum_{\{r_A\}}\beta^j\Big(  g_A|m_j(\{r_A\}),s_3\rangle\langle m_j(\{r_A\}),s_3|g_A
+g_A|m_j(\{r_A\}),s_4\rangle\langle m_j(\{r_A\}),s_4|g_A\Big)\bigg].
\end{aligned}
\end{equation}
The eigenvalue $\lambda_{i}$ and degeneracy $D_{i}$ are
\begin{equation}
\lambda_j=\frac{|G_{\bar A}|\beta^j}{2|G|(1+\beta)^{m_A}},\quad D_j=2{m_A\choose j}\frac{|G|}{|G_{\bar A}|},
\end{equation}
and the entropy for partition-1 is 
\begin{equation}
\begin{aligned}
S_{A1}&=\sum_{j=0}^{m_A}D_j\lambda_j\log \lambda_j\\
% &=\sum_{j=0}^{m_A}\frac{{m_A\choose j}\beta^j}{(1+\beta)^{m_A}}\log \frac{|G_{\bar A}|\beta^j}{2|G|(1+\beta)^{m_A}}\\
% &=\log{\frac{|G|}{G_{\bar A}}}+\log 2+m_Af(\beta)\\
&=(|A|+|\partial A|)\log 2+m_Af(\beta).
\end{aligned}
\end{equation}

\section{Topological entropy of Model-2}\label{sec:appendixqm}
To calculate TEE, we need to define some variables that can label the configurations with $e$ and $m$ particles. Like the variables defined in Eq. \eqref{eq:variable1},  we need to define two sets of variables:
\begin{equation}
\begin{aligned}
n_{v/p}&\equiv\# \text{ of vertices/plaquettes on the whole lattice,}\\
m_{v/p}&\equiv\#\text{ of vertices/plaquettes totally inside } A, \\
%&\equiv\#\text{vertices/plaquettes with all four links belonging to } A\\
2k_{e/m}&\equiv\#\text{ of $e/m$ particles on the whole lattice},\\
j_{e/m}&\equiv\# \text{ of $e/m$ particles totally inside $A$ },\\
% j'_{e/m}&\equiv 2k_{e/m}-j_{e/m}\\
% \{r\}&\equiv \text{configuration of $e$ particles}\\
% \{p_m\}&\equiv \text{configuration of $m$ particles}\\ 
k_{e/m1}+j_{e/m}&\equiv\# \text{ of $e/m$ particles totally inside $B_1 \cup A$}, \\
k_{e/m2}+j_{e/m}&\equiv\# \text{ of $e/m$ particles totally inside $B_2\cup A$}, \\
M_{v/p1}+m_{v/p}&\equiv \# \text{ of vertices/plaquettes totally inside $B_1\cup A$},\\
M_{v/p2}+m_{v/p}&\equiv \# \text{ of vertices/plaquettes totally inside $B_2\cup A$},\\
C(g,\{p_m\})&\equiv\# \text{ of $B_p$ operators acting on plaquettes with $m$ particles},\\
Q_{1(2)}(g) &\equiv \left\{  \begin{array}{ll} 0,& \text{ if $g$ contains the product of all $A_{v}$ operators acting on $B_{1(2)}$},\\
1,& \text{ otherwise} \end{array}\right\}
\end{aligned}
\end{equation}
$j'_{e/m}=2k_{e/m}-j_{e/m}$. $\{r_{e}\}$ and $\{p_m\}$ give the distribution of $e$ and $m$ particles. The latter five definitions only apply to partition-1 where $\bar A=B_1\cup B_2$. 
There are some constraints:
\begin{equation}
\begin{aligned}
0\leq j_{e}\leq m_v,\quad 0\leq k_{e/p1}\leq M_{v/p1},\quad 0\leq k_{e/p2}\leq M_{e/p2},\quad j_{e/m}+k_{e/m1}+k_{e/m2}=2k_{e/m}.
\end{aligned}
\end{equation}
Throughout the calculation, we always take the thermodynamic limit, 
\begin{equation}
n_{v/p},m_{v/p},n_{v/p}-m_{v/p},M_{v/p_1},M_{v/p_2}\gg \beta_{e/m}^{-1}.
\label{eq:thermolimit}
\end{equation}
Here, we take the Levin-Wen scheme.

\subsection{$h_{x}=h_{z}=0$}
For nonperturbed case $h_{x}=h_{z}=0$, we choose one of the steady states $\rho_{ss}=\frac{1}{|G|}\sum_{g\in G}\sum_{\tilde{g}\in G}g|\Uparrow\rangle\langle\Uparrow|g\tilde{g}$, and calculate the reduced density matrix,
\begin{equation}
\begin{aligned}
\rho_A=\text{Tr}_{\bar A}\rho_{ss}=\frac{1}{|G|}\sum_{g\in G}\sum_{\tilde{g}\in G_A}g_A|\Uparrow\rangle\langle\Uparrow|g_A \tilde{g}_{A}=\frac{|G_{\bar A}|}{|G|}\sum_{g\in G/G_{\bar A}}\sum_{\tilde{g}\in G_A}g_A|\Uparrow\rangle\langle\Uparrow|g_A\tilde{g}_{A}.
\end{aligned}
\label{unp}
\end{equation}
Following the standard procedure and with the condition that $\text{Tr}(\rho_{A})=1$, we have
\begin{equation}
    \begin{aligned}
        &\rho_A^n=\left(\frac{|G_A||G_{\bar A}|}{|G|}\right)^{n-1}\rho_A,\\
& S_A=\lim_{n\rightarrow 1}\frac{1}{1-n}\log\text{Tr}(\rho_A^n)=\log\frac{|G|}{|G_A||G_{\bar A}|}.
    \end{aligned}
\end{equation}
Using $|G|=2^{n-1},|G_A|=2^{|A|+p_{\bar A}-1},|G_{\bar A}|=2^{|\bar A|+p_A-1}$, we finally get
\begin{equation}
    S_A=\left(|\partial A|+1-p_A-p_{\bar A}\right)\log 2.
\end{equation}
Then the topological entanglement entropy is
\begin{equation}
\begin{aligned}
S_{\text{topo}}=&(p_{A_{1}}+p_{\bar A_{1}}-p_{A_2}-p_{\bar{A}_2}\\
& -p_{A_3}-p_{\bar A_3}+p_{A_4}+p_{\bar A_4})\log 2\\
=&2\log 2.
\end{aligned}
\end{equation}

\subsection{$h_{z}=0$, \ $h_{x}\neq0$}
Similar to the calculation of Model-1 in Sec. \ref{sec:appendixkp}, we only need to arbitrarily take one of the degenerate steady states, 
\begin{equation}
\rho_{ss}= \frac{1}{T'_{m}}\sum_{g\in G'}\sum_{g'\in G'}g(\sum_{k,\{r\}}\beta_{m}^{2k}|m_{2k}(\{r\})\rangle \langle m_{2k}(\{r\})|)g', 
\end{equation}
and the form of $\rho_A=\text{Tr}_{\bar A}\rho_{ss}$ would depend on whether $\bar A$ is connected. We first discuss the situation when there is only one connected piece of $\bar A$, as is the case for partition-$2,3,4$. Then the reduced density matrix is
\begin{equation}
\begin{aligned}
\rho_A&=\frac{1}{T'_m}\text{Tr}_{\bar A}\sum_{\{j_{m},j'_{m}\}}\beta_m^{j_{m}+j'_{m}}\sum_{\{r_A\}}\sum_{\{r_{\bar A}\}}\sum_{g,g'\in G'} g|m_{j_{m}}(\{r_A\})m_{j'_{m}}(\{r_{\bar A}\})\rangle\langle m_{j_{m}}(\{r_A\})m_{j'_{m}}(\{r_{\bar A}\})|g'\\
&=\frac{|G_{\bar A}|}{T_m}\sum_{j_{m}}\beta_m^{j_{m}}\sum_{j'_{m}} {n_p-m_p\choose j'_{m}} \beta_m^{j'_{m}} \sum_{\{r_A\}}\sum_{g\in G/G_{\bar{A}}}\sum_{\tilde{g}\in G_A}g_A|m_{j_{m}}(\{r_A\})\rangle\langle m_{j_{m}}(\{r_A\})|g_A\tilde{g}\\
&=\frac{|G_{\bar A}|}{|G|(1+\beta_m)^{m_p}}\sum_{j_{m}}\beta_{m}^{j_{m}}\sum_{\{r_A\}}\sum_{g\in G/G_{\bar{A}}}\sum_{\tilde{g}\in G_A}g_A|m_{j_{m}}(\{r_A\})\rangle\langle m_{j_{m}}(\{r_A\})|g_A\tilde{g}.
\end{aligned}
\end{equation}
The last line is in the thermodynamic limit in Eq. \eqref{eq:thermolimit} and $\rho^{n}_{A}$ is
\begin{equation}
    \rho_A^n=\left[\frac{|G_A||G_{\bar A}|}{|G|(1+\beta_m)^{m_v}}\right]^{n-1}\frac{|G_{\bar A}|}{|G|(1+\beta_m)^{m_v}}\sum_{j_{m}} \beta_m^{nj_{m}}\sum_{\{r_A\}}\sum_{g\in G/G_{\bar{A}}}\sum_{\tilde{g}\in G_A}g_A|m_{j_{m}}(\{r_A\})\rangle\langle m_{j_{m}}(\{r_A\})|g_A\tilde{g}.
\end{equation}
Then, for partition-2, 3, and 4, it is easy to get the entropy,
\begin{equation}
    \begin{aligned}
& \text{Tr} (\rho_A^n)=\left[\frac{|G_A||G_{\bar A}|}{|G|(1+\beta_m)^{m_p}}\right]^{n-1}\left(\frac{1+\beta_m^{n}}{1+\beta_m}\right)^{m_p},\\
& S_A=\lim_{n\rightarrow 1}\frac{1}{1-n}\log\text{Tr}(\rho_A^n)=\log\frac{|G|}{|G_A||G_{\bar A}|}+m_p f(\beta_m).\label{partition2}
    \end{aligned}
\end{equation}

Next, we calculate the subsystem entropy of partition-1. Because $\bar A$ has two disconnected regions, there are four topologically inequivalent scenarios as illustrated in Fig. \ref{fig:fourclasses}, resulting in a more complicated $\rho_A$:

\begin{equation}
\begin{aligned}
\rho_A =\frac{|G_{\bar A}|}{2|G|(1+\beta_m)^{m_p}}\sum_{g\in G/G_{\bar A}}\sum_{\tilde{g}\in G_A}g_A\bigg\{&\sum_{j\text{ odd}}\sum_{\{r_A\}}\sum_{s=s_1,s_2}\beta_m^j|m_j(\{r_A\}),s\rangle\langle m_j(\{r_A\}),s|\\
+&\sum_{j\text{ even}}\sum_{\{r_A\}}\sum_{s=s_3,s_4}\beta_m^j|m_j(\{r_A\}),s\rangle\langle m_j(\{r_A\}),s|\bigg\}g_A\tilde{g},\\
\end{aligned}
\label{partition1}
\end{equation}
and the following are similar,
\begin{equation}
    \begin{aligned}
        \rho_A^n&=\left[\frac{|G_A||G_{\bar A}|}{2|G|(1+\beta_m)^{m_p}} \right]^{n-1}\frac{|G_{\bar A}|}{2|G|(1+\beta_m)^{m_p}}\sum_{\substack{g\in G/G_{\bar A} \\ \tilde{g}\in G_A}}g_A\sum_{j\text{ odd}}\sum_{\{r_A\}}\sum_{s=s_1,s_2}\beta_m^{nj}|m_j(\{r_A\}),s\rangle\langle m_j(\{r_A\}),s|g_A\tilde{g}\\
&+\left[\frac{|G_A||G_{\bar A}|}{2|G|(1+\beta_m)^{m_p}} \right]^{n-1}\frac{|G_{\bar A}|}{2|G|(1+\beta_m)^{m_p}}\sum_{\substack{g\in G/G_{\bar A} \\ \tilde{g}\in G_A}}g_A\sum_{j\text{ even}}\sum_{\{r_A\}}\sum_{s=s_3,s_4}\beta_m^{nj}|m_j(\{r_A\}),s\rangle\langle m_j(\{r_A\}),s|g_A\tilde{g}.\\
    \end{aligned}
\end{equation}
and the entropy is a little different which is caused by four different states,
\begin{equation}
    \begin{aligned}
        &\text{Tr}(\rho_A^n)=\left[\frac{|G_A||G_{\bar A}|}{2|G|(1+\beta_m)^{m_p}} \right]^{n-1}\left(\frac{1+\beta_m^ n}{1+\beta}\right)^{m_p},\\
& S_{A}=\lim_{n\rightarrow 1}\frac{1}{1-n}\log\text{Tr}(\rho_A^n)=\log\frac{|G|}{|G_A||G_{\bar A}|}+m_p f(\beta_m)+\log 2.
    \end{aligned}
\end{equation}
With the results in partition-1 and partition-2, 3, 4, we can straightforwardly get the topological entropy:
\begin{equation}
S_{\text{topo}}=2\log 2+(-m_{1p}+m_{2p}+m_{3p}-m_{4p})f(\beta_m)-\log 2=\log 2.
\end{equation}
which is half  of the topological entropy of the unperturbed Model-${2}$, and identical to that of model-${1}$. One can notice that this is closely related to the reduction of steady-state degeneracy from 16 to $4=\sqrt{16}$. This result again shows the rationality of generalizing the topological entropy defined by Levin and Wen to steady states of open systems. We learn that the topological entropy is vulnerable to the fluctuation of $m$ particles. The remaining topological entropy as well as the topological degeneracy is due to the fact that the $e$ particles remain unaffected. Therefore, we expect the topological entropy to vanish completely once we also perturb the $e$ particles by turning on $h_z$. In the following, we show this is indeed the case.

\subsection{$h_{x}\neq0,\ h_{z}\neq0$}
For $h_x,h_z\neq0$, remind the steady state:
\begin{equation}
\begin{aligned}
\rho_{em} &= \frac{1}{T'_{em}}\sum_{\mu,\nu}\sum_{k_e,\{r_e\}}\beta_{e}^{2k_e} \prod_{j=1}^{k_e}S^{z}_{t_{j}}\sum_{g,g'\in G}g\left(\sum_{k_m,\{p_m\}}\beta_{m}^{2k_m}   W_x^\mu {W_y}^\nu\prod_{i=1}^{k_m}S^{x}_{\tilde{t}_{i}}|\Uparrow\rangle \langle \Uparrow|\prod_{i=1}^{k_m}S^{x}_{\tilde{t}_{i}}W_x^\mu W_y^\nu\right)g'\prod_{j=1}^{k_e}S^{z}_{t_{j}}.\\
%&= \frac{1}{T'_{em}}\sum_{k_m,\{p_m\}}\beta_{m}^{2k_m}\sum_{g,g'\in G}(-1)^{C(g,\{p_m\})}g\sum_{k_e,\{r_e\}}\beta_{e}^{2k_e}\sum_{\mu,\nu}W_x^\mu W_y^\nu|e_{2k_e}(\{r_e\})\rangle\langle e_{2k_e}(\{r_e\})|{W_x}^\mu {W_y}^\nu g'(-1)^{C(g',\{p_m\})}
\end{aligned}
\end{equation}
Here $\{r_{e}\}$ and $\{p_{m}\}$ give the positions of $e$ and $m$ particles. To calculate the reduced density matrix, we should simplify the steady state. Since the string operator, $S^{z}_{t_{j}}$ is diagonal in the $\sigma^{z}$ basis, we can move $S^{z}_{t_{j}}$ through $gW^{\mu}_{x}W^{\nu}_{y}S^{x}_{\tilde{t}_{i}}$ and $S^{z}_{t_{j}}|\Uparrow\rangle=|\Uparrow\rangle$,  
\begin{equation}
\begin{aligned}
\rho_{em} &= \frac{1}{T'_{em}}\sum_{k_e,\{r_e\}}\beta_{e}^{2k_e}\sum_{g,g'\in G}(-1)^{C(g,\{r_{e}\})}g\sum_{k_m,\{p_{m}\}}\beta_{m}^{2k_m}\sum_{\mu,\nu}W_x^\mu W_y^\nu|m_{2k_m}(\{p_m\})\rangle\langle m_{2k_m}(\{p_m\})|{W_x}^\mu {W_y}^\nu g'(-1)^{C(g',\{p_{m}\})}\\
&=\frac{1}{T'_{em}}\sum_{k_e,\{r_{e}\}}\beta_{e}^{2k_e}\sum_{g,\tilde{g}\in G}(-1)^{C(\tilde{g},\{r_{e}\})}g\sum_{\mu,\nu=0,1}\sum_{k_m,\{p_{m}\}}\beta_{m}^{2k_m}W_x^\mu W_y^\nu|m_{2k_m}(\{p_m\})\rangle\langle m_{2k_m}(\{p_m\})|W_x^\mu W_y^\nu g\tilde{g},
\end{aligned}
\end{equation}
where $(-1)^{C(g,\{r_{e}\})}$ is the phase factor arising from permuting $S^{z}_{t_{j}}$ and $g$, and $C(g,\{r_{e}\})$ is the number of $A_{v}$ operators acting on vertices with $e$ particles.

Firstly, we consider partition-2, 3, and 4,
\begin{equation}
\begin{aligned}
\rho_A= \frac{|G_{\bar A}|}{T_{e}}\frac{1}{(1+\beta_m)^{m_p}}\sum_{k_e,\{r_{m}\}}\beta_{e}^{2k_e}\sum_{g\in G/G_{\bar{A}}}\sum_{\tilde{g}\in G_A}(-1)^{C(\tilde{g},\{r_{e}\})}\sum_{j_m,\{p_{m}\}}\beta_{m}^{j_m}g_A|m_{j_m}(\{p_{m}\})\rangle\langle m_{j_m}(\{p_{m}\})| g_A\tilde{g}.
\end{aligned}
\end{equation}
The sum of the contribution of different $m$ particle configurations is similar to the last case. Now we need to work out the summation over $e$ particle configurations, which does not generate new matrix elements of $\rho_A$, but only changes the coefficients of states,
\begin{equation}
\begin{aligned}
\sum_{k_e,\{r_{e}\}}\beta_{e}^{2k_e}(-1)^{C(\tilde{g},\{r_{e}\})}=\sum_{j_e}\sum_{j'_e}\sum_{\{r_{A}\}}\sum_{\{r_{\bar{A}}\}}\beta_{e}^{j_{e}+j'_{e}}(-1)^{C(\tilde{g},\{r_{e}\})}=\sum_{j_e}\sum_{\{r_{A}\}}\beta_e^{j_e}\frac{(1+\beta_e)^{n_e-m_e} } {2} (-1)^{C(\tilde{g},\{r_{A}\})}.
\end{aligned}
\end{equation}
Here $r_{A}$ gives the position of $e$ particle in area $A$. Since $\tilde{g}\in G_A$ only contains vertex operators in $A$, in the last step, we use 
\begin{equation}
{C(\tilde{g},\{r_e\})}={C(\tilde{g},\{r_{A}\})}.
\label{C}
\end{equation}
Thus for the partition-2, 3 and 4, 
\begin{equation}
\begin{aligned}
&\rho_A=\frac{1}{(1+\beta_e)^{m_v}(1+\beta_m)^{m_p}}\frac{|G_{\bar A}|}{|G|}\sum_{j_m,j_e}\sum_{\{r_A\},\{p_{A}\}}\beta_m^{j_m}\beta_e^{j_e}\sum_{\substack{g\in G/G_{\bar A}\\\tilde{g}\in G_A}}(-1)^{C(\tilde{g},\{p_{A}\})}g_A |m_{j_m}(\{p_A\})\rangle\langle m_{j_m}(\{r_A\})| g_A\tilde{g},\\
%& \rho_A^n=\left[   \frac{|G_A||G_{\bar A}|} {(1+\beta_e)^{m_v}(1+\beta_m)^{m_p} |G| }\right]^{n-1}  \frac{|G_{\bar A}|} {(1+\beta_e)^{m_v}(1+\beta_m)^{m_p} |G|}\sum_{j_e,j_m}\sum_{\{r_A\},\{p_{mA}\}}\beta_e^{nj_e}\beta_m^{nj_m}(-1)^{C(\tilde{g},\{p_{mA}\})}\\
% &\text{Tr}(\rho_A^n)= \left[   \frac{|G_A||G_{\bar A}|} {(1+\beta_e)^{m_v}(1+\beta_m)^{m_p} |G| }\right]^{n-1} \left(\frac{1+\beta_e^n}{1+\beta_e}\right)^{m_v}
% \left(\frac{1+\beta_m^n}{1+\beta_m}\right)^{m_p}\\
% & S_A=\lim_{n\rightarrow 1}\frac{1}{1-n}\log\text{Tr}(\rho_A^n)=\log\frac{|G|}{|G_A||G_{\bar A}|}+m_v f(\beta_e)+m_p f(\beta_m)\\
\end{aligned}
\end{equation}
and finally, we get
\begin{equation}
    \begin{aligned}
        &\text{Tr}(\rho_A^n)= \left[   \frac{|G_A||G_{\bar A}|} {(1+\beta_e)^{m_v}(1+\beta_m)^{m_p} |G| }\right]^{n-1} \left(\frac{1+\beta_e^n}{1+\beta_e}\right)^{m_v}
\left(\frac{1+\beta_m^n}{1+\beta_m}\right)^{m_p},\\
& S_A=\lim_{n\rightarrow 1}\frac{1}{1-n}\log\text{Tr}(\rho_A^n)=\log\frac{|G|}{|G_A||G_{\bar A}|}+m_v f(\beta_e)+m_p f(\beta_m).\label{Partition2}
    \end{aligned}
\end{equation}

Now we turn to partition-1. Again, we need to sum over the states corresponding to the topologically inequivalent configuration of $e$ particles in Fig. \ref{fig:fourclasses},

\begin{equation}
\begin{aligned}
\rho_A&=\frac{|G_{\bar A}|}{(1+\beta_e)^{m_v}(1+\beta_m)^{m_p}|G|}\sum_{j_e,\{r_{A}\}}\beta_e^{j_e}\sum_{\substack{g\in G/G_{\bar A}\\\tilde{g}\in G_A}} (-1)^{C(\tilde{g},\{r_{A}\})}g_A\sum_{\substack{j_m\text{ odd}\\p_{A}}} \sum_{s=s_1,s_2}\beta_m^{j_m} |m_{j_m}(\{p_A\}),s\rangle\langle m_{j_m}(\{p_A\}),s|g_A\tilde{g}\\
&+\frac{|G_{\bar A}|}{(1+\beta_e)^{m_v}(1+\beta_m)^{m_p}|G|}\sum_{j_e,\{r_{A}\}}\beta_e^{j_e}\sum_{\substack{g\in G/G_{\bar A}\\\tilde{g}\in G_A}} (-1)^{C(\tilde{g},\{r_{A}\})}g_A\sum_{\substack{j_m\text{ even}\\p_{A}}} \sum_{s=s_1,s_2}\beta_m^{j_m} |m_{j_m}(\{p_A\}),s\rangle\langle m_{j_m}(\{p_A\}),s|g_A\tilde{g}.\\
% &\bigg\{\sum_{j_m\text{ odd}} \sum_{s=s_1,s_2} \sum_{\{p_A\}} \beta_m^{j_m} |m_{j_m}(\{p_A\}),s\rangle\langle m_{j_m}(\{p_A\}),s| +
% \sum_{j_m\text{ even}} \sum_{s=s_3,s_4} \sum_{\{p_A\}} \beta_m^{j_m} |m_{j_m}(\{p_A\}),s\rangle\langle m_{j_m}(\{p_A\}),s|\bigg\}
% g_A\tilde{g}
\label{rhoA}
\end{aligned}
\end{equation}
Similarly, the configuration of $e$ particles also has four scenarios similar to those in Fig. \ref{fig:fourclasses}: $1.$ $j_e$ odd, $k_{e1}$ odd, $k_{e2}$ even; $2.$ $j_e$ odd, $k_{e1}$ even, $k_{e2}$ odd; $3.$ $j_e$ even, $k_{e1}$ even, $k_{e2}$ even; $4.$ $j_m$ even, $k_{e1}$ odd, $k_{e2}$ odd. The complication this brings is not new states (new matrix elements in $\rho_A$) but affects the sign of each term. More explicitly, in this case, Eq. \eqref{C}  is no longer true because $\tilde{g}$ in general contains products of all $A_{v}$ operators in $B_1/B_2$. Therefore, we need to deal with the summation of $e$-particle configurations more carefully,
\begin{equation}
\begin{aligned}
\sum_{k_e,\{r_e\}}\beta_{e}^{2k_e}(-1)^{C(\tilde{g},\{r_{e}\})}=&\sum_{j_e=\text{ odd}}\bigg\{\sum_{\substack{k_{e1}=\text{ odd} \\k_{e2}=\text{ even}}} \sum_{\{r_{A}\}}\beta_e^{j_e+k_{e1}+k_{e2}}{M_{v1}\choose k_{e1}}{M_{v2}\choose k_{e2}}(-1)^{C(\tilde{g},\{r_{A}\})+Q_1(\tilde{g})}\\
&\quad\qquad +\sum_{\substack{k_{e1} \text{ even} \\k_{e2}\text{ odd}}} \sum_{\{r_{A}\}}\beta_r^{j_e+k_{e1}+k_{e2}}{M_{v1}\choose k_{e1}}{M_{v2}\choose k_{e2}}(-1)^{C(\tilde{g},\{r_{A}\})+Q_2(\tilde{g})}\bigg\}\\
+&\sum_{j_e=\text{ even}} \bigg\{\sum_{\substack{k_{e1}=\text{ even} \\k_{e2}=\text{ even}}} \sum_{\{r_{A}\}}\beta_e^{j_e+k_{e1}+k_{e2}}{M_{v1}\choose k_{e1}}{M_{v2}\choose k_{e2}}(-1)^{C(\tilde{g},\{r_{A}\})}\\
&\quad\qquad+\sum_{\substack{k_{e1}=\text{ odd} \\k_{e2}=\text{ odd}}} \sum_{\{r_{A}\}}\beta_e^{j_e+k_{e1}+k_{e2}}{M_{v1}\choose k_{e1}}{M_{v2}\choose k_{e2}}(-1)^{C(\tilde{g},\{r_{A}\})+Q_1(\tilde{g})+Q_2(\tilde{g})}\bigg\}.
\end{aligned}
\end{equation}

In the thermodynamic limit,
\begin{equation}
    \begin{aligned}
        \sum_{k_e,\{r_e\}}\beta_{e}^{2k_e}(-1)^{C(\tilde{g},\{r_e\})}=&\sum_{ j_e\text{ odd}}\sum_{\{r_{A}\}}\beta_e^{j_e}\frac{(1+\beta_e)^{M_{v1}}}{2}\frac{(1+\beta_e)^{M_{v2}}}{2}
(-1)^{C(\tilde{g},\{r_{A}\})}\left[(-1)^{Q_1(\tilde{g})}+(-1)^{Q_2(\tilde{g})}\right]\\
+&\sum_{ j_e\text{ even}}\sum_{\{r_{A}\}}\beta_e^{j_e}\frac{(1+\beta_e)^{M_{v1}}}{2}\frac{(1+\beta_e)^{M_{v2}}}{2}
(-1)^{C(\tilde{g},\{r_{A}\})}\left[1+(-1)^{Q_1(\tilde{g})+Q_2(\tilde{g})}\right]\\
=&\sum_{ j_e\text{ odd}}\sum_{\{r_{A}\}}\beta_e^{j_e}\frac{(1+\beta_e)^{n_v-m_v}}{4}
(-1)^{C(\tilde{g},\{r_{A}\})}\left[(-1)^{Q_1(\tilde{g})}+(-1)^{Q_2(\tilde{g})}\right]\\
+&\sum_{j_e\text{ even}}\sum_{\{r_{A}\}}\beta_e^{j_e}\frac{(1+\beta_e)^{n_v-m_v}}{4}
(-1)^{C(\tilde{g},\{r_{A}\})}\left[1+(-1)^{Q_1(\tilde{g})+Q_2(\tilde{g})}\right].
    \end{aligned}
\end{equation}
Substitute this into Eq. \eqref{rhoA} , and we get 

\begin{equation}
\begin{aligned}
\rho_A=\frac{|G_{\bar A}|}{4(1+\beta_e)^{m_v}(1+\beta_m)^{m_p}|G|}&\\
\sum_{\substack{g\in G/G_{\bar A}\\ \tilde{g}\in G_A}}
\sum_{ \substack{j_e\text{ odd}\\ \{r_{A}\} }}\beta_e^{j_e}(-1)^{C(\tilde{g},\{r_{A}\})}\left[(-1)^{Q_1(\tilde{g})}+(-1)^{Q_2(\tilde{g})}\right]&\sum_{ \substack{j_e\text{ even}\\ \{r_{A}\}}}\beta_e^{j_e}(-1)^{C(\tilde{g},\{r_{A}\})}\left[1+(-1)^{Q_1(\tilde{g})+Q_2(\tilde{g})}\right]  \\
g_A\bigg\{\sum_{\substack{j_m\text{ odd}\\p_{A}}}\sum_{s=s_1,s_2} \beta_m^{j_m} |m_{j_m}(\{p_A\}),s\rangle\langle m_{j_m}(\{p_A\}),s| 
+&\sum_{\substack{j_m\text{ even}\\p_{A}}} \sum_{s=s_3,s_4} \beta_m^{j_m} |m_{j_m}(\{p_A\}),s\rangle\langle m_{j_m}(\{p_A\}),s|\bigg\}
g_A\tilde{g}.\label{rhoA2}
\end{aligned}
\end{equation}
Finally, for partition-1, $S_{A}$ is
\begin{equation}
    \begin{aligned}
%         &\rho_A^n=\left[\frac{|G_A||G_{\bar A}|}{4(1+\beta_e)^{m_v}(1+\beta_m)^{m_p}|G|}\right]^{n-1}\frac{|G_{\bar A}|}{4(1+\beta_e)^{m_v}(1+\beta_m)^{m_p}|G|}
% \sum_{g\in G/G_{\bar A}}\sum_{\tilde{g}\in G_A} \bigg\{\sum_{ \substack{0\leq j_m\leq m_p\\j_m\text{ odd} } }\sum_{\{p_{mA}\}}\\
% &\beta_m^{j_m}(-1)^{C(\tilde{g},\{p_{mA}\})}\left[(-1)^{Q_1(\tilde{g})}+(-1)^{Q_2(\tilde{g})}\right]\sum_{ \substack{0\leq j_m\leq m_p\\j_m\text{ even} } }\sum_{\{p_{mA}\}}\beta_m^{j_m}(-1)^{C(\tilde{g},\{p_{mA}\})}\left[1+(-1)^{Q_1(\tilde{g})+Q_2(\tilde{g})}\right]\bigg\} g_A\\
% &\bigg\{\sum_{j_e\text{ odd}} \sum_{s=s_1,s_2} \sum_{\{r_A\}} \beta_e^{j_e} |e_{j_e}(\{r_A\}),s\rangle\langle e_{j_e}(\{r_A\}),s| +
% \sum_{j_e\text{ even}} \sum_{s=s_3,s_4} \sum_{\{r_A\}} \beta_e^{j_e} |e_{j_e}(\{r_A\}),s\rangle\langle e_{j_e}(\{r_A\}),s|\bigg\}
% g_A\tilde{g}\\
&\text{Tr}(\rho_A^n)=\left[\frac{|G_A||G_{\bar A}|}{4(1+\beta_e)^{m_v}(1+\beta_m)^{m_p}|G|}\right]^{n-1}
\left(\frac{1+\beta_e^n}{1+\beta_e}\right)^{m_v} \left(\frac{1+\beta_m^n}{1+\beta_m}\right)^{m_p},\\
& S_A=\lim_{n\rightarrow 1}\frac{1}{1-n}\log\text{Tr}(\rho_A^n)=\log\frac{|G|}{|G_A||G_{\bar A}|}+m_v f(\beta_e)+m_p f(\beta_m)-2\log 2.\\
    \end{aligned}
\end{equation}
Together with Eq. \eqref{Partition2}, we can easily get the topological entropy:
\begin{equation}
S_{\text{topo}}=2\log 2+(-m_{v1}+m_{v2}+m_{v3}-m_{v4})f(\beta_e)-\log 2+(-m_{p1}+m_{p2}+m_{p3}-m_{p4})f(\beta_m)-\log 2=0.
\end{equation}
That is, the topological entropy immediately drops to zero under perturbation on both gauge sectors in the thermodynamic limit, as expected from the fact that the topological degeneracy is also completely lifted.

\end{widetext}

\bibliographystyle{unsrt}
\bibliography{DTO}

\begin{thebibliography}{10}

\bibitem{wen1990topological}
Xiao-Gang Wen.
\newblock Topological orders in rigid states.
\newblock {\em International Journal of Modern Physics B}, 4(02):239--271,
  1990.

\bibitem{wen1990ground}
Xiao-Gang Wen and Qian Niu.
\newblock Ground-state degeneracy of the fractional quantum hall states in the
  presence of a random potential and on high-genus riemann surfaces.
\newblock {\em Physical Review B}, 41(13):9377, 1990.

\bibitem{wen2017colloquium}
Xiao-Gang Wen.
\newblock Colloquium: Zoo of quantum-topological phases of matter.
\newblock {\em Reviews of Modern Physics}, 89(4):041004, 2017.

\bibitem{trebst2007breakdown}
Simon Trebst, Philipp Werner, Matthias Troyer, Kirill Shtengel, and Chetan
  Nayak.
\newblock Breakdown of a topological phase: Quantum phase transition in a loop
  gas model with tension.
\newblock {\em Physical review letters}, 98(7):070602, 2007.

\bibitem{hamma2008adiabatic}
Alioscia Hamma and Daniel~A Lidar.
\newblock Adiabatic preparation of topological order.
\newblock {\em Physical review letters}, 100(3):030502, 2008.

\bibitem{tupitsyn2010topological}
IS~Tupitsyn, Alexei Kitaev, NV~Prokof’Ev, and PCE Stamp.
\newblock Topological multicritical point in the phase diagram of the toric
  code model and three-dimensional lattice gauge higgs model.
\newblock {\em Physical Review B}, 82(8):085114, 2010.

\bibitem{kitaev2006topological}
Alexei Kitaev and John Preskill.
\newblock Topological entanglement entropy.
\newblock {\em Physical review letters}, 96(11):110404, 2006.

\bibitem{levin2006detecting}
Michael Levin and Xiao-Gang Wen.
\newblock Detecting topological order in a ground state wave function.
\newblock {\em Physical review letters}, 96(11):110405, 2006.

\bibitem{kitaev2003fault}
A~Yu Kitaev.
\newblock Fault-tolerant quantum computation by anyons.
\newblock {\em Annals of Physics}, 303(1):2--30, 2003.

\bibitem{dennis2002topological}
Eric Dennis, Alexei Kitaev, Andrew Landahl, and John Preskill.
\newblock Topological quantum memory.
\newblock {\em Journal of Mathematical Physics}, 43(9):4452--4505, 2002.

\bibitem{fan2023diagnostics}
Ruihua Fan, Yimu Bao, Ehud Altman, and Ashvin Vishwanath.
\newblock Diagnostics of mixed-state topological order and breakdown of quantum
  memory.
\newblock {\em arXiv preprint arXiv:2301.05689}, 2023.

\bibitem{bao2023mixed}
Yimu Bao, Ruihua Fan, Ashvin Vishwanath, and Ehud Altman.
\newblock Mixed-state topological order and the errorfield double formulation
  of decoherence-induced transitions.
\newblock {\em arXiv preprint arXiv:2301.05687}, 2023.

\bibitem{lee2023quantum}
Jong~Yeon Lee, Chao-Ming Jian, and Cenke Xu.
\newblock Quantum criticality under decoherence or weak measurement.
\newblock {\em arXiv preprint arXiv:2301.05238}, 2023.

\bibitem{wang2023intrinsic}
Zijian Wang, Zhengzhi Wu, and Zhong Wang.
\newblock Intrinsic mixed-state topological order without quantum memory.
\newblock {\em arXiv preprint arXiv:2307.13758}, 2023.

\bibitem{kraus2008preparation}
Barbara Kraus, Hans~P B{\"u}chler, Sebastian Diehl, Adrian Kantian, Andrea
  Micheli, and Peter Zoller.
\newblock Preparation of entangled states by quantum markov processes.
\newblock {\em Physical Review A}, 78(4):042307, 2008.

\bibitem{diehl2010dissipation}
Sebastian Diehl, W~Yi, AJ~Daley, and P~Zoller.
\newblock Dissipation-induced d-wave pairing of fermionic atoms in an optical
  lattice.
\newblock {\em Physical review letters}, 105(22):227001, 2010.

\bibitem{diehl2011topology}
Sebastian Diehl, Enrique Rico, Mikhail~A Baranov, and Peter Zoller.
\newblock Topology by dissipation in atomic quantum wires.
\newblock {\em Nature Physics}, 7(12):971--977, 2011.

\bibitem{kastoryano2011dissipative}
Michael~James Kastoryano, Florentin Reiter, and Anders~S{\o}ndberg S{\o}rensen.
\newblock Dissipative preparation of entanglement in optical cavities.
\newblock {\em Physical review letters}, 106(9):090502, 2011.

\bibitem{reiter2016scalable}
Florentin Reiter, David Reeb, and Anders~S S{\o}rensen.
\newblock Scalable dissipative preparation of many-body entanglement.
\newblock {\em Physical review letters}, 117(4):040501, 2016.

\bibitem{wang2023topologically}
Zijian Wang, Xu-Dong Dai, He-Ran Wang, and Zhong Wang.
\newblock Topologically ordered steady states in open quantum systems.
\newblock {\em arXiv preprint arXiv:2306.12482}, 2023.

\bibitem{Note1}
Here we make the implicit assumption that that the system has translation
  symmetry, which excludes special cases with non-Hermitian skin effect, where
  there can be a finite Liouvillian gap even if the relaxation time is
  divergent. In other words, $\alpha $ in Fig.\ref {fig:degenerate_gapless} can
  be zero in such special cases.

\bibitem{hamma2005ground}
Alioscia Hamma, Radu Ionicioiu, and Paolo Zanardi.
\newblock Ground state entanglement and geometric entropy in the kitaev model.
\newblock {\em Physics Letters A}, 337(1-2):22--28, 2005.

\bibitem{hamma2005string}
Alioscia Hamma, Paolo Zanardi, and Xiao-Gang Wen.
\newblock String and membrane condensation on three-dimensional lattices.
\newblock {\em Physical Review B}, 72(3):035307, 2005.

\bibitem{nussinov2008autocorrelations}
Zohar Nussinov and Gerardo Ortiz.
\newblock Autocorrelations and thermal fragility of anyonic loops in
  topologically quantum ordered systems.
\newblock {\em Physical Review B}, 77(6):064302, 2008.

\bibitem{reiss2019quantum}
David~A Reiss and Kai~P Schmidt.
\newblock Quantum robustness and phase transitions of the 3d toric code in a
  field.
\newblock {\em SciPost Physics}, 6(6):078, 2019.

\bibitem{lindblad1976generators}
Goran Lindblad.
\newblock On the generators of quantum dynamical semigroups.
\newblock {\em Communications in Mathematical Physics}, 48(2):119--130, 1976.

\bibitem{glauber1963time}
Roy~J Glauber.
\newblock Time-dependent statistics of the ising model.
\newblock {\em Journal of mathematical physics}, 4(2):294--307, 1963.

\bibitem{castelnovo2007finiteT}
Claudio Castelnovo and Claudio Chamon.
\newblock Entanglement and topological entropy of the toric code at finite
  temperature.
\newblock {\em Physical Review B}, 76(18):184442, 2007.

\bibitem{castelnovo2007classical}
Claudio Castelnovo and Claudio Chamon.
\newblock Topological order and topological entropy in classical systems.
\newblock {\em Physical Review B}, 76(17):174416, 2007.

\bibitem{Note2}
$\lfloor x \rfloor =\protect \text {max}\{z\in Z|z\leq x \}$ and $\lceil x
  \rceil =\protect \text {min}\{z\in Z|z\geq x\}$.

\bibitem{verstraete2009quantum}
Frank Verstraete, Michael~M Wolf, and J~Ignacio~Cirac.
\newblock Quantum computation and quantum-state engineering driven by
  dissipation.
\newblock {\em Nature physics}, 5(9):633--636, 2009.

\bibitem{kogut1979introduction}
John~B Kogut.
\newblock An introduction to lattice gauge theory and spin systems.
\newblock {\em Reviews of Modern Physics}, 51(4):659, 1979.

\bibitem{savary2016quantum}
Lucile Savary and Leon Balents.
\newblock Quantum spin liquids: a review.
\newblock {\em Reports on Progress in Physics}, 80(1):016502, 2016.

\bibitem{gregor2011diagnosing}
K~Gregor, David~A Huse, R~Moessner, and Shivaji~Lal Sondhi.
\newblock Diagnosing deconfinement and topological order.
\newblock {\em New Journal of Physics}, 13(2):025009, 2011.

\bibitem{Note3}
Here we implicitly make the assumption that that the system has translation
  symmetry, which excludes special cases such as systems with boundary
  dissipation or skin effect.

\bibitem{henkel1984statistical}
Malte Henkel.
\newblock Statistical mechanics of the 2d quantum xy model in a transverse
  field.
\newblock {\em Journal of Physics A: Mathematical and General}, 17(14):L795,
  1984.

\bibitem{bray2002theory}
Alan~J Bray.
\newblock Theory of phase-ordering kinetics.
\newblock {\em Advances in Physics}, 51(2):481--587, 2002.

\bibitem{spirin2001freezing}
V~Spirin, PL~Krapivsky, and S~Redner.
\newblock Freezing in ising ferromagnets.
\newblock {\em Physical Review E}, 65(1):016119, 2001.

\bibitem{cai2013algebraic}
Zi~Cai and Thomas Barthel.
\newblock Algebraic versus exponential decoherence in dissipative many-particle
  systems.
\newblock {\em Phys. Rev. Lett.}, 111:150403, Oct 2013.

\end{thebibliography}
\end{document}